\documentclass[longauth]{aa}

\usepackage{graphicx}
\usepackage{natbib}

\usepackage[section]{placeins}

\usepackage{txfonts}
\usepackage{hyperref}
\hypersetup{
    colorlinks=true,
    linkcolor=blue,
    filecolor=magenta,      
    urlcolor=blue,
    citecolor=blue
}

\makeatletter
\renewcommand*\aa@pageof{, page \thepage{} of \pageref*{LastPage}}
\makeatother

\usepackage{euclid}
\usepackage{orcidlink}
\let\orcid\orcidlink

\newcommand{\Msun}{\ensuremath{{M_\odot}}\xspace}

\renewcommand*{\ha}{\ensuremath{\mathrm{H}\alpha}}
\newcommand{\ew}{\ensuremath{\rm EW(\ha)}\xspace}

\newcommand{\wl}{\ensuremath{\lambda}}
\newcommand{\dd}{{\rm d}}

\newcommand{\zstar}{\ensuremath{Z_\ast}\xspace}
\newcommand{\mstar}{\ensuremath{M_\ast}\xspace}
\newcommand{\mass}{\ensuremath{\mstar(t)}\xspace}
\newcommand{\metal}{\ensuremath{Z(t)}\xspace}
\newcommand{\massnormof}[1]{\ensuremath{\widetilde{M}_\ast(#1)}\xspace}
\newcommand{\massnorm}{\massnormof{t}}
\newcommand{\massnormssp}{\ensuremath{\widetilde{M}_{\rm SSP}}\xspace}
\newcommand{\balbreak}{\ensuremath{D_n(4000)}\xspace}

\newcommand{\re}{\ensuremath{R_{\rm e}}\xspace}

\newcommand{\sfr}{\ensuremath{{\rm SFR}}\xspace}
\newcommand{\ssfr}{\ensuremath{{\rm sSFR}}\xspace}

\newcommand{\avssfr}{\ensuremath{\ssfr_{\log(\tau)}}\xspace}
\newcommand{\avssfrval}[1]{\ensuremath{\ssfr_{#1}}\xspace}
\newcommand{\logavssfrval}[1]{\ensuremath{\log_{10}\left(\avssfrval{#1} / \rm yr^{-1}\right)}\xspace}

\newcommand{\besta}{\texttt{BESTA}\xspace}

\defcitealias{Casado+15}{C15}
\defcitealias{Corcho-Caballero+23a}{CC23a}
\defcitealias{Corcho-Caballero+23b}{CC23b}
\defcitealias{Corcho-Caballero+21a}{CC21a}
\defcitealias{Corcho-Caballero+21b}{CC21b}
\defcitealias{Corcho-Caballero+20}{CC20}

\begin{document}

%
%
\title{Euclid Quick Data Release (Q1)}
\subtitle{A probabilistic classification of quenched galaxies}
\author{Euclid Collaboration: P.~Corcho-Caballero\orcid{0000-0001-6327-7080}\thanks{\email{p.corcho.caballero@rug.nl}}\inst{\ref{aff1}}
\and Y.~Ascasibar\orcid{0000-0003-1577-2479}\inst{\ref{aff2},\ref{aff3}}
\and G.~Verdoes~Kleijn\orcid{0000-0001-5803-2580}\inst{\ref{aff1}}
\and C.~C.~Lovell\orcid{0000-0001-7964-5933}\inst{\ref{aff4}}
\and G.~De~Lucia\orcid{0000-0002-6220-9104}\inst{\ref{aff5}}
\and C.~Cleland\orcid{0009-0002-1769-1437}\inst{\ref{aff6}}
\and F.~Fontanot\orcid{0000-0003-4744-0188}\inst{\ref{aff5},\ref{aff7}}
\and C.~Tortora\orcid{0000-0001-7958-6531}\inst{\ref{aff8}}
\and L.~V.~E.~Koopmans\orcid{0000-0003-1840-0312}\inst{\ref{aff1}}
\and S.~Eales\inst{\ref{aff9}}
\and T.~Moutard\orcid{0000-0002-3305-9901}\inst{\ref{aff10}}
\and C.~Laigle\orcid{0009-0008-5926-818X}\inst{\ref{aff11}}
\and A.~Nersesian\orcid{0000-0001-6843-409X}\inst{\ref{aff12},\ref{aff13}}
\and F.~Shankar\orcid{0000-0001-8973-5051}\inst{\ref{aff14}}
\and M.~Dunn\orcid{0000-0001-5374-1644}\inst{\ref{aff15},\ref{aff16},\ref{aff17}}
\and N.~Aghanim\orcid{0000-0002-6688-8992}\inst{\ref{aff18}}
\and B.~Altieri\orcid{0000-0003-3936-0284}\inst{\ref{aff10}}
\and A.~Amara\inst{\ref{aff19}}
\and S.~Andreon\orcid{0000-0002-2041-8784}\inst{\ref{aff20}}
\and H.~Aussel\orcid{0000-0002-1371-5705}\inst{\ref{aff21}}
\and C.~Baccigalupi\orcid{0000-0002-8211-1630}\inst{\ref{aff7},\ref{aff5},\ref{aff22},\ref{aff23}}
\and M.~Baldi\orcid{0000-0003-4145-1943}\inst{\ref{aff24},\ref{aff25},\ref{aff26}}
\and A.~Balestra\orcid{0000-0002-6967-261X}\inst{\ref{aff27}}
\and S.~Bardelli\orcid{0000-0002-8900-0298}\inst{\ref{aff25}}
\and P.~Battaglia\orcid{0000-0002-7337-5909}\inst{\ref{aff25}}
\and A.~Biviano\orcid{0000-0002-0857-0732}\inst{\ref{aff5},\ref{aff7}}
\and A.~Bonchi\orcid{0000-0002-2667-5482}\inst{\ref{aff28}}
\and D.~Bonino\orcid{0000-0002-3336-9977}\inst{\ref{aff29}}
\and E.~Branchini\orcid{0000-0002-0808-6908}\inst{\ref{aff30},\ref{aff31},\ref{aff20}}
\and M.~Brescia\orcid{0000-0001-9506-5680}\inst{\ref{aff32},\ref{aff8}}
\and J.~Brinchmann\orcid{0000-0003-4359-8797}\inst{\ref{aff33},\ref{aff34}}
\and G.~Ca\~nas-Herrera\orcid{0000-0003-2796-2149}\inst{\ref{aff35},\ref{aff36},\ref{aff37}}
\and V.~Capobianco\orcid{0000-0002-3309-7692}\inst{\ref{aff29}}
\and C.~Carbone\orcid{0000-0003-0125-3563}\inst{\ref{aff38}}
\and J.~Carretero\orcid{0000-0002-3130-0204}\inst{\ref{aff39},\ref{aff40}}
\and S.~Casas\orcid{0000-0002-4751-5138}\inst{\ref{aff41}}
\and F.~J.~Castander\orcid{0000-0001-7316-4573}\inst{\ref{aff42},\ref{aff43}}
\and M.~Castellano\orcid{0000-0001-9875-8263}\inst{\ref{aff44}}
\and G.~Castignani\orcid{0000-0001-6831-0687}\inst{\ref{aff25}}
\and S.~Cavuoti\orcid{0000-0002-3787-4196}\inst{\ref{aff8},\ref{aff45}}
\and K.~C.~Chambers\orcid{0000-0001-6965-7789}\inst{\ref{aff46}}
\and A.~Cimatti\inst{\ref{aff47}}
\and C.~Colodro-Conde\inst{\ref{aff15}}
\and G.~Congedo\orcid{0000-0003-2508-0046}\inst{\ref{aff48}}
\and C.~J.~Conselice\orcid{0000-0003-1949-7638}\inst{\ref{aff49}}
\and L.~Conversi\orcid{0000-0002-6710-8476}\inst{\ref{aff50},\ref{aff10}}
\and Y.~Copin\orcid{0000-0002-5317-7518}\inst{\ref{aff51}}
\and A.~Costille\inst{\ref{aff52}}
\and F.~Courbin\orcid{0000-0003-0758-6510}\inst{\ref{aff53},\ref{aff54}}
\and H.~M.~Courtois\orcid{0000-0003-0509-1776}\inst{\ref{aff55}}
\and M.~Cropper\orcid{0000-0003-4571-9468}\inst{\ref{aff56}}
\and A.~Da~Silva\orcid{0000-0002-6385-1609}\inst{\ref{aff57},\ref{aff58}}
\and H.~Degaudenzi\orcid{0000-0002-5887-6799}\inst{\ref{aff59}}
\and A.~M.~Di~Giorgio\orcid{0000-0002-4767-2360}\inst{\ref{aff60}}
\and C.~Dolding\orcid{0009-0003-7199-6108}\inst{\ref{aff56}}
\and H.~Dole\orcid{0000-0002-9767-3839}\inst{\ref{aff18}}
\and F.~Dubath\orcid{0000-0002-6533-2810}\inst{\ref{aff59}}
\and X.~Dupac\inst{\ref{aff10}}
\and A.~Ealet\orcid{0000-0003-3070-014X}\inst{\ref{aff51}}
\and S.~Escoffier\orcid{0000-0002-2847-7498}\inst{\ref{aff61}}
\and M.~Farina\orcid{0000-0002-3089-7846}\inst{\ref{aff60}}
\and R.~Farinelli\inst{\ref{aff25}}
\and F.~Faustini\orcid{0000-0001-6274-5145}\inst{\ref{aff28},\ref{aff44}}
\and S.~Ferriol\inst{\ref{aff51}}
\and F.~Finelli\orcid{0000-0002-6694-3269}\inst{\ref{aff25},\ref{aff62}}
\and S.~Fotopoulou\orcid{0000-0002-9686-254X}\inst{\ref{aff63}}
\and M.~Frailis\orcid{0000-0002-7400-2135}\inst{\ref{aff5}}
\and E.~Franceschi\orcid{0000-0002-0585-6591}\inst{\ref{aff25}}
\and M.~Fumana\orcid{0000-0001-6787-5950}\inst{\ref{aff38}}
\and S.~Galeotta\orcid{0000-0002-3748-5115}\inst{\ref{aff5}}
\and K.~George\orcid{0000-0002-1734-8455}\inst{\ref{aff64}}
\and B.~Gillis\orcid{0000-0002-4478-1270}\inst{\ref{aff48}}
\and C.~Giocoli\orcid{0000-0002-9590-7961}\inst{\ref{aff25},\ref{aff26}}
\and J.~Gracia-Carpio\inst{\ref{aff65}}
\and B.~R.~Granett\orcid{0000-0003-2694-9284}\inst{\ref{aff20}}
\and A.~Grazian\orcid{0000-0002-5688-0663}\inst{\ref{aff27}}
\and F.~Grupp\inst{\ref{aff65},\ref{aff64}}
\and L.~Guzzo\orcid{0000-0001-8264-5192}\inst{\ref{aff66},\ref{aff20},\ref{aff67}}
\and S.~Gwyn\orcid{0000-0001-8221-8406}\inst{\ref{aff68}}
\and S.~V.~H.~Haugan\orcid{0000-0001-9648-7260}\inst{\ref{aff69}}
\and W.~Holmes\inst{\ref{aff70}}
\and I.~M.~Hook\orcid{0000-0002-2960-978X}\inst{\ref{aff71}}
\and F.~Hormuth\inst{\ref{aff72}}
\and A.~Hornstrup\orcid{0000-0002-3363-0936}\inst{\ref{aff73},\ref{aff74}}
\and P.~Hudelot\inst{\ref{aff11}}
\and K.~Jahnke\orcid{0000-0003-3804-2137}\inst{\ref{aff75}}
\and M.~Jhabvala\inst{\ref{aff76}}
\and E.~Keih\"anen\orcid{0000-0003-1804-7715}\inst{\ref{aff77}}
\and S.~Kermiche\orcid{0000-0002-0302-5735}\inst{\ref{aff61}}
\and A.~Kiessling\orcid{0000-0002-2590-1273}\inst{\ref{aff70}}
\and B.~Kubik\orcid{0009-0006-5823-4880}\inst{\ref{aff51}}
\and K.~Kuijken\orcid{0000-0002-3827-0175}\inst{\ref{aff37}}
\and M.~K\"ummel\orcid{0000-0003-2791-2117}\inst{\ref{aff64}}
\and M.~Kunz\orcid{0000-0002-3052-7394}\inst{\ref{aff78}}
\and H.~Kurki-Suonio\orcid{0000-0002-4618-3063}\inst{\ref{aff79},\ref{aff80}}
\and Q.~Le~Boulc'h\inst{\ref{aff81}}
\and A.~M.~C.~Le~Brun\orcid{0000-0002-0936-4594}\inst{\ref{aff82}}
\and D.~Le~Mignant\orcid{0000-0002-5339-5515}\inst{\ref{aff52}}
\and S.~Ligori\orcid{0000-0003-4172-4606}\inst{\ref{aff29}}
\and P.~B.~Lilje\orcid{0000-0003-4324-7794}\inst{\ref{aff69}}
\and V.~Lindholm\orcid{0000-0003-2317-5471}\inst{\ref{aff79},\ref{aff80}}
\and I.~Lloro\orcid{0000-0001-5966-1434}\inst{\ref{aff83}}
\and G.~Mainetti\orcid{0000-0003-2384-2377}\inst{\ref{aff81}}
\and D.~Maino\inst{\ref{aff66},\ref{aff38},\ref{aff67}}
\and E.~Maiorano\orcid{0000-0003-2593-4355}\inst{\ref{aff25}}
\and O.~Mansutti\orcid{0000-0001-5758-4658}\inst{\ref{aff5}}
\and S.~Marcin\inst{\ref{aff84}}
\and O.~Marggraf\orcid{0000-0001-7242-3852}\inst{\ref{aff85}}
\and M.~Martinelli\orcid{0000-0002-6943-7732}\inst{\ref{aff44},\ref{aff86}}
\and N.~Martinet\orcid{0000-0003-2786-7790}\inst{\ref{aff52}}
\and F.~Marulli\orcid{0000-0002-8850-0303}\inst{\ref{aff87},\ref{aff25},\ref{aff26}}
\and R.~Massey\orcid{0000-0002-6085-3780}\inst{\ref{aff88}}
\and S.~Maurogordato\inst{\ref{aff89}}
\and E.~Medinaceli\orcid{0000-0002-4040-7783}\inst{\ref{aff25}}
\and S.~Mei\orcid{0000-0002-2849-559X}\inst{\ref{aff6},\ref{aff90}}
\and M.~Melchior\inst{\ref{aff91}}
\and Y.~Mellier\inst{\ref{aff92},\ref{aff11}}
\and M.~Meneghetti\orcid{0000-0003-1225-7084}\inst{\ref{aff25},\ref{aff26}}
\and E.~Merlin\orcid{0000-0001-6870-8900}\inst{\ref{aff44}}
\and G.~Meylan\inst{\ref{aff93}}
\and A.~Mora\orcid{0000-0002-1922-8529}\inst{\ref{aff94}}
\and M.~Moresco\orcid{0000-0002-7616-7136}\inst{\ref{aff87},\ref{aff25}}
\and L.~Moscardini\orcid{0000-0002-3473-6716}\inst{\ref{aff87},\ref{aff25},\ref{aff26}}
\and R.~Nakajima\orcid{0009-0009-1213-7040}\inst{\ref{aff85}}
\and C.~Neissner\orcid{0000-0001-8524-4968}\inst{\ref{aff95},\ref{aff40}}
\and S.-M.~Niemi\inst{\ref{aff35}}
\and J.~W.~Nightingale\orcid{0000-0002-8987-7401}\inst{\ref{aff96}}
\and C.~Padilla\orcid{0000-0001-7951-0166}\inst{\ref{aff95}}
\and S.~Paltani\orcid{0000-0002-8108-9179}\inst{\ref{aff59}}
\and F.~Pasian\orcid{0000-0002-4869-3227}\inst{\ref{aff5}}
\and W.~J.~Percival\orcid{0000-0002-0644-5727}\inst{\ref{aff97},\ref{aff98},\ref{aff99}}
\and V.~Pettorino\inst{\ref{aff35}}
\and G.~Polenta\orcid{0000-0003-4067-9196}\inst{\ref{aff28}}
\and M.~Poncet\inst{\ref{aff100}}
\and L.~A.~Popa\inst{\ref{aff101}}
\and L.~Pozzetti\orcid{0000-0001-7085-0412}\inst{\ref{aff25}}
\and F.~Raison\orcid{0000-0002-7819-6918}\inst{\ref{aff65}}
\and R.~Rebolo\orcid{0000-0003-3767-7085}\inst{\ref{aff15},\ref{aff102},\ref{aff16}}
\and A.~Renzi\orcid{0000-0001-9856-1970}\inst{\ref{aff103},\ref{aff104}}
\and J.~Rhodes\orcid{0000-0002-4485-8549}\inst{\ref{aff70}}
\and G.~Riccio\inst{\ref{aff8}}
\and E.~Romelli\orcid{0000-0003-3069-9222}\inst{\ref{aff5}}
\and M.~Roncarelli\orcid{0000-0001-9587-7822}\inst{\ref{aff25}}
\and R.~Saglia\orcid{0000-0003-0378-7032}\inst{\ref{aff64},\ref{aff65}}
\and Z.~Sakr\orcid{0000-0002-4823-3757}\inst{\ref{aff105},\ref{aff106},\ref{aff107}}
\and A.~G.~S\'anchez\orcid{0000-0003-1198-831X}\inst{\ref{aff65}}
\and D.~Sapone\orcid{0000-0001-7089-4503}\inst{\ref{aff108}}
\and B.~Sartoris\orcid{0000-0003-1337-5269}\inst{\ref{aff64},\ref{aff5}}
\and J.~A.~Schewtschenko\orcid{0000-0002-4913-6393}\inst{\ref{aff48}}
\and P.~Schneider\orcid{0000-0001-8561-2679}\inst{\ref{aff85}}
\and T.~Schrabback\orcid{0000-0002-6987-7834}\inst{\ref{aff109}}
\and M.~Scodeggio\inst{\ref{aff38}}
\and A.~Secroun\orcid{0000-0003-0505-3710}\inst{\ref{aff61}}
\and G.~Seidel\orcid{0000-0003-2907-353X}\inst{\ref{aff75}}
\and S.~Serrano\orcid{0000-0002-0211-2861}\inst{\ref{aff43},\ref{aff110},\ref{aff42}}
\and P.~Simon\inst{\ref{aff85}}
\and C.~Sirignano\orcid{0000-0002-0995-7146}\inst{\ref{aff103},\ref{aff104}}
\and G.~Sirri\orcid{0000-0003-2626-2853}\inst{\ref{aff26}}
\and L.~Stanco\orcid{0000-0002-9706-5104}\inst{\ref{aff104}}
\and J.~Steinwagner\orcid{0000-0001-7443-1047}\inst{\ref{aff65}}
\and P.~Tallada-Cresp\'{i}\orcid{0000-0002-1336-8328}\inst{\ref{aff39},\ref{aff40}}
\and A.~N.~Taylor\inst{\ref{aff48}}
\and I.~Tereno\inst{\ref{aff57},\ref{aff111}}
\and N.~Tessore\orcid{0000-0002-9696-7931}\inst{\ref{aff112}}
\and S.~Toft\orcid{0000-0003-3631-7176}\inst{\ref{aff113},\ref{aff114}}
\and R.~Toledo-Moreo\orcid{0000-0002-2997-4859}\inst{\ref{aff115}}
\and F.~Torradeflot\orcid{0000-0003-1160-1517}\inst{\ref{aff40},\ref{aff39}}
\and I.~Tutusaus\orcid{0000-0002-3199-0399}\inst{\ref{aff106}}
\and L.~Valenziano\orcid{0000-0002-1170-0104}\inst{\ref{aff25},\ref{aff62}}
\and J.~Valiviita\orcid{0000-0001-6225-3693}\inst{\ref{aff79},\ref{aff80}}
\and T.~Vassallo\orcid{0000-0001-6512-6358}\inst{\ref{aff64},\ref{aff5}}
\and A.~Veropalumbo\orcid{0000-0003-2387-1194}\inst{\ref{aff20},\ref{aff31},\ref{aff30}}
\and Y.~Wang\orcid{0000-0002-4749-2984}\inst{\ref{aff116}}
\and J.~Weller\orcid{0000-0002-8282-2010}\inst{\ref{aff64},\ref{aff65}}
\and A.~Zacchei\orcid{0000-0003-0396-1192}\inst{\ref{aff5},\ref{aff7}}
\and G.~Zamorani\orcid{0000-0002-2318-301X}\inst{\ref{aff25}}
\and F.~M.~Zerbi\inst{\ref{aff20}}
\and I.~A.~Zinchenko\orcid{0000-0002-2944-2449}\inst{\ref{aff64}}
\and E.~Zucca\orcid{0000-0002-5845-8132}\inst{\ref{aff25}}
\and V.~Allevato\orcid{0000-0001-7232-5152}\inst{\ref{aff8}}
\and M.~Ballardini\orcid{0000-0003-4481-3559}\inst{\ref{aff117},\ref{aff118},\ref{aff25}}
\and M.~Bolzonella\orcid{0000-0003-3278-4607}\inst{\ref{aff25}}
\and E.~Bozzo\orcid{0000-0002-8201-1525}\inst{\ref{aff59}}
\and C.~Burigana\orcid{0000-0002-3005-5796}\inst{\ref{aff119},\ref{aff62}}
\and R.~Cabanac\orcid{0000-0001-6679-2600}\inst{\ref{aff106}}
\and A.~Cappi\inst{\ref{aff25},\ref{aff89}}
\and D.~Di~Ferdinando\inst{\ref{aff26}}
\and J.~A.~Escartin~Vigo\inst{\ref{aff65}}
\and L.~Gabarra\orcid{0000-0002-8486-8856}\inst{\ref{aff120}}
\and M.~Huertas-Company\orcid{0000-0002-1416-8483}\inst{\ref{aff15},\ref{aff17},\ref{aff121},\ref{aff122}}
\and J.~Mart\'{i}n-Fleitas\orcid{0000-0002-8594-569X}\inst{\ref{aff94}}
\and S.~Matthew\orcid{0000-0001-8448-1697}\inst{\ref{aff48}}
\and N.~Mauri\orcid{0000-0001-8196-1548}\inst{\ref{aff47},\ref{aff26}}
\and R.~B.~Metcalf\orcid{0000-0003-3167-2574}\inst{\ref{aff87},\ref{aff25}}
\and A.~Pezzotta\orcid{0000-0003-0726-2268}\inst{\ref{aff123},\ref{aff65}}
\and M.~P\"ontinen\orcid{0000-0001-5442-2530}\inst{\ref{aff79}}
\and C.~Porciani\orcid{0000-0002-7797-2508}\inst{\ref{aff85}}
\and I.~Risso\orcid{0000-0003-2525-7761}\inst{\ref{aff124}}
\and V.~Scottez\inst{\ref{aff92},\ref{aff125}}
\and M.~Sereno\orcid{0000-0003-0302-0325}\inst{\ref{aff25},\ref{aff26}}
\and M.~Tenti\orcid{0000-0002-4254-5901}\inst{\ref{aff26}}
\and M.~Viel\orcid{0000-0002-2642-5707}\inst{\ref{aff7},\ref{aff5},\ref{aff23},\ref{aff22},\ref{aff126}}
\and M.~Wiesmann\orcid{0009-0000-8199-5860}\inst{\ref{aff69}}
\and Y.~Akrami\orcid{0000-0002-2407-7956}\inst{\ref{aff127},\ref{aff128}}
\and S.~Alvi\orcid{0000-0001-5779-8568}\inst{\ref{aff117}}
\and I.~T.~Andika\orcid{0000-0001-6102-9526}\inst{\ref{aff129},\ref{aff130}}
\and S.~Anselmi\orcid{0000-0002-3579-9583}\inst{\ref{aff104},\ref{aff103},\ref{aff131}}
\and M.~Archidiacono\orcid{0000-0003-4952-9012}\inst{\ref{aff66},\ref{aff67}}
\and F.~Atrio-Barandela\orcid{0000-0002-2130-2513}\inst{\ref{aff132}}
\and K.~Benson\inst{\ref{aff56}}
\and D.~Bertacca\orcid{0000-0002-2490-7139}\inst{\ref{aff103},\ref{aff27},\ref{aff104}}
\and M.~Bethermin\orcid{0000-0002-3915-2015}\inst{\ref{aff133}}
\and A.~Blanchard\orcid{0000-0001-8555-9003}\inst{\ref{aff106}}
\and L.~Blot\orcid{0000-0002-9622-7167}\inst{\ref{aff134},\ref{aff131}}
\and S.~Borgani\orcid{0000-0001-6151-6439}\inst{\ref{aff135},\ref{aff7},\ref{aff5},\ref{aff22},\ref{aff126}}
\and M.~L.~Brown\orcid{0000-0002-0370-8077}\inst{\ref{aff49}}
\and S.~Bruton\orcid{0000-0002-6503-5218}\inst{\ref{aff136}}
\and A.~Calabro\orcid{0000-0003-2536-1614}\inst{\ref{aff44}}
\and F.~Caro\inst{\ref{aff44}}
\and C.~S.~Carvalho\inst{\ref{aff111}}
\and T.~Castro\orcid{0000-0002-6292-3228}\inst{\ref{aff5},\ref{aff22},\ref{aff7},\ref{aff126}}
\and F.~Cogato\orcid{0000-0003-4632-6113}\inst{\ref{aff87},\ref{aff25}}
\and A.~R.~Cooray\orcid{0000-0002-3892-0190}\inst{\ref{aff137}}
\and O.~Cucciati\orcid{0000-0002-9336-7551}\inst{\ref{aff25}}
\and S.~Davini\orcid{0000-0003-3269-1718}\inst{\ref{aff31}}
\and F.~De~Paolis\orcid{0000-0001-6460-7563}\inst{\ref{aff138},\ref{aff139},\ref{aff140}}
\and G.~Desprez\orcid{0000-0001-8325-1742}\inst{\ref{aff1}}
\and A.~D\'iaz-S\'anchez\orcid{0000-0003-0748-4768}\inst{\ref{aff141}}
\and J.~J.~Diaz\inst{\ref{aff17}}
\and S.~Di~Domizio\orcid{0000-0003-2863-5895}\inst{\ref{aff30},\ref{aff31}}
\and J.~M.~Diego\orcid{0000-0001-9065-3926}\inst{\ref{aff142}}
\and A.~Enia\orcid{0000-0002-0200-2857}\inst{\ref{aff24},\ref{aff25}}
\and Y.~Fang\inst{\ref{aff64}}
\and A.~G.~Ferrari\orcid{0009-0005-5266-4110}\inst{\ref{aff26}}
\and P.~G.~Ferreira\orcid{0000-0002-3021-2851}\inst{\ref{aff120}}
\and A.~Finoguenov\orcid{0000-0002-4606-5403}\inst{\ref{aff79}}
\and A.~Fontana\orcid{0000-0003-3820-2823}\inst{\ref{aff44}}
\and A.~Franco\orcid{0000-0002-4761-366X}\inst{\ref{aff139},\ref{aff138},\ref{aff140}}
\and K.~Ganga\orcid{0000-0001-8159-8208}\inst{\ref{aff6}}
\and J.~Garc\'ia-Bellido\orcid{0000-0002-9370-8360}\inst{\ref{aff127}}
\and T.~Gasparetto\orcid{0000-0002-7913-4866}\inst{\ref{aff5}}
\and V.~Gautard\inst{\ref{aff143}}
\and E.~Gaztanaga\orcid{0000-0001-9632-0815}\inst{\ref{aff42},\ref{aff43},\ref{aff4}}
\and F.~Giacomini\orcid{0000-0002-3129-2814}\inst{\ref{aff26}}
\and F.~Gianotti\orcid{0000-0003-4666-119X}\inst{\ref{aff25}}
\and G.~Gozaliasl\orcid{0000-0002-0236-919X}\inst{\ref{aff144},\ref{aff79}}
\and M.~Guidi\orcid{0000-0001-9408-1101}\inst{\ref{aff24},\ref{aff25}}
\and C.~M.~Gutierrez\orcid{0000-0001-7854-783X}\inst{\ref{aff145}}
\and A.~Hall\orcid{0000-0002-3139-8651}\inst{\ref{aff48}}
\and W.~G.~Hartley\inst{\ref{aff59}}
\and S.~Hemmati\orcid{0000-0003-2226-5395}\inst{\ref{aff146}}
\and H.~Hildebrandt\orcid{0000-0002-9814-3338}\inst{\ref{aff147}}
\and J.~Hjorth\orcid{0000-0002-4571-2306}\inst{\ref{aff148}}
\and J.~J.~E.~Kajava\orcid{0000-0002-3010-8333}\inst{\ref{aff149},\ref{aff150}}
\and Y.~Kang\orcid{0009-0000-8588-7250}\inst{\ref{aff59}}
\and V.~Kansal\orcid{0000-0002-4008-6078}\inst{\ref{aff151},\ref{aff152}}
\and D.~Karagiannis\orcid{0000-0002-4927-0816}\inst{\ref{aff117},\ref{aff153}}
\and K.~Kiiveri\inst{\ref{aff77}}
\and C.~C.~Kirkpatrick\inst{\ref{aff77}}
\and S.~Kruk\orcid{0000-0001-8010-8879}\inst{\ref{aff10}}
\and J.~Le~Graet\orcid{0000-0001-6523-7971}\inst{\ref{aff61}}
\and L.~Legrand\orcid{0000-0003-0610-5252}\inst{\ref{aff154},\ref{aff155}}
\and M.~Lembo\orcid{0000-0002-5271-5070}\inst{\ref{aff117},\ref{aff118}}
\and F.~Lepori\orcid{0009-0000-5061-7138}\inst{\ref{aff156}}
\and G.~Leroy\orcid{0009-0004-2523-4425}\inst{\ref{aff157},\ref{aff88}}
\and G.~F.~Lesci\orcid{0000-0002-4607-2830}\inst{\ref{aff87},\ref{aff25}}
\and J.~Lesgourgues\orcid{0000-0001-7627-353X}\inst{\ref{aff41}}
\and L.~Leuzzi\orcid{0009-0006-4479-7017}\inst{\ref{aff87},\ref{aff25}}
\and T.~I.~Liaudat\orcid{0000-0002-9104-314X}\inst{\ref{aff158}}
\and S.~J.~Liu\orcid{0000-0001-7680-2139}\inst{\ref{aff60}}
\and A.~Loureiro\orcid{0000-0002-4371-0876}\inst{\ref{aff159},\ref{aff160}}
\and J.~Macias-Perez\orcid{0000-0002-5385-2763}\inst{\ref{aff161}}
\and G.~Maggio\orcid{0000-0003-4020-4836}\inst{\ref{aff5}}
\and M.~Magliocchetti\orcid{0000-0001-9158-4838}\inst{\ref{aff60}}
\and E.~A.~Magnier\orcid{0000-0002-7965-2815}\inst{\ref{aff46}}
\and C.~Mancini\orcid{0000-0002-4297-0561}\inst{\ref{aff38}}
\and F.~Mannucci\orcid{0000-0002-4803-2381}\inst{\ref{aff162}}
\and R.~Maoli\orcid{0000-0002-6065-3025}\inst{\ref{aff163},\ref{aff44}}
\and C.~J.~A.~P.~Martins\orcid{0000-0002-4886-9261}\inst{\ref{aff164},\ref{aff33}}
\and L.~Maurin\orcid{0000-0002-8406-0857}\inst{\ref{aff18}}
\and M.~Miluzio\inst{\ref{aff10},\ref{aff165}}
\and P.~Monaco\orcid{0000-0003-2083-7564}\inst{\ref{aff135},\ref{aff5},\ref{aff22},\ref{aff7}}
\and C.~Moretti\orcid{0000-0003-3314-8936}\inst{\ref{aff23},\ref{aff126},\ref{aff5},\ref{aff7},\ref{aff22}}
\and G.~Morgante\inst{\ref{aff25}}
\and K.~Naidoo\orcid{0000-0002-9182-1802}\inst{\ref{aff4}}
\and A.~Navarro-Alsina\orcid{0000-0002-3173-2592}\inst{\ref{aff85}}
\and S.~Nesseris\orcid{0000-0002-0567-0324}\inst{\ref{aff127}}
\and F.~Passalacqua\orcid{0000-0002-8606-4093}\inst{\ref{aff103},\ref{aff104}}
\and K.~Paterson\orcid{0000-0001-8340-3486}\inst{\ref{aff75}}
\and L.~Patrizii\inst{\ref{aff26}}
\and A.~Pisani\orcid{0000-0002-6146-4437}\inst{\ref{aff61},\ref{aff166}}
\and D.~Potter\orcid{0000-0002-0757-5195}\inst{\ref{aff156}}
\and S.~Quai\orcid{0000-0002-0449-8163}\inst{\ref{aff87},\ref{aff25}}
\and M.~Radovich\orcid{0000-0002-3585-866X}\inst{\ref{aff27}}
\and P.-F.~Rocci\inst{\ref{aff18}}
\and S.~Sacquegna\orcid{0000-0002-8433-6630}\inst{\ref{aff138},\ref{aff139},\ref{aff140}}
\and M.~Sahl\'en\orcid{0000-0003-0973-4804}\inst{\ref{aff167}}
\and D.~B.~Sanders\orcid{0000-0002-1233-9998}\inst{\ref{aff46}}
\and E.~Sarpa\orcid{0000-0002-1256-655X}\inst{\ref{aff23},\ref{aff126},\ref{aff22}}
\and C.~Scarlata\orcid{0000-0002-9136-8876}\inst{\ref{aff168}}
\and J.~Schaye\orcid{0000-0002-0668-5560}\inst{\ref{aff37}}
\and A.~Schneider\orcid{0000-0001-7055-8104}\inst{\ref{aff156}}
\and D.~Sciotti\orcid{0009-0008-4519-2620}\inst{\ref{aff44},\ref{aff86}}
\and E.~Sellentin\inst{\ref{aff169},\ref{aff37}}
\and L.~C.~Smith\orcid{0000-0002-3259-2771}\inst{\ref{aff170}}
\and S.~A.~Stanford\orcid{0000-0003-0122-0841}\inst{\ref{aff171}}
\and K.~Tanidis\orcid{0000-0001-9843-5130}\inst{\ref{aff120}}
\and G.~Testera\inst{\ref{aff31}}
\and R.~Teyssier\orcid{0000-0001-7689-0933}\inst{\ref{aff166}}
\and S.~Tosi\orcid{0000-0002-7275-9193}\inst{\ref{aff30},\ref{aff31},\ref{aff20}}
\and A.~Troja\orcid{0000-0003-0239-4595}\inst{\ref{aff103},\ref{aff104}}
\and M.~Tucci\inst{\ref{aff59}}
\and C.~Valieri\inst{\ref{aff26}}
\and A.~Venhola\orcid{0000-0001-6071-4564}\inst{\ref{aff172}}
\and D.~Vergani\orcid{0000-0003-0898-2216}\inst{\ref{aff25}}
\and G.~Verza\orcid{0000-0002-1886-8348}\inst{\ref{aff173}}
\and P.~Vielzeuf\orcid{0000-0003-2035-9339}\inst{\ref{aff61}}
\and N.~A.~Walton\orcid{0000-0003-3983-8778}\inst{\ref{aff170}}
\and J.~R.~Weaver\orcid{0000-0003-1614-196X}\inst{\ref{aff174}}
\and J.~G.~Sorce\orcid{0000-0002-2307-2432}\inst{\ref{aff175},\ref{aff18}}}
                                                                                   
\institute{Kapteyn Astronomical Institute, University of Groningen, PO Box 800, 9700 AV Groningen, The Netherlands\label{aff1}
\and
Departamento de F\'isica Te\'orica, Facultad de Ciencias, Universidad Aut\'onoma de Madrid, 28049 Cantoblanco, Madrid, Spain\label{aff2}
\and
Centro de Investigaci\'{o}n Avanzada en F\'isica Fundamental (CIAFF), Facultad de Ciencias, Universidad Aut\'{o}noma de Madrid, 28049 Madrid, Spain\label{aff3}
\and
Institute of Cosmology and Gravitation, University of Portsmouth, Portsmouth PO1 3FX, UK\label{aff4}
\and
INAF-Osservatorio Astronomico di Trieste, Via G. B. Tiepolo 11, 34143 Trieste, Italy\label{aff5}
\and
Universit\'e Paris Cit\'e, CNRS, Astroparticule et Cosmologie, 75013 Paris, France\label{aff6}
\and
IFPU, Institute for Fundamental Physics of the Universe, via Beirut 2, 34151 Trieste, Italy\label{aff7}
\and
INAF-Osservatorio Astronomico di Capodimonte, Via Moiariello 16, 80131 Napoli, Italy\label{aff8}
\and
School of Physics and Astronomy, Cardiff University, The Parade, Cardiff, CF24 3AA, UK\label{aff9}
\and
ESAC/ESA, Camino Bajo del Castillo, s/n., Urb. Villafranca del Castillo, 28692 Villanueva de la Ca\~nada, Madrid, Spain\label{aff10}
\and
Institut d'Astrophysique de Paris, UMR 7095, CNRS, and Sorbonne Universit\'e, 98 bis boulevard Arago, 75014 Paris, France\label{aff11}
\and
STAR Institute, University of Li{\`e}ge, Quartier Agora, All\'ee du six Ao\^ut 19c, 4000 Li\`ege, Belgium\label{aff12}
\and
Sterrenkundig Observatorium, Universiteit Gent, Krijgslaan 281 S9, 9000 Gent, Belgium\label{aff13}
\and
School of Physics \& Astronomy, University of Southampton, Highfield Campus, Southampton SO17 1BJ, UK\label{aff14}
\and
Instituto de Astrof\'{\i}sica de Canarias, V\'{\i}a L\'actea, 38205 La Laguna, Tenerife, Spain\label{aff15}
\and
Universidad de La Laguna, Departamento de Astrof\'{\i}sica, 38206 La Laguna, Tenerife, Spain\label{aff16}
\and
Instituto de Astrof\'isica de Canarias (IAC); Departamento de Astrof\'isica, Universidad de La Laguna (ULL), 38200, La Laguna, Tenerife, Spain\label{aff17}
\and
Universit\'e Paris-Saclay, CNRS, Institut d'astrophysique spatiale, 91405, Orsay, France\label{aff18}
\and
School of Mathematics and Physics, University of Surrey, Guildford, Surrey, GU2 7XH, UK\label{aff19}
\and
INAF-Osservatorio Astronomico di Brera, Via Brera 28, 20122 Milano, Italy\label{aff20}
\and
Universit\'e Paris-Saclay, Universit\'e Paris Cit\'e, CEA, CNRS, AIM, 91191, Gif-sur-Yvette, France\label{aff21}
\and
INFN, Sezione di Trieste, Via Valerio 2, 34127 Trieste TS, Italy\label{aff22}
\and
SISSA, International School for Advanced Studies, Via Bonomea 265, 34136 Trieste TS, Italy\label{aff23}
\and
Dipartimento di Fisica e Astronomia, Universit\`a di Bologna, Via Gobetti 93/2, 40129 Bologna, Italy\label{aff24}
\and
INAF-Osservatorio di Astrofisica e Scienza dello Spazio di Bologna, Via Piero Gobetti 93/3, 40129 Bologna, Italy\label{aff25}
\and
INFN-Sezione di Bologna, Viale Berti Pichat 6/2, 40127 Bologna, Italy\label{aff26}
\and
INAF-Osservatorio Astronomico di Padova, Via dell'Osservatorio 5, 35122 Padova, Italy\label{aff27}
\and
Space Science Data Center, Italian Space Agency, via del Politecnico snc, 00133 Roma, Italy\label{aff28}
\and
INAF-Osservatorio Astrofisico di Torino, Via Osservatorio 20, 10025 Pino Torinese (TO), Italy\label{aff29}
\and
Dipartimento di Fisica, Universit\`a di Genova, Via Dodecaneso 33, 16146, Genova, Italy\label{aff30}
\and
INFN-Sezione di Genova, Via Dodecaneso 33, 16146, Genova, Italy\label{aff31}
\and
Department of Physics "E. Pancini", University Federico II, Via Cinthia 6, 80126, Napoli, Italy\label{aff32}
\and
Instituto de Astrof\'isica e Ci\^encias do Espa\c{c}o, Universidade do Porto, CAUP, Rua das Estrelas, PT4150-762 Porto, Portugal\label{aff33}
\and
Faculdade de Ci\^encias da Universidade do Porto, Rua do Campo de Alegre, 4150-007 Porto, Portugal\label{aff34}
\and
European Space Agency/ESTEC, Keplerlaan 1, 2201 AZ Noordwijk, The Netherlands\label{aff35}
\and
Institute Lorentz, Leiden University, Niels Bohrweg 2, 2333 CA Leiden, The Netherlands\label{aff36}
\and
Leiden Observatory, Leiden University, Einsteinweg 55, 2333 CC Leiden, The Netherlands\label{aff37}
\and
INAF-IASF Milano, Via Alfonso Corti 12, 20133 Milano, Italy\label{aff38}
\and
Centro de Investigaciones Energ\'eticas, Medioambientales y Tecnol\'ogicas (CIEMAT), Avenida Complutense 40, 28040 Madrid, Spain\label{aff39}
\and
Port d'Informaci\'{o} Cient\'{i}fica, Campus UAB, C. Albareda s/n, 08193 Bellaterra (Barcelona), Spain\label{aff40}
\and
Institute for Theoretical Particle Physics and Cosmology (TTK), RWTH Aachen University, 52056 Aachen, Germany\label{aff41}
\and
Institute of Space Sciences (ICE, CSIC), Campus UAB, Carrer de Can Magrans, s/n, 08193 Barcelona, Spain\label{aff42}
\and
Institut d'Estudis Espacials de Catalunya (IEEC),  Edifici RDIT, Campus UPC, 08860 Castelldefels, Barcelona, Spain\label{aff43}
\and
INAF-Osservatorio Astronomico di Roma, Via Frascati 33, 00078 Monteporzio Catone, Italy\label{aff44}
\and
INFN section of Naples, Via Cinthia 6, 80126, Napoli, Italy\label{aff45}
\and
Institute for Astronomy, University of Hawaii, 2680 Woodlawn Drive, Honolulu, HI 96822, USA\label{aff46}
\and
Dipartimento di Fisica e Astronomia "Augusto Righi" - Alma Mater Studiorum Universit\`a di Bologna, Viale Berti Pichat 6/2, 40127 Bologna, Italy\label{aff47}
\and
Institute for Astronomy, University of Edinburgh, Royal Observatory, Blackford Hill, Edinburgh EH9 3HJ, UK\label{aff48}
\and
Jodrell Bank Centre for Astrophysics, Department of Physics and Astronomy, University of Manchester, Oxford Road, Manchester M13 9PL, UK\label{aff49}
\and
European Space Agency/ESRIN, Largo Galileo Galilei 1, 00044 Frascati, Roma, Italy\label{aff50}
\and
Universit\'e Claude Bernard Lyon 1, CNRS/IN2P3, IP2I Lyon, UMR 5822, Villeurbanne, F-69100, France\label{aff51}
\and
Aix-Marseille Universit\'e, CNRS, CNES, LAM, Marseille, France\label{aff52}
\and
Institut de Ci\`{e}ncies del Cosmos (ICCUB), Universitat de Barcelona (IEEC-UB), Mart\'{i} i Franqu\`{e}s 1, 08028 Barcelona, Spain\label{aff53}
\and
Instituci\'o Catalana de Recerca i Estudis Avan\c{c}ats (ICREA), Passeig de Llu\'{\i}s Companys 23, 08010 Barcelona, Spain\label{aff54}
\and
UCB Lyon 1, CNRS/IN2P3, IUF, IP2I Lyon, 4 rue Enrico Fermi, 69622 Villeurbanne, France\label{aff55}
\and
Mullard Space Science Laboratory, University College London, Holmbury St Mary, Dorking, Surrey RH5 6NT, UK\label{aff56}
\and
Departamento de F\'isica, Faculdade de Ci\^encias, Universidade de Lisboa, Edif\'icio C8, Campo Grande, PT1749-016 Lisboa, Portugal\label{aff57}
\and
Instituto de Astrof\'isica e Ci\^encias do Espa\c{c}o, Faculdade de Ci\^encias, Universidade de Lisboa, Campo Grande, 1749-016 Lisboa, Portugal\label{aff58}
\and
Department of Astronomy, University of Geneva, ch. d'Ecogia 16, 1290 Versoix, Switzerland\label{aff59}
\and
INAF-Istituto di Astrofisica e Planetologia Spaziali, via del Fosso del Cavaliere, 100, 00100 Roma, Italy\label{aff60}
\and
Aix-Marseille Universit\'e, CNRS/IN2P3, CPPM, Marseille, France\label{aff61}
\and
INFN-Bologna, Via Irnerio 46, 40126 Bologna, Italy\label{aff62}
\and
School of Physics, HH Wills Physics Laboratory, University of Bristol, Tyndall Avenue, Bristol, BS8 1TL, UK\label{aff63}
\and
Universit\"ats-Sternwarte M\"unchen, Fakult\"at f\"ur Physik, Ludwig-Maximilians-Universit\"at M\"unchen, Scheinerstrasse 1, 81679 M\"unchen, Germany\label{aff64}
\and
Max Planck Institute for Extraterrestrial Physics, Giessenbachstr. 1, 85748 Garching, Germany\label{aff65}
\and
Dipartimento di Fisica "Aldo Pontremoli", Universit\`a degli Studi di Milano, Via Celoria 16, 20133 Milano, Italy\label{aff66}
\and
INFN-Sezione di Milano, Via Celoria 16, 20133 Milano, Italy\label{aff67}
\and
NRC Herzberg, 5071 West Saanich Rd, Victoria, BC V9E 2E7, Canada\label{aff68}
\and
Institute of Theoretical Astrophysics, University of Oslo, P.O. Box 1029 Blindern, 0315 Oslo, Norway\label{aff69}
\and
Jet Propulsion Laboratory, California Institute of Technology, 4800 Oak Grove Drive, Pasadena, CA, 91109, USA\label{aff70}
\and
Department of Physics, Lancaster University, Lancaster, LA1 4YB, UK\label{aff71}
\and
Felix Hormuth Engineering, Goethestr. 17, 69181 Leimen, Germany\label{aff72}
\and
Technical University of Denmark, Elektrovej 327, 2800 Kgs. Lyngby, Denmark\label{aff73}
\and
Cosmic Dawn Center (DAWN), Denmark\label{aff74}
\and
Max-Planck-Institut f\"ur Astronomie, K\"onigstuhl 17, 69117 Heidelberg, Germany\label{aff75}
\and
NASA Goddard Space Flight Center, Greenbelt, MD 20771, USA\label{aff76}
\and
Department of Physics and Helsinki Institute of Physics, Gustaf H\"allstr\"omin katu 2, 00014 University of Helsinki, Finland\label{aff77}
\and
Universit\'e de Gen\`eve, D\'epartement de Physique Th\'eorique and Centre for Astroparticle Physics, 24 quai Ernest-Ansermet, CH-1211 Gen\`eve 4, Switzerland\label{aff78}
\and
Department of Physics, P.O. Box 64, 00014 University of Helsinki, Finland\label{aff79}
\and
Helsinki Institute of Physics, Gustaf H{\"a}llstr{\"o}min katu 2, University of Helsinki, Helsinki, Finland\label{aff80}
\and
Centre de Calcul de l'IN2P3/CNRS, 21 avenue Pierre de Coubertin 69627 Villeurbanne Cedex, France\label{aff81}
\and
Laboratoire d'etude de l'Univers et des phenomenes eXtremes, Observatoire de Paris, Universit\'e PSL, Sorbonne Universit\'e, CNRS, 92190 Meudon, France\label{aff82}
\and
SKA Observatory, Jodrell Bank, Lower Withington, Macclesfield, Cheshire SK11 9FT, UK\label{aff83}
\and
University of Applied Sciences and Arts of Northwestern Switzerland, School of Computer Science, 5210 Windisch, Switzerland\label{aff84}
\and
Universit\"at Bonn, Argelander-Institut f\"ur Astronomie, Auf dem H\"ugel 71, 53121 Bonn, Germany\label{aff85}
\and
INFN-Sezione di Roma, Piazzale Aldo Moro, 2 - c/o Dipartimento di Fisica, Edificio G. Marconi, 00185 Roma, Italy\label{aff86}
\and
Dipartimento di Fisica e Astronomia "Augusto Righi" - Alma Mater Studiorum Universit\`a di Bologna, via Piero Gobetti 93/2, 40129 Bologna, Italy\label{aff87}
\and
Department of Physics, Institute for Computational Cosmology, Durham University, South Road, Durham, DH1 3LE, UK\label{aff88}
\and
Universit\'e C\^{o}te d'Azur, Observatoire de la C\^{o}te d'Azur, CNRS, Laboratoire Lagrange, Bd de l'Observatoire, CS 34229, 06304 Nice cedex 4, France\label{aff89}
\and
CNRS-UCB International Research Laboratory, Centre Pierre Bin\'etruy, IRL2007, CPB-IN2P3, Berkeley, USA\label{aff90}
\and
University of Applied Sciences and Arts of Northwestern Switzerland, School of Engineering, 5210 Windisch, Switzerland\label{aff91}
\and
Institut d'Astrophysique de Paris, 98bis Boulevard Arago, 75014, Paris, France\label{aff92}
\and
Institute of Physics, Laboratory of Astrophysics, Ecole Polytechnique F\'ed\'erale de Lausanne (EPFL), Observatoire de Sauverny, 1290 Versoix, Switzerland\label{aff93}
\and
Aurora Technology for European Space Agency (ESA), Camino bajo del Castillo, s/n, Urbanizacion Villafranca del Castillo, Villanueva de la Ca\~nada, 28692 Madrid, Spain\label{aff94}
\and
Institut de F\'{i}sica d'Altes Energies (IFAE), The Barcelona Institute of Science and Technology, Campus UAB, 08193 Bellaterra (Barcelona), Spain\label{aff95}
\and
School of Mathematics, Statistics and Physics, Newcastle University, Herschel Building, Newcastle-upon-Tyne, NE1 7RU, UK\label{aff96}
\and
Waterloo Centre for Astrophysics, University of Waterloo, Waterloo, Ontario N2L 3G1, Canada\label{aff97}
\and
Department of Physics and Astronomy, University of Waterloo, Waterloo, Ontario N2L 3G1, Canada\label{aff98}
\and
Perimeter Institute for Theoretical Physics, Waterloo, Ontario N2L 2Y5, Canada\label{aff99}
\and
Centre National d'Etudes Spatiales -- Centre spatial de Toulouse, 18 avenue Edouard Belin, 31401 Toulouse Cedex 9, France\label{aff100}
\and
Institute of Space Science, Str. Atomistilor, nr. 409 M\u{a}gurele, Ilfov, 077125, Romania\label{aff101}
\and
Consejo Superior de Investigaciones Cientificas, Calle Serrano 117, 28006 Madrid, Spain\label{aff102}
\and
Dipartimento di Fisica e Astronomia "G. Galilei", Universit\`a di Padova, Via Marzolo 8, 35131 Padova, Italy\label{aff103}
\and
INFN-Padova, Via Marzolo 8, 35131 Padova, Italy\label{aff104}
\and
Institut f\"ur Theoretische Physik, University of Heidelberg, Philosophenweg 16, 69120 Heidelberg, Germany\label{aff105}
\and
Institut de Recherche en Astrophysique et Plan\'etologie (IRAP), Universit\'e de Toulouse, CNRS, UPS, CNES, 14 Av. Edouard Belin, 31400 Toulouse, France\label{aff106}
\and
Universit\'e St Joseph; Faculty of Sciences, Beirut, Lebanon\label{aff107}
\and
Departamento de F\'isica, FCFM, Universidad de Chile, Blanco Encalada 2008, Santiago, Chile\label{aff108}
\and
Universit\"at Innsbruck, Institut f\"ur Astro- und Teilchenphysik, Technikerstr. 25/8, 6020 Innsbruck, Austria\label{aff109}
\and
Satlantis, University Science Park, Sede Bld 48940, Leioa-Bilbao, Spain\label{aff110}
\and
Instituto de Astrof\'isica e Ci\^encias do Espa\c{c}o, Faculdade de Ci\^encias, Universidade de Lisboa, Tapada da Ajuda, 1349-018 Lisboa, Portugal\label{aff111}
\and
Department of Physics and Astronomy, University College London, Gower Street, London WC1E 6BT, UK\label{aff112}
\and
Cosmic Dawn Center (DAWN)\label{aff113}
\and
Niels Bohr Institute, University of Copenhagen, Jagtvej 128, 2200 Copenhagen, Denmark\label{aff114}
\and
Universidad Polit\'ecnica de Cartagena, Departamento de Electr\'onica y Tecnolog\'ia de Computadoras,  Plaza del Hospital 1, 30202 Cartagena, Spain\label{aff115}
\and
Infrared Processing and Analysis Center, California Institute of Technology, Pasadena, CA 91125, USA\label{aff116}
\and
Dipartimento di Fisica e Scienze della Terra, Universit\`a degli Studi di Ferrara, Via Giuseppe Saragat 1, 44122 Ferrara, Italy\label{aff117}
\and
Istituto Nazionale di Fisica Nucleare, Sezione di Ferrara, Via Giuseppe Saragat 1, 44122 Ferrara, Italy\label{aff118}
\and
INAF, Istituto di Radioastronomia, Via Piero Gobetti 101, 40129 Bologna, Italy\label{aff119}
\and
Department of Physics, Oxford University, Keble Road, Oxford OX1 3RH, UK\label{aff120}
\and
Universit\'e PSL, Observatoire de Paris, Sorbonne Universit\'e, CNRS, LERMA, 75014, Paris, France\label{aff121}
\and
Universit\'e Paris-Cit\'e, 5 Rue Thomas Mann, 75013, Paris, France\label{aff122}
\and
INAF - Osservatorio Astronomico di Brera, via Emilio Bianchi 46, 23807 Merate, Italy\label{aff123}
\and
INAF-Osservatorio Astronomico di Brera, Via Brera 28, 20122 Milano, Italy, and INFN-Sezione di Genova, Via Dodecaneso 33, 16146, Genova, Italy\label{aff124}
\and
ICL, Junia, Universit\'e Catholique de Lille, LITL, 59000 Lille, France\label{aff125}
\and
ICSC - Centro Nazionale di Ricerca in High Performance Computing, Big Data e Quantum Computing, Via Magnanelli 2, Bologna, Italy\label{aff126}
\and
Instituto de F\'isica Te\'orica UAM-CSIC, Campus de Cantoblanco, 28049 Madrid, Spain\label{aff127}
\and
CERCA/ISO, Department of Physics, Case Western Reserve University, 10900 Euclid Avenue, Cleveland, OH 44106, USA\label{aff128}
\and
Technical University of Munich, TUM School of Natural Sciences, Physics Department, James-Franck-Str.~1, 85748 Garching, Germany\label{aff129}
\and
Max-Planck-Institut f\"ur Astrophysik, Karl-Schwarzschild-Str.~1, 85748 Garching, Germany\label{aff130}
\and
Laboratoire Univers et Th\'eorie, Observatoire de Paris, Universit\'e PSL, Universit\'e Paris Cit\'e, CNRS, 92190 Meudon, France\label{aff131}
\and
Departamento de F{\'\i}sica Fundamental. Universidad de Salamanca. Plaza de la Merced s/n. 37008 Salamanca, Spain\label{aff132}
\and
Universit\'e de Strasbourg, CNRS, Observatoire astronomique de Strasbourg, UMR 7550, 67000 Strasbourg, France\label{aff133}
\and
Center for Data-Driven Discovery, Kavli IPMU (WPI), UTIAS, The University of Tokyo, Kashiwa, Chiba 277-8583, Japan\label{aff134}
\and
Dipartimento di Fisica - Sezione di Astronomia, Universit\`a di Trieste, Via Tiepolo 11, 34131 Trieste, Italy\label{aff135}
\and
California Institute of Technology, 1200 E California Blvd, Pasadena, CA 91125, USA\label{aff136}
\and
Department of Physics \& Astronomy, University of California Irvine, Irvine CA 92697, USA\label{aff137}
\and
Department of Mathematics and Physics E. De Giorgi, University of Salento, Via per Arnesano, CP-I93, 73100, Lecce, Italy\label{aff138}
\and
INFN, Sezione di Lecce, Via per Arnesano, CP-193, 73100, Lecce, Italy\label{aff139}
\and
INAF-Sezione di Lecce, c/o Dipartimento Matematica e Fisica, Via per Arnesano, 73100, Lecce, Italy\label{aff140}
\and
Departamento F\'isica Aplicada, Universidad Polit\'ecnica de Cartagena, Campus Muralla del Mar, 30202 Cartagena, Murcia, Spain\label{aff141}
\and
Instituto de F\'isica de Cantabria, Edificio Juan Jord\'a, Avenida de los Castros, 39005 Santander, Spain\label{aff142}
\and
CEA Saclay, DFR/IRFU, Service d'Astrophysique, Bat. 709, 91191 Gif-sur-Yvette, France\label{aff143}
\and
Department of Computer Science, Aalto University, PO Box 15400, Espoo, FI-00 076, Finland\label{aff144}
\and
Instituto de Astrof\'\i sica de Canarias, c/ Via Lactea s/n, La Laguna 38200, Spain. Departamento de Astrof\'\i sica de la Universidad de La Laguna, Avda. Francisco Sanchez, La Laguna, 38200, Spain\label{aff145}
\and
Caltech/IPAC, 1200 E. California Blvd., Pasadena, CA 91125, USA\label{aff146}
\and
Ruhr University Bochum, Faculty of Physics and Astronomy, Astronomical Institute (AIRUB), German Centre for Cosmological Lensing (GCCL), 44780 Bochum, Germany\label{aff147}
\and
DARK, Niels Bohr Institute, University of Copenhagen, Jagtvej 155, 2200 Copenhagen, Denmark\label{aff148}
\and
Department of Physics and Astronomy, Vesilinnantie 5, 20014 University of Turku, Finland\label{aff149}
\and
Serco for European Space Agency (ESA), Camino bajo del Castillo, s/n, Urbanizacion Villafranca del Castillo, Villanueva de la Ca\~nada, 28692 Madrid, Spain\label{aff150}
\and
ARC Centre of Excellence for Dark Matter Particle Physics, Melbourne, Australia\label{aff151}
\and
Centre for Astrophysics \& Supercomputing, Swinburne University of Technology,  Hawthorn, Victoria 3122, Australia\label{aff152}
\and
Department of Physics and Astronomy, University of the Western Cape, Bellville, Cape Town, 7535, South Africa\label{aff153}
\and
DAMTP, Centre for Mathematical Sciences, Wilberforce Road, Cambridge CB3 0WA, UK\label{aff154}
\and
Kavli Institute for Cosmology Cambridge, Madingley Road, Cambridge, CB3 0HA, UK\label{aff155}
\and
Department of Astrophysics, University of Zurich, Winterthurerstrasse 190, 8057 Zurich, Switzerland\label{aff156}
\and
Department of Physics, Centre for Extragalactic Astronomy, Durham University, South Road, Durham, DH1 3LE, UK\label{aff157}
\and
IRFU, CEA, Universit\'e Paris-Saclay 91191 Gif-sur-Yvette Cedex, France\label{aff158}
\and
Oskar Klein Centre for Cosmoparticle Physics, Department of Physics, Stockholm University, Stockholm, SE-106 91, Sweden\label{aff159}
\and
Astrophysics Group, Blackett Laboratory, Imperial College London, London SW7 2AZ, UK\label{aff160}
\and
Univ. Grenoble Alpes, CNRS, Grenoble INP, LPSC-IN2P3, 53, Avenue des Martyrs, 38000, Grenoble, France\label{aff161}
\and
INAF-Osservatorio Astrofisico di Arcetri, Largo E. Fermi 5, 50125, Firenze, Italy\label{aff162}
\and
Dipartimento di Fisica, Sapienza Universit\`a di Roma, Piazzale Aldo Moro 2, 00185 Roma, Italy\label{aff163}
\and
Centro de Astrof\'{\i}sica da Universidade do Porto, Rua das Estrelas, 4150-762 Porto, Portugal\label{aff164}
\and
HE Space for European Space Agency (ESA), Camino bajo del Castillo, s/n, Urbanizacion Villafranca del Castillo, Villanueva de la Ca\~nada, 28692 Madrid, Spain\label{aff165}
\and
Department of Astrophysical Sciences, Peyton Hall, Princeton University, Princeton, NJ 08544, USA\label{aff166}
\and
Theoretical astrophysics, Department of Physics and Astronomy, Uppsala University, Box 516, 751 37 Uppsala, Sweden\label{aff167}
\and
Minnesota Institute for Astrophysics, University of Minnesota, 116 Church St SE, Minneapolis, MN 55455, USA\label{aff168}
\and
Mathematical Institute, University of Leiden, Einsteinweg 55, 2333 CA Leiden, The Netherlands\label{aff169}
\and
Institute of Astronomy, University of Cambridge, Madingley Road, Cambridge CB3 0HA, UK\label{aff170}
\and
Department of Physics and Astronomy, University of California, Davis, CA 95616, USA\label{aff171}
\and
Space physics and astronomy research unit, University of Oulu, Pentti Kaiteran katu 1, FI-90014 Oulu, Finland\label{aff172}
\and
Center for Computational Astrophysics, Flatiron Institute, 162 5th Avenue, 10010, New York, NY, USA\label{aff173}
\and
Department of Astronomy, University of Massachusetts, Amherst, MA 01003, USA\label{aff174}
\and
Univ. Lille, CNRS, Centrale Lille, UMR 9189 CRIStAL, 59000 Lille, France\label{aff175}}
%
%
 \abstract 
 {
Investigating what drives the quenching of star formation in galaxies is key to understanding their evolution.
The \Euclid mission will provide rich spatial and spectral data from optical to infrared wavelengths for millions of galaxies, and enable precise measurements of their star formation histories.
Using the first Euclid Quick Data Release (Q1), we developed a probabilistic classification framework that combines the average specific star-formation rate (\avssfr) inferred over two timescales ($\tau={10^8,10^9}$ yr) to categorise galaxies as `ageing' (secularly evolving), `quenched' (recently halted star formation), or `retired' (dominated by old stars).
We validated this methodology using synthetic observations from the IllustrisTNG simulation.
Two classification methods were employed: a probabilistic approach, which integrates posterior distributions, and a model-driven method, which optimises sample purity and completeness using IllustrisTNG.
At $z<0.1$ and $\mstar \gtrsim 3\times10^{8}\,\Msun$, we obtain \Euclid class fractions of 68--72\%, 8--17\%, and 14--19\% for ageing, quenched, and retired populations, respectively, which is consistent with previous studies.
Ageing and retired galaxies dominate at the low- and high-mass end, respectively, while quenched galaxies surpass the retired fraction for $M_\ast \lesssim 10^{10}\,\Msun$.
The evolution with redshift shows increasing and decreasing fractions of ageing and retired galaxies, respectively.
The fraction of quenched systems shows a weaker dependence on stellar mass and redshift, varying between 5 and 15$\%$.
We find tentative evidence that more massive galaxies usually undergo quenching episodes at earlier times with respect to their low-mass counterparts.
We analysed the mass-size-metallicity relation for each population.
Ageing galaxies generally exhibit disc morphologies and low metallicities.
Retired galaxies show compact structures and enhanced chemical enrichment, while quenched galaxies form an intermediate population that is more compact and chemically evolved than ageing systems.
Despite potential selection biases, this work demonstrates \Euclid's great potential for elucidating the physical nature of the quenching mechanisms that govern galaxy evolution.
}
%
%
\keywords{Galaxies: general -- Galaxies: evolution -- Galaxies: fundamental parameters -- Galaxies: star formation -- Galaxies: stellar content}
%
%
\titlerunning{A probabilistic classification of ageing, quenched, and retired galaxies}
\authorrunning{Euclid Collaboration: P. Corcho-Caballero et al.}
   
\maketitle
%
   
\section{\label{sc:Intro}Introduction}

One of the current central questions in galaxy evolution is
to identify the main physical processes that govern the specific star-formation rate ($\ssfr\equiv\sfr/\mstar$) and drive the transition from blue star-forming galaxies to red quiescent systems \citep[e.g.][]{Faber+07, Peng+10, Schawinski+14, Casado+15, Tacchella+16, Moutard+16, Belli+19, Tacchella+22}.
Different approaches have been proposed in the literature to classify these processes in order to gain a deeper understanding of their potential role.
Based on their origin (the so-called `nature versus nurture' debate), two broad categories have been proposed: internally triggered mechanisms (`nature'), such as negative feedback from active galactic nuclei (AGN), supernovae-driven winds \citep[][]{Crenshaw+03, DiMatteo+05, Croton+06, Sawala+10, Cheung+16, Fitts+17}, which can expel or heat gas, or morpho-kinematic related effects that prevent gas cloud fragmentation \citep[e.g.][]{Bigiel+08, Martig+09, Gensior+20}, versus
environmentally driven processes (`nurture') such as ram-pressure stripping \citep[able to remove part of or even all the gas reservoir; see e.g.][]{Gunn&Gott72, Boselli&Gavazzi06, Brown+17}, starvation \citep[the suppression of gas infall; see e.g.][]{Larson+80, Wetzel+13, Peng+15}, which leads to a suppression of star formation, or galaxy interactions \citep[][]{Moore+96, Bialas+15}.
These mechanisms often act simultaneously or sequentially, and their relative importance may depend on galaxy mass, environment, and redshift.
For example, both theoretical predictions and observational evidence suggest that low-mass galaxies are more strongly affected by environmental effects, whereas massive systems are likely dominated by internal processes like AGN feedback \citep{Peng+10, DeLucia+12, Corcho-Caballero+23b}.

An alternative and complementary way to distinguish between these physical mechanisms is by their characteristic timescales.
In the classical picture \citep{Kormendy04}, fast, violent processes are expected to occur on a free-fall (dynamical) scale, $t_{\rm dyn} \sim 1/\sqrt{G\rho}$, where $\rho$ is a representative density of the galactic halo.
Several of the aforementioned mechanisms, such as galactic outflows, harassment, or ram-pressure stripping, fall into this category.
In contrast, other processes, including starvation or morphological transformations, occur on much longer (`secular') timescales and require several galactic rotations for their effect on the SFR to be noticed \citep[e.g.][]{Wright+19, Walters+22}.

A third and final distinction concerns the magnitude of the effect.
Some physical mechanisms contribute to `regulate' the efficiency of gas accretion, cooling, and/or conversion into stars through interactions with the interstellar medium \citep[e.g.][]{Booth+09, Hopkins+12}, which leads to smooth or mildly oscillatory star formation histories (SFHs).
In particular, any negative `feedback' process whose intensity scales with the star formation activity (e.g. supernovae-driven winds) will drive the system towards a steady-state `bathtub' mode where the SFR is proportional to the gas infall rate \citep[e.g.][and references therein]{Bouche+10, Maiolino&Mannucci19}, but it will never be able to halt it completely unless the cold gas supply is shut down.
On the other hand, the strength of other `suppressing' mechanisms is not tied to the SFR, and thus they may reduce it to negligible levels, by extinguishing the gas supply (e.g. starvation, ram-pressure stripping), preventing it from forming stars (e.g. morphological quenching), or both.

In this work, we define `quenching' as a process capable of terminating -- or significantly suppressing -- the star formation of a galaxy on a short timescale (i.e. less than about 1 Gyr).
In contrast, we use the term `ageing' to denote the continuous secular evolution of a galaxy, which encompasses different evolutionary stages from star-forming (including short-lived star-burst episodes) to quiescent phases, driven by the steady consumption of its gas reservoir through uninterrupted star formation \citep[e.g.][]{Casado+15, Tacchella+16}.
All galaxies, blue and red, undergo ageing as their sSFR gradually decreases with time.
Discriminating between galaxies that are merely ageing and systems that are also affected by slow suppressing processes \citep[often referred to as `slow quenching'; see e.g.][]{Schawinski+14, Moutard+16, Belli+19, Tacchella+22} on timescales $\gtrsim 1$~Gyr is challenging and, in practice, often impossible.
We consider that all these secularly evolving systems belong to the ageing category.
At the red end, it is particularly challenging to reconstruct the evolutionary path a galaxy has followed over cosmic time: presently, red objects that have gradually evolved to quiescence appear spectroscopically similar to those that experienced a sudden quenching event in the distant past \citep[e.g.][]{Corcho-Caballero+23a}.
We refer to those systems as `retired'.

Consequently, we focus on identifying galaxies that show clear evidence of recent quenching and distinguishing them from the ageing and retired populations.
To achieve this goal,
we previously developed an empirical diagnostic tool based on several observational samples of nearby galaxies \citep[][hereafter \citetalias{Casado+15}, \citetalias{Corcho-Caballero+21b}, and \citetalias{Corcho-Caballero+23a}]{Casado+15, Corcho-Caballero+21b, Corcho-Caballero+23a}.
The ageing diagram (AD) combines two proxies for star-formation, sensitive to different timescales, to probe the derivative of the recent SFH during the last 1--3 Gyr.
Specifically, we used the equivalent width of the $\ha$ line (\ew) to trace star formation over the last $10^{7}$ years, while we employed optical colours \citepalias[$u-r$ and $g-r$ in][respectively]{Casado+15, Corcho-Caballero+21b} or the $4000\,\AA$ break \citepalias{Corcho-Caballero+23a}  to trace sSFRs over the last billion years.

Systems with smoothly varying SFHs form a sequence characterised by a tight correlation between these proxies.
In contrast, galaxies that recently experienced quenching display suppressed \ew\ due to a lack of O and B stars but retain a relatively blue stellar continuum dominated by intermediate-age populations (e.g. A-type stars).
We introduced two demarcation lines within the AD to classify galaxies into four domains similar in spirit to the `star-forming', `young quiescent', and `old-quiescent' populations proposed by \citet{Moutard+18}: ageing galaxies (AGs) undergoing secular evolution, undetermined galaxies (UGs) with unclear classifications, quenched galaxies (QGs), which show evidence of recent quenching events ($\lesssim$ 1 Gyr), and retired galaxies (RGs) at the red end of the diagram where ageing and quenched sequences converge.
Inferring the recent SFH in these galaxies is extremely difficult, and one cannot in general discern which path (ageing or quenching) they have followed to reach the retired class.

Numerous studies have sought to identify quenching in the Universe.
Early efforts often relied on UV-to-IR photometric measurements to distinguish between star-forming and quiescent galaxies \citep[e.g.][]{Williams+09, Schawinski+14}.
However, these methods can struggle to clearly separate the two populations, particularly if red systems are assumed to always represent quenched galaxies \citep[see e.g.][for alternative interpretations]{Abramson+16}.
Closer to our approach, some previous studies discriminated between fast and slow evolutionary modes combining colours and/or spectral features \citep[e.g.][]{Moutard+16, Moutard+18, Moutard+20, Cleland+21}.
Finally, recent studies have characterised the time derivative, or simply tracked recent changes, of the SFH in galaxies \citep{Martin+17, Merlin+18, Merlin+19, Jimenez-Lopez+22, Weibel+22, Aufort+24}.
These works often used mock SFHs (either based on analytical models or simulations) to derive synthetic observables, such as broad-band colours or spectral features -- e.g. \ew, $\rm EW(H\delta)$, and \balbreak~-- combined with regression techniques to infer SFRs over different timescales.

The European Space Agency’s \Euclid mission \citep{EuclidSkyOverview} offers an unparalleled opportunity to investigate galaxy quenching across cosmic time.
The multi-wavelength high-quality photometric and spectroscopic data, across optical-to-IR bands, provided by the Wide \citep{Scaramella-EP1} and Deep \citep{EuclidSkyOverview} surveys, will enable a detailed reconstruction of the SFHs of galaxies across an unprecedented range of redshifts and stellar masses \citep[e.g.][]{Bisigello-EP23, EP-Enia, Abdurrouf+25}.
A key strength of \Euclid lies in its high-resolution imaging capabilities, which will allow for an in-depth exploration of the connection between galaxy SFHs and their optical morphologies.

In this paper, we utilise the first Euclid Quick Data Release \citep[Q1;][]{Q1-TP001} to characterise the SFHs of galaxies and employ a probabilistic framework inspired by previous work using the AD.
In Sect.~\ref{sec:data}, we describe the selection of the \Euclid galaxy sample and the numerical simulations from the IllustrisTNG suite that we used to calibrate our methods.
The classification scheme and the Bayesian inference of the SFHs is presented in Sect.~\ref{sec:classification} and~\ref{sec:bayes_inference}, respectively.
The validation of our method using IllustrisTNG synthetic data is discussed in Sects.~\ref{sec:results_tng} to ~\ref{sec:kde}.
Results from applying the classification framework to the \Euclid sample are described in Sect.~\ref{sec:results_euclid}, while a discussion in terms of stellar mass and evolution with redshift are presented in Sect.~\ref{sec:discussion_ad_fractions}, and Sect.~\ref{sec:discussion_redshift_evol}, respectively.
Throughout this work, we adopt a flat $\Lambda$CDM cosmology, with $H_0=70~ \rm km\,s^{-1}\,Mpc^{-1}$ and $\Omega_{\rm m} = 0.28$.

\section{\label{sec:data}Data}

In this work, we use a combination of ground- and space-based photometry from the \Euclid Quick Release 1 (Sect.~\ref{sec:euc_data}), together with synthetic photometry derived from the IllustrisTNG model (Sect.~\ref{sec:illustris}), devoted to tailoring the proposed classification scheme and evaluating its performance.

\subsection{\label{sec:euc_data}Euclid}

The sample of \Euclid galaxies is selected from the three Euclid Deep Fields (EDFs) that form part of Q1 \citet{Q1-TP001}. 
This work uses image data from the \Euclid NISP instrument \citep[\YE, \JE, and \HE; see][]{EuclidSkyNISP} as well as external data from ground-based surveys.
EDF-N $ugriz$ optical imaging data is provided by the Ultraviolet Near Infrared Optical Northern Survey (UNIONS; Gwyn et al. in prep.).
The dataset comprises observations performed with the Canada-France-Hawaii Telescope (CFHT) in the $u$ and $r$ bands, $i$-band data are provided by the Panchromatic Survey Telescope and Rapid Response system \citep[Pan-STARRS;][]{Chambers+16}, whereas $g$- and $z$-band imaging is acquired by two programmes using Subaru Hyper Suprime-Cam \citep[HSC;][]{Miyazaki+18}: Wide Imaging with Subaru
Hyper Suprime-Cam Euclid Sky (WISHES) and Waterloo-Hawaii-IfA $g$-band Survey (WHIGS), respectively.
In the southern hemisphere, imaging data in the $griz$ bands are provided by the Dark Energy Survey \citep{Abbott+18, Abbott+21}, as well as additional observations performed with the DECam instrument at the Blanco Telescope.
For additional details on the survey design of \Euclid, see \citet{EuclidSkyOverview}.

The present work uses template-fitting photometry (\texttt{TEMPLFIT}) extracted using \texttt{T-PHOT} by the MER processing function \citep{Q1-TP004}.
Fluxes are computed by convolving the shape of the detected source in the VIS band with the corresponding PSF model at the different bands, and fitting the surface brightness profiles \citep[see][for details]{Merlin-EP25}.
We apply a series of cuts to maximise the quality of the $ugriz\YE\JE\HE$ photometry:
\begin{itemize}
        \item \texttt{VIS\_DET = 1} (i.e. the source must be detected on the \IE band);
        \item \texttt{FLAG\_\$BAND\_TEMPLFIT < 4} (i.e. reject sources with saturated pixels or close to tile borders);
        \item A homogeneous signal-to-noise ratio (S/N) threshold applied to $ugriz\YE\JE\HE$ bands, defined as \texttt{FLUX\_\$BAND\_TEMPLFIT / FLUXERR\_\$BAND\_TEMPLFIT > 30}, equivalent to $\Delta {\rm mag}\approx0.036$;
    \item \texttt{SPURIOUS\_FLAG = 0} (i.e. discard potential artefacts);
    \item \texttt{POINT\_LIKE\_PROB < 0.2} (i.e. remove misclassified stars).
\end{itemize}
See \citet{Q1-TP004} for further details on the flags.

Additionally, we require the sources in our sample to have publicly available spectroscopy-based redshift estimates.
In EDF-N, we cross-matched the sources with catalogues available from the DESI early data relase \citep{DESI+24} and the Sloan Digital Sky Survey DR16 \citep{Ahumada+20}.
The sample selected from the EDF-S and EDF-F fields results from a crossmatch with multiple spectroscopic campaigns: 2dFGRS \citep{Colless+01}; 2dFLenS \citep{Blake+16}; 3D-HST GOODs \citep{Brammer+12}; OzDES \citep{Lidman+20}; PRIMUS \citep{Coil+11}; and VVDS \citep{lefevre+13}.
We applied several additional quality assurance cuts to prevent outliers with unreliable photometry or spectroscopic redshift:
\begin{itemize}
        \item redshift cut: $0.001 < z < 1.3$;
        \item colour cut: $0 < g-r < 2$;
        \item absolute magnitude cut: $-16 > M_r > -24.5$.
\end{itemize}

The resulting sample comprises $9699$ and $15\,941$ galaxies across the EDF-N and EDF-S+EDF-F fields, respectively.
The northern sample is primarily composed of sources at $0 \lesssim z\lesssim0.4$, with red sequence galaxies dropping out at $z\gtrsim0.3$.
On the other hand, the EDF-S+EDF-F sample extends to higher redshifts, $0.2 \lesssim z\lesssim 0.8$.
Appendix~\ref{appendix:sample_completeness} provides a discussion regarding the characterisation of the sample completeness and the statistical volume-correction applied to the sample in Sects.~\ref{sec:results_euclid},~\ref{sec:discussion_ad_fractions}, and~\ref{sec:discussion_redshift_evol}.

\subsection{\label{sec:illustris}IllustrisTNG}

To test and validate our methodology, we build a synthetic sample using the IllustrisTNG simulations \citep{Naiman+18, Marinacci+18,Springel+18, Pillepich+18b, Nelson+18}.
This suite comprises a series of cosmological magneto-hydrodynamical simulations, run with the moving-mesh \texttt{AREPO} code \citep{Springel10}, which model a vast range of physical processes such as gas cooling and heating, star-formation, stellar evolution and chemical enrichment, SN feedback, BH growth, or AGN feedback \citep[see][for details]{Weinberger+18, Pillepich+18a}.
In this work, we use the publicly available results from the TNG100-1 run at $z = 0$, which consists of a cubic volume with a box length of about 107 Mpc, with dark matter and baryonic mass resolutions of $7.5\times10^6$~\Msun and $1.4\times10^6$~\Msun, respectively. 
We select all non-flagged sub-haloes (i.e. rejecting those objects believed to be numerical artefacts) with total stellar mass within two effective radii, defined as the 3D comoving radius containing half of the stellar mass, in the range $10^9<M_*/$\Msun$<10^{12}$, comprising \num{18367} sources.

For each simulated galaxy, we compute its star-formation history by accounting for all stellar particles located within two effective radii, defined in terms of the stellar component of each subhalo.
The SFHs are used to predict observed-frame synthetic spectral energy distribution (SED) at redshift 0.0, 0.3, and 0.6 by means of the Population Synthesis Toolkit\footnote{\url{https://population-synthesis-toolkit.readthedocs.io}} \citep[\texttt{PST},][see also next section]{Corcho-Caballero+25}.
Using the SFH obtained from the snapshot at $z=0$ to compute the SED at prior times neglects the effects of mergers and accretion, which are not particularly relevant for our purposes (the SFHs are equally representative, and the number of galaxies between $z=0$ and $z=0.6$ changes up to $5\%$).

Particles are treated as simple stellar populations (SSPs), assuming a universal \citet{Kroupa+01} initial mass function.
In each redshift bin, their ages and metallicities are used to interpolate the SED templates, using a cloud-in-cell approach in terms of $\log(t_{\rm age} / \rm yr)$ and $\log(Z_{\ast} / Z_\odot)$, from the \texttt{PyPopStar} SSP library \citep{Millan-Irigoyen+21}, which provides 106 models spanning 23 ages from $10^5$ to $1.5 \times 10^{10}$ yr and four metallicities $Z_{\ast} = \{0.004,\ 0.008,\ 0.02,\ 0.05\}$.
Particles whose values lie outside the grid are assigned the nearest corresponding value.

To make the mock sample more realistic, we adopt a \citet{Cardelli+89} dust extinction law with $R_V=3.1$.
For each galaxy and redshift bin, the $V$-band magnitude extinction $A_V$ is sampled from an exponential probability distribution with a mean value of 0.3 mag, truncated at $A_V=3$.
While this is an oversimplification, probably far from the intrinsic distribution of dust in galaxies, known to be dependent on fundamental quantities such as chemical composition and gas abundance \citep[e.g.,][]{Li+19}, our ultimate goal is to test the ability of our method to recover the input extinction together with the intrinsic SFH.

We computed fluxes in the $ugriz\YE\JE\HE$ bands by multiplying the synthetic spectra with the corresponding filter sensitivity curves\footnote{Available through the Spanish Virtual Observatory Filter profile service at \url{http://svo2.cab.inta-csic.es/theory/fps/}.}.
For the sake of simplicity, only the effective throughputs of UNIONS filters corresponding to EDF-N data were used to produce the synthetic photometry.
Finally, random Gaussian noise is added to the computed fluxes to model a S/N of 30, consistent with the minimum threshold imposed to the observational sample.

\section{\label{sec:methods}Characterising star-formation histories}
\subsection{\label{sec:classification}Classification}

In previous studies, quenched galaxies were identified using observational proxies for the specific star-formation rate, such as spectral features like \ew\ and \balbreak, along with broad-band colours, which are sensitive to star formation on different timescales.
In this work, we work directly with the average specific star-formation rate, \avssfr, computed over various timescales, $\tau$, defined as:
\begin{equation}
        \label{eq:ssfr_tau}
        \avssfr(t)
    \equiv
    \frac{ \frac{1}{\tau} \int_{t-\tau}^{t} \sfr(t')\ \dd t' }{ \int_{0}^{t} \sfr(t')\ \dd t' }
    \approx
    \frac{1}{\tau} \frac{ \mstar(t) - \mstar(t-\tau) }{\mstar(t)},
\end{equation}
where $\sfr(t)$ denotes the instantaneous star-formation rate as function of cosmic time, and $\mstar(t)$ represents the cumulative stellar mass formed in the galaxy, which is not exactly equal to the current stellar mass because of stellar mass loss.
In other words, $\tau\,\avssfr$ is the fraction of stellar mass formed in the last $\tau$ years.
Note also that this description only takes into account the formation time of the stars and does not discriminate in situ star-formation from accretion in mergers.

While \citetalias{Corcho-Caballero+23a} considered very short timescales (around $20$ Myr) to capture the fraction of massive O and B stars responsible for $\ha$, broadband optical and infrared photometry has limited sensitivity to such young populations.
Consequently, we adopt timescales of $100$ Myr ($\avssfrval{8}$) and $1000$ Myr ($\avssfrval{9}$) as fiducial values, providing a more robust framework for our analysis.

\begin{figure}
        \includegraphics[width=\linewidth]{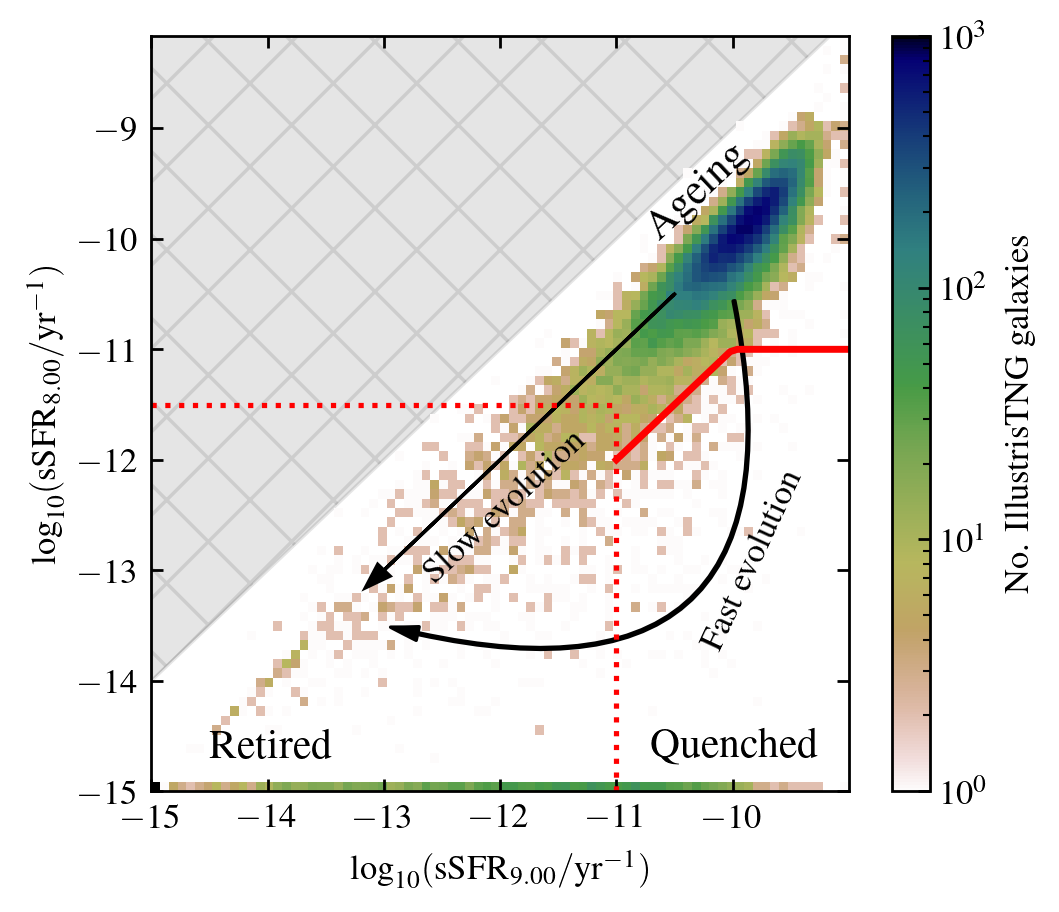}
        \caption{Distribution of IllustrisTNG galaxies across the plane defined by the average specific star-formation measured over the last $10^8$ (\avssfrval{8}) and $10^9$ yr (\avssfrval{9}).
        Quenched galaxies are found below the solid red line, whereas the dotted line delimits the retired domain.
        Ageing systems are located above both regions.
        The grey-shaded area denotes the forbidden region of the parameter space.}
        \label{fig:tng_ssfr8_ssfr9}
\end{figure}

Figure~\ref{fig:tng_ssfr8_ssfr9} shows the distribution of IllustrisTNG galaxies in the \avssfrval{8} versus \avssfrval{9} plane, which serves as a basis for distinguishing between different evolutionary stages.
This parameter space can be interpreted in terms of the existence of three domains: ageing galaxies, featuring a relatively stable star-formation activity across both timescales (roughly following a one-to-one relation); quenched galaxies, which have experienced a recent and rapid suppression of star-formation (i.e. very low levels of sSFR at present, while still showing significant star-forming activity over larger timescales); and retired systems, dominated by old stellar populations regardless of their star-formation history (i.e. both formerly ageing or quenched systems).

The classification criteria are defined as follows.
Quenched galaxies (QGs) satisfy
\begin{equation}
        \label{eq:quenched_line}
        {\rm QGs} : \begin{cases}
                \logavssfrval{9} > -11.0, \\
        \logavssfrval{8} < -11.0, \\
                \logavssfrval{8} < \logavssfrval{9} - 1, \\
        \end{cases} 
\end{equation}
whereas retired galaxies meet the conditions
\begin{equation}
        \label{eq:retired_line}
        {\rm RGs} : \begin{cases}
                \logavssfrval{9} < -11.0, \\
                \logavssfrval{8} < -11.5, \\
        \end{cases} 
\end{equation}
and systems that do not fall into either of these categories are classified as ageing galaxies (AGs).
This classification shares similarities with previous studies in the literature \citep[e.g.,][]{Moutard+18, Quai+18, Owers+19, Belli+19, Cleland+21, Tacchella+22, Corcho-Caballero+23a}.
For instance, compared to \citet{Moutard+18}, most AGs correspond to the authors' definition of `star-forming' galaxies, QGs align with their `green-valley' and `quiescent' populations, and RGs are broadly consistent with their `old quiescent' category (see Appendix \ref{appendix:class_comparison} for a more detailed comparison).
The key distinction between the two approaches lies in the definition of quenching.
While \citet{Moutard+18} considers all transitions from star-forming to quiescent as (fast or slow) quenching, we interpret quenching as a specific process occurring under particular physical conditions in a short timescale, such as a mergers \citep[e.g.,][]{Ellison+24}, potentially triggering strong AGN kinetic feedback \citep[e.g.,][]{Quai+21}.

The division between retired galaxies and the other regions is set by the observational limit where specific star formation rates can be reliably measured.
Retired galaxies exhibit extremely low \avssfr values at both timescales, making it challenging to distinguish between systems that underwent quenching events more than 1 Gyr ago, and those that evolved purely in a secular fashion or under the influence of milder `regulating' or `suppressing' mechanisms.
In the simulations, substituting the mass resolution into Eq.~\eqref{eq:retired_line} yields the minimum stellar mass of an IllustrisTNG retired galaxy
\begin{equation}
    M_{\rm min} = \frac{\rm SFR_{8,\,min}}{\avssfrval{8}(\rm RGs) } = \frac{1.4\times10^6\,\Msun / 10^8\,{\rm yr}}{3\times10^{-12}\,{\rm yr^{-1}}}\approx 5\times10^{9}\Msun.
\end{equation}

\subsection{\label{sec:bayes_inference}Bayesian inference of physical parameters}

The values of \avssfr of Illustris-TNG and \Euclid galaxies are inferred from their synthetic and observed fluxes using \besta\footnote{\url{https://besta.readthedocs.io}} (Bayesian Estimator for Stellar Population Analysis, Corcho-Caballero et al., in prep.), a Python-based Bayesian framework for deriving physical properties from observational data. \besta integrates the \texttt{PST} (Population Synthesis Toolkit), and \texttt{CosmoSIS}\footnote{\url{https://cosmosis.readthedocs.io}} \citep[Cosmological Survey Inference System,][]{Zuntz+15} libraries devoted for highly flexible stellar population synthesis, and Monte Carlo sampling techniques, respectively.

The \texttt{PST} library is designed to provide a user-friendly interface for working with simple stellar population (SSP) models and synthesizing a variety of observable quantities such as spectra or photometry from different prescriptions of the star-formation and chemical enrichment histories.
On the other hand, \texttt{CosmoSIS} is a framework, originally devoted for cosmological parameter estimations, that brings together a wide diversity of Bayesian inference methods in a modular architecture.

We infer the values of \avssfr by describing the star formation history of galaxies using a non-analytic model.
The fraction of the total stellar mass formed by cosmic time $t$, denoted as $\massnorm\equiv \mass / M_{\ast}(t_{\rm obs})$, where $t_{\rm obs}$ is the age of the Universe at the time of observation, is parametrised as a monotonic piecewise function, with boundary conditions $\massnormof{0} = 0$ and $\massnormof{t_{\rm obs}} = 1$.
This function is related to \avssfr (Eq.~\ref{eq:ssfr_tau}) by the expression
 \begin{equation}
        \label{eq:mass_frac}
        \widetilde{M}_\ast(t_{\rm obs} - \tau) = 1 - \tau \, \avssfr
 .\end{equation}

During the sampling, Eq.~\eqref{eq:mass_frac} is evaluated at fixed lookback times $\tau\equiv t_{\rm obs} -  t = 0.1$, 0.3, 0.5, 1, 3, and 5 Gyr.
The values of $\tau$ are chosen to roughly align with the age ranges where SSPs exhibit the largest differences in optical/IR colours \citep{Millan-Irigoyen+21}.
The values of \avssfr are sampled from a log-uniform prior distribution within $-14 < \log_{10}(\avssfr / \mathrm{yr}\inv) < 1/\tau$, for each value of $\tau$, rejecting solutions that do not yield a monotonically increasing \massnorm.
For each sample of \avssfr, we estimate \massnorm by interpolating the 8 points (six values of $\tau$ + boundary conditions) using a monotonic cubic spline\footnote{The monotonic Piecewise Cubic Hermite Interpolating Polynomial (PCHIP) implemented in \texttt{scipy}.}.

The chemical evolution of the stellar content of galaxies is modelled by assuming that the metallicity of stars formed at a given time, $Z(t)$, is proportional to the mass growth history of the galaxy \citep[e.g.,][]{Zibetti+17}, following
 \begin{equation}
    \label{eq:metallicity_history}
        \metal = Z(t_0) \, \frac{\mstar(t)}{\mstar(t_0)} = Z(t_0) \, \widetilde{M}_\ast(t),
 \end{equation}
 where $Z(t_0)$, the metallicity of stars formed at $t_0$, is sampled using a uniform prior distribution between $0.25 < Z(t_0) / Z_\odot < 4$.
 For the sake of clarity,~Eq.~\eqref{eq:metallicity_history} is related to the mass-weighted average stellar metallicity, $Z_\ast$, by the expression
 \begin{equation}
     Z_\ast(t_0) \equiv \frac{\int_0^{t_0} \, \sfr(t') \, Z(t') \, \dd t'}{\int_0^{t_0} \, \sfr(t') \, \dd t'} = \frac{Z(t_0)}{2}.
 \end{equation}

To build composite spectra and associated photometric fluxes resulting from the model described above, \massnorm and \metal are evaluated at the grid edges defined by the \texttt{PyPopStar} SSPs \citep[][see Sect.~\ref{sec:illustris} for more details]{Millan-Irigoyen+21}.
This allows us to estimate the relative contribution of each SSP, $\massnormssp(t, Z)$ for any given SSP age and metallicity, and the resulting spectral energy distribution (SED), in units of specific luminosity per wavelength unit and stellar mass, is computed as
\begin{equation}
        \label{eq:l_lambda}
        L_\lambda(\lambda) = \sum L_{\rm SSP}(t, Z, \lambda) \, \massnormssp(t, Z),
\end{equation}
where $L_{\rm SSP}$ denotes the SSP model SEDs.

Dust extinction is included by assuming a single dust screen, using the \citet{Cardelli+89} extinction law with a fixed value of $R_V=3.1$.
The extinction values given by $A_V$, are also sampled from a uniform distribution ranging from 0 to 3.0 mag.

Finally, the predicted specific flux per frequency unit on each photometric band, $\hat{f}_{\nu, i}$, is given by the expression
\begin{equation}
\begin{split}
        \hat{f}_{\nu, i}        = \frac{\int_0^{\infty} 10^{-0.4 \, A_V} \,F_\wl\left[\lambda\, \left(1+z\right)^{-1}\right] \, S_{i}(\wl) \,\wl\, \dd\wl}{\int_0^{\infty} \frac{3631 {\rm Jy}}{c\,\wl} S_{i}(\wl)\,\dd\wl}, \\
    F_\wl\left[\lambda\, \left(1+z\right)^{-1}\right]=\frac{L_\wl\left[\lambda\, \left(1+z\right)^{-1}\right]}{(1 + z)\,4\pi d_{\rm L}^2},
    \end{split}
\end{equation}
where $S_{i}$ is the transmission function of the filter $i$, $z$ is the source redshift, and $d_{\rm L}$ is the luminosity distance evaluated at $z$.

Given a set of flux measurements, $f_{\nu, i}$, and assuming Gaussian uncertainties, the log-likelihood associated with a given model can be computed as
\begin{equation}
        \ln \mathcal{L}(f | \theta) = \sum_i -0.5 \left(\frac{f_{\nu, i} - \alpha \, \hat{f}_{\nu, i}}{\sigma(f_{\nu, i})}\right)^2,
\end{equation}
where $\theta$ denotes the vector of parameters used in the model, $\alpha$ is a normalisation constant between the observed and the predicted fluxes that corresponds to $\mstar(t_0)$, computed using the mean value of $f_{\nu, i} / \hat{f}_{\nu, i}$, and $\sigma(f_{\nu, i})$ represent the flux uncertainty estimates associated with each band $i$.

Given a prior, $P\left(\theta\right)$, Bayes' theorem determines that the posterior probability distribution of a given model, $P\left(\theta | f\right)$, can be estimated as
\begin{equation}
        P\left(\theta | f\right) = \frac{\mathcal{L}\left(f | \theta\right) \, P\left(\theta\right)}{\int \mathcal{L}\left(f | \theta\right) \, P\left(\theta\right)\dd \theta} \propto \mathcal{L}\left(f | \theta\right)\, P\left(\theta\right),
\end{equation}
where the denominator denotes the Bayesian evidence, that for the purposes of this work can be simply treated as a proportionality constant.

To efficiently explore the posterior probability distribution, we use the \texttt{max-like} and \texttt{emcee} \citep{Foreman-Mackey+13} samplers available in \texttt{CosmoSIS}.
The former performs an initial minimisation of the problem, trying to locate the maximum of the posterior distribution, whereas the second sampler is a form of Monte-Carlo Markov chain that uses an ensemble of walkers to explore the parameter space.
The resulting chain of parameters values are used to estimate the 9-dimensional posterior PDF via a Gaussian kernel density estimator (KDE) and derive the percentile values of \avssfr, $Z(t_0)$, $A_V$, and $\mstar(t_0)$.
The parameters used in the modelling of the observed photometry and associated priors are summarised in Table~\ref{tab:fitting_parameters}.
The redshift is fixed to the spectroscopic measurement, and stellar masses are based on the $\Lambda$CDM luminosity distance. Note that the latter factor does not affect the sSFR.

\begin{table}
        \caption{Parameters used in \besta for the SFH inference.}
        \label{tab:fitting_parameters}
        \centering
        \begin{tabular}{cll}
        \hline\hline
                Parameter & Prior & Description\\
                \hline
                $\log_{10}(\ssfr_{8.00} / \rm yr^{-1})$ & unif($-14$, $-8.00$) & Last 0.10 Gyr\\
                $\log_{10}(\ssfr_{8.48} / \rm yr^{-1})$ & unif($-14$, $-8.48$) & Last 0.3 Gyr\\
                $\log_{10}(\ssfr_{8.70} / \rm yr^{-1})$ & unif($-14$, $-8.70$) & Last 0.5 Gyr\\
                $\log_{10}(\ssfr_{9.00} / \rm yr^{-1})$ & unif($-14$, $-9.00$) & Last 1 Gyr\\
                $\log_{10}(\ssfr_{9.48} / \rm yr^{-1})$ & unif($-14$, $-9.48$) & Last 3 Gyr\\
                $\log_{10}(\ssfr_{9.70} / \rm yr^{-1})$ & unif($-14$, $-9.70$) & Last 5 Gyr\\
                $A_V$ & unif(0.0, 2.5) & Dust extinction\\
                $Z(t_0)$ & unif(0.005, 0.08) & Present metallicity\\
                \hline
        \end{tabular}
\end{table}

\section{\label{sec:results}Results}

\subsection{\label{sec:results_tng}Accuracy of the sSFR estimates}

First, we utilise the results obtained from fitting IllustrisTNG synthetic photometry to benchmark the reliability of recovering the correct values of \avssfr.
In order to minimise the impact of catastrophic outliers, we remove from our sample those galaxies with a poor fit quality.
This is done by comparing the maximum value of the posterior probability distribution of each galaxy with the total distribution.
We discard sources with values lower than the 5th percentile of the distribution (i.e. 5\% of the total sample).
Although no clear correlation is observed between the physical properties of the galaxy sample and the fit results, we find that fits fail more frequently for galaxies with high extinction ($A_V > 0.8$).

\begin{figure*}[htbp!]
        \includegraphics[width=\linewidth]{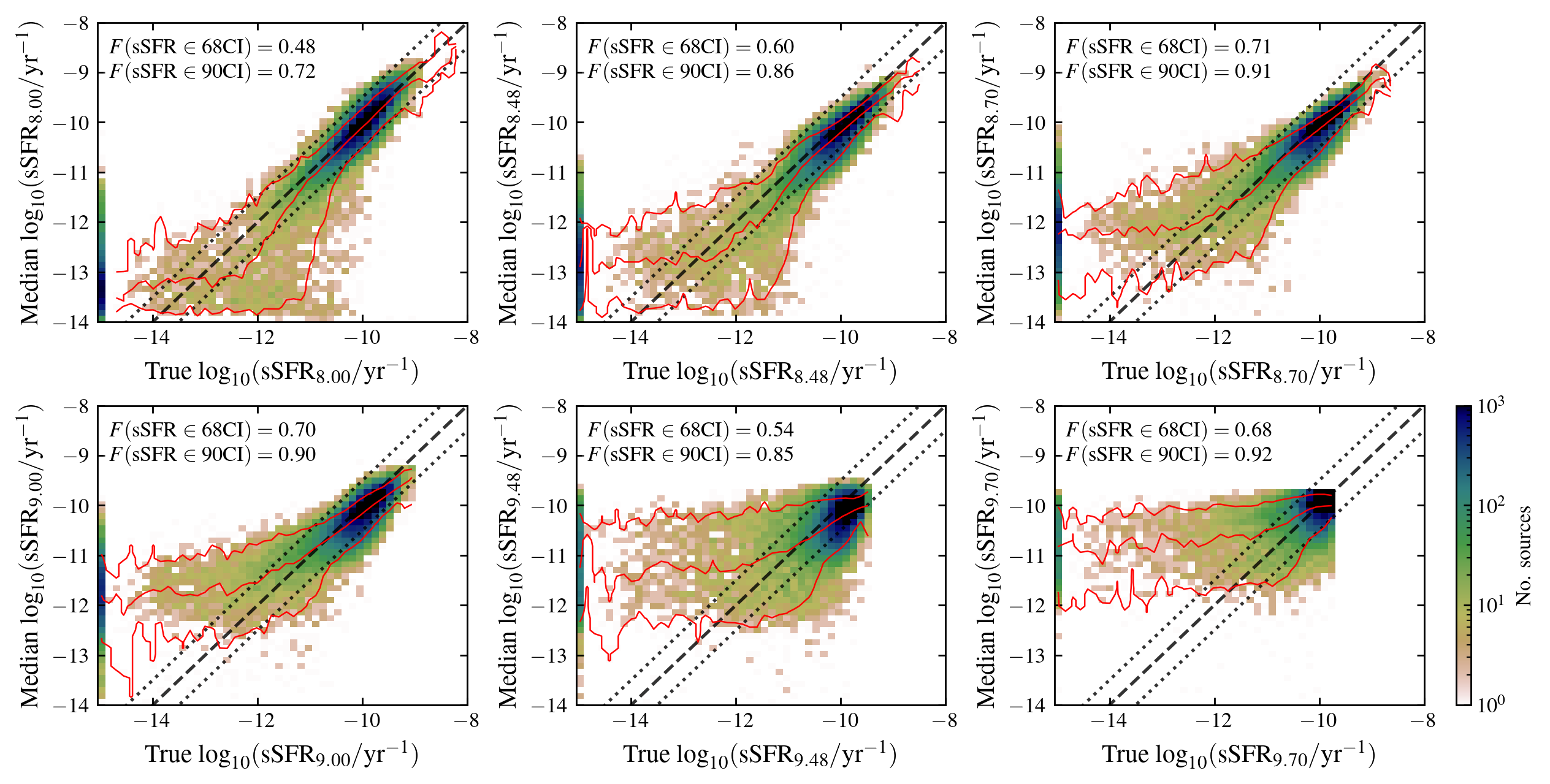}
        \caption{IllustrisTNG \avssfr true values versus the median value recovered by \besta.
    Coloured maps denote the number of sources in each bin.
    Bottom to top red lines of every panel correspond to the running 5, 50, and 95 percentiles of \avssfr as a function of the true value.
    Dashed and dotted black lines illustrate the one-to-one and 0.5 dex offset lines, respectively.
    Each panel includes at the top-right corner the fraction of sources whose true value lies within the 68 and 90\% estimated credible intervals, respectively.}
        \label{fig:illustris_ssfr_recovered_values}
\end{figure*}

The true input values of \avssfr versus the median value recovered by \besta are shown in Fig.~\ref{fig:illustris_ssfr_recovered_values}, for the combined sample that includes the realisations at $z=0.0$, $0.3$, and $0.6$ (see Appendix~\ref{appendix:sfh_inference} for plots at each redshift bin).
For any true value of \avssfr, the red lines denote the 5th, 50th, and 95th running percentiles of the median \avssfr retrieved by \besta, illustrating where the bulk of the distribution lies.
Overall, the results are satisfactory across all timescales ($\tau$), although we notice some systematic bias.
For short timescales (top row panels), \besta occasionally underestimates the true value of \avssfr, leading to an artificial population of galaxies with $\avssfrval{8/8.48}\simeq 10^{-14}\,\rm yr^{-1}$.
Conversely, when the true value is consistent with 0, i.e. no stellar particle was formed in the last $\tau$ Gyr (artificially set to $\ssfr=10^{-15}\,\rm yr^{-1}$ for visualisation purposes), the inferred median values typically stay close to the prior limit at $10^{-14}~\rm yr^{-1}$, but they also present an extended tail towards higher values of \avssfr up to about $10^{-11}~\rm yr^{-1}$.
This clearly highlights how difficult is to distinguish between strictly 0 and the $\simeq 1\%$ level in terms of the mass fraction formed over a given timescale using optical/IR photometry \citep[e.g.,][]{Salvador-Rusinol+20}.

\begin{figure*}[htbp!]
        \includegraphics[width=\linewidth]{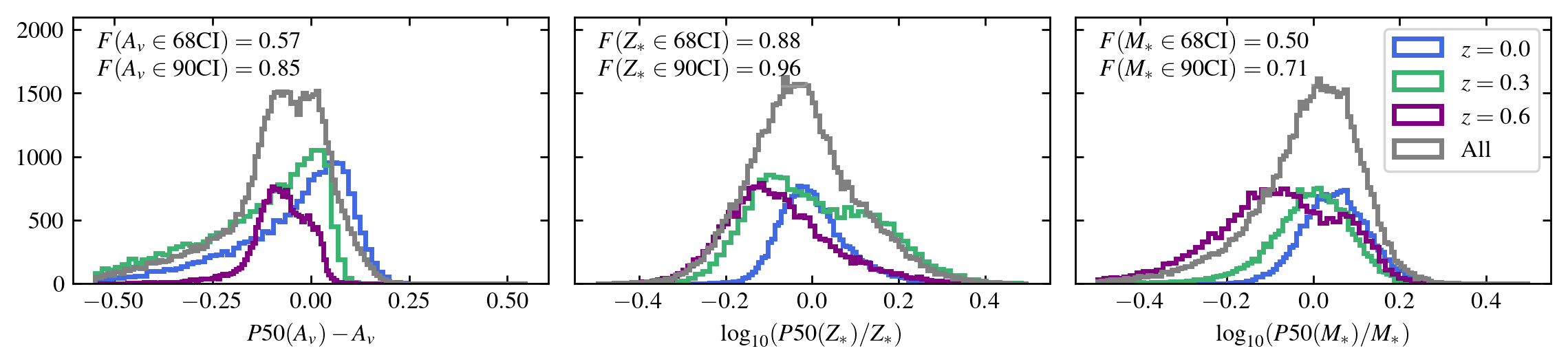}
        \caption{Statistical distribution of the difference between the input IllustrisTNG values of dust extinction, $A_V$, stellar metallicity, $\log_{10}(Z_\ast / Z_\odot)$, and total stellar mass, $\log_{10}(\mstar/ \rm \Msun)$, versus the median value recovered by \besta.
        Each panel includes the fraction of sources whose true value lies within the 68 and 90\% estimated credible intervals, respectively.}
        \label{fig:illustris_physical_prop_recovered_values}
\end{figure*}

At longer timescales ($\tau>1$ Gyr), \besta tends to overestimate \avssfr for values below $10^{-11}~\rm yr^{-1}$.
The resulting median values typically lie in the range $10^{-12}$--$10^{-10}~\rm yr^{-1}$, which corresponds to formed mass fractions of $0.003$--$0.3$ and $0.005$--$0.5$ at lookback times of 3 and 5 Gyr, respectively, in contrast to the true value of 0.
In general, we observe that extreme outliers, i.e. that values significantly above/below the one-to-one line, primarily emerge as a consequence of over/under-estimating the dust extinction and/or metallicity by approximately $\gtrsim0.2$ dex.

Given the intrinsic degeneracy between the parameters of our model, that may lead to an erroneous classification, it is of the utmost importance to not only rely on the median/mean/mode estimates of the resulting posterior PDF, but to make use of the whole distribution to account for the large uncertainties.
To that end, Fig.~\ref{fig:illustris_ssfr_recovered_values} also includes the fraction of sources whose real value of \avssfr lies within the 68 and 90\% credible intervals estimated from the posterior PDF.
This number depends on how well-calibrated the credible intervals are to the actual posterior probabilities and can be used as a proxy of the ability of the model to capture the complexity of the data.
To estimate the fractions, only sources with $\avssfr > 10^{-14}~\rm yr^{-1}$ have been considered.
As expected from a reasonably well-calibrated estimate of the uncertainties associated with \avssfr, both intervals properly account for $\simeq68$\% and $\simeq90$\% of the sample.

In Fig.~\ref{fig:illustris_physical_prop_recovered_values}, we show the ratio between the median value of the other physical properties included in the model -- dust extinction, $A_V$, stellar metallicity $\log_{10}(Z_\ast / Z_\odot)$, and total stellar mass, $\log_{10}(\mstar/ \rm \Msun)$ -- and the true input value, for the three samples considered in this work (at $z=0$, $0.3$, and $0.6$, respectively).
The three quantities are roughly recovered by \besta with a reasonable degree of accuracy (about $0.2$ dex), although the biggest challenge is to effectively inferring $A_V$.
In a significant fraction of cases, $A_V$ is underestimated, resulting into an increase of the metallicity for compensating the change of colours.
As a result of changing the mass-to-light ratio of the underlying stellar population, the final stellar mass estimate is also affected by a few percent.
Nevertheless, the posterior probability distribution still seems to properly account for the intrinsic uncertainties as indicated by the fraction of sources within the 68 and 90 credible intervals.

\begin{figure*}
        \includegraphics[width=\linewidth]{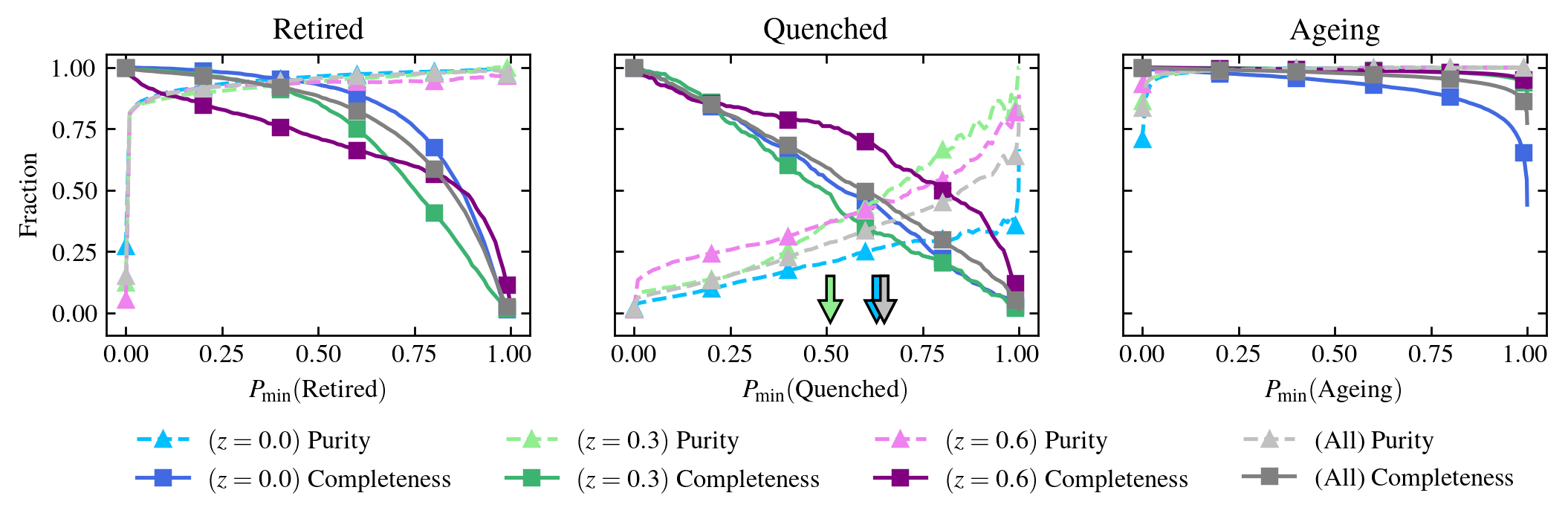}
        \caption{Retired (left), quenched (middle), and ageing (right) galaxy classification purity and completeness, as a function of the minimum probability threshold used to perform the classification. Each redshift sample at $z=0$, $z=0.3$, and $z=0.6$ is denoted by the blue, green, and purple lines, respectively. Coloured arrows denote the value of $P_{\rm min}(\rm Quenched)$ that maximises the $F$-score at each redshift (see Sect.~\ref{sec:quenched_selection}).}
        \label{fig:tng_vanilla_classes}
\end{figure*}

It is worth remarking that the use of the median value is adopted only for its simplicity in terms of visualisation purposes.
Throughout the remaining sections the goal is to make use of the entire posterior PDF in order to maximise the information content.
In addition, it is also important to point out that our fitting procedure assumes the same basic ingredients (IMF, SSPs, extinction law) as the mock observations. This is of course an ideal case, and therefore our estimated uncertainties should be regarded as a lower limit \citep[see e.g.,][for an exhaustive discussion]{Conroy13}.

\subsection{Probabilistic AD classification\label{sec:ad_prob}}

In this section we present our classification scheme mainly devoted for discriminating between slow (ageing) and fast (Quenching) evolution.
Here, we describe the method based on analysing the posterior probability distribution of each galaxy.
In contrast, Sect.~\ref{sec:kde} introduces an alternative, model-driven approach that maximises classification completeness and purity by leveraging the predictive power of the IllustrisTNG simulation.

We estimate the probability that a galaxy belongs to the ageing, quenched and retired populations ($i$) by integrating the marginalised posterior PDF, $P(\avssfrval{9}, \avssfrval{8} | f)$, over the regions $D(i)$ defined by the AD classification (described in Sect.~~\ref{sec:classification}), as specified by Eqs.~\eqref{eq:quenched_line} and~\eqref{eq:retired_line}
\begin{equation}
        P(i) = \frac{\iint_{D(i)} P(\avssfrval{9}, \avssfrval{8} | f) \, \dd \log_{10}\left(\frac{\avssfrval{9}}{\rm yr^{-1}}\right)\, \dd\log_{10}\left(\frac{\avssfrval{8}}{\rm yr^{-1}}\right)}{\int P(\theta | f)\,\dd \theta},
\end{equation}
where the denominator represents the Bayesian evidence for the model, and $P(i)$ indicates the probability of the galaxy being classified as ageing, quenched, or retired, and $\sum P(i) = 1$.

In Fig.~\ref{fig:tng_vanilla_classes} we show the purity and completeness fraction achieved when classifying the retired (left), quenched (middle), and ageing (right) populations by imposing a minimum probability threshold, $P_{\rm min}$, on the sample of IllustrisTNG galaxies.
For each class, purity and completeness are defined as the number of true positives divided by the total number of observed (true and false) positives, $N_{\rm TP}/(N_{\rm TP} + N_{\rm FP})$, and the ratio between true positives divided by the number of true positives and false negatives, $N_{\rm TP}/(N_{\rm TP} + N_{\rm FN})$, respectively.
To explore the effects of redshift on the performance of the classification, the sample has been split into the three redshift bins under consideration: $z=0.0$, $0.3$, and $0.6$.
Both ageing and retired populations are successfully classified with purity scores above 80\% at all redshift bins.
The completeness of both classes presents a stronger dependence with redshift. For a given probability threshold, retired galaxies tend to be more under-represented at higher values of $z$, whereas constraining the fraction of ageing systems at $z\simeq0$ results more challenging than at high redshift.

On the other hand, estimating the population of quenched galaxies presents more difficulties.
At $z=0$, the population of quenched galaxies is poorly identified, and it includes a significant number of false positives, that strongly limit the purity of the sample to $\simeq30\%$ even when applying a threshold cut of $P_{\rm min}({\rm Quenched}) > 0.75$.
When considering the same cut at higher values of $z$, this bias is significantly alleviated, and the sample purity increases to values between 50\% and 75\%.
This is expected since optical photometry becomes more sensitive to the recent star-formation episodes due to the more prominent contribution of rest-frame UV light (mostly supported by young stellar populations).

\subsection{Selection of quenched galaxies\label{sec:quenched_selection}}

\begin{figure}[htbp!]
        \includegraphics[width=\linewidth]{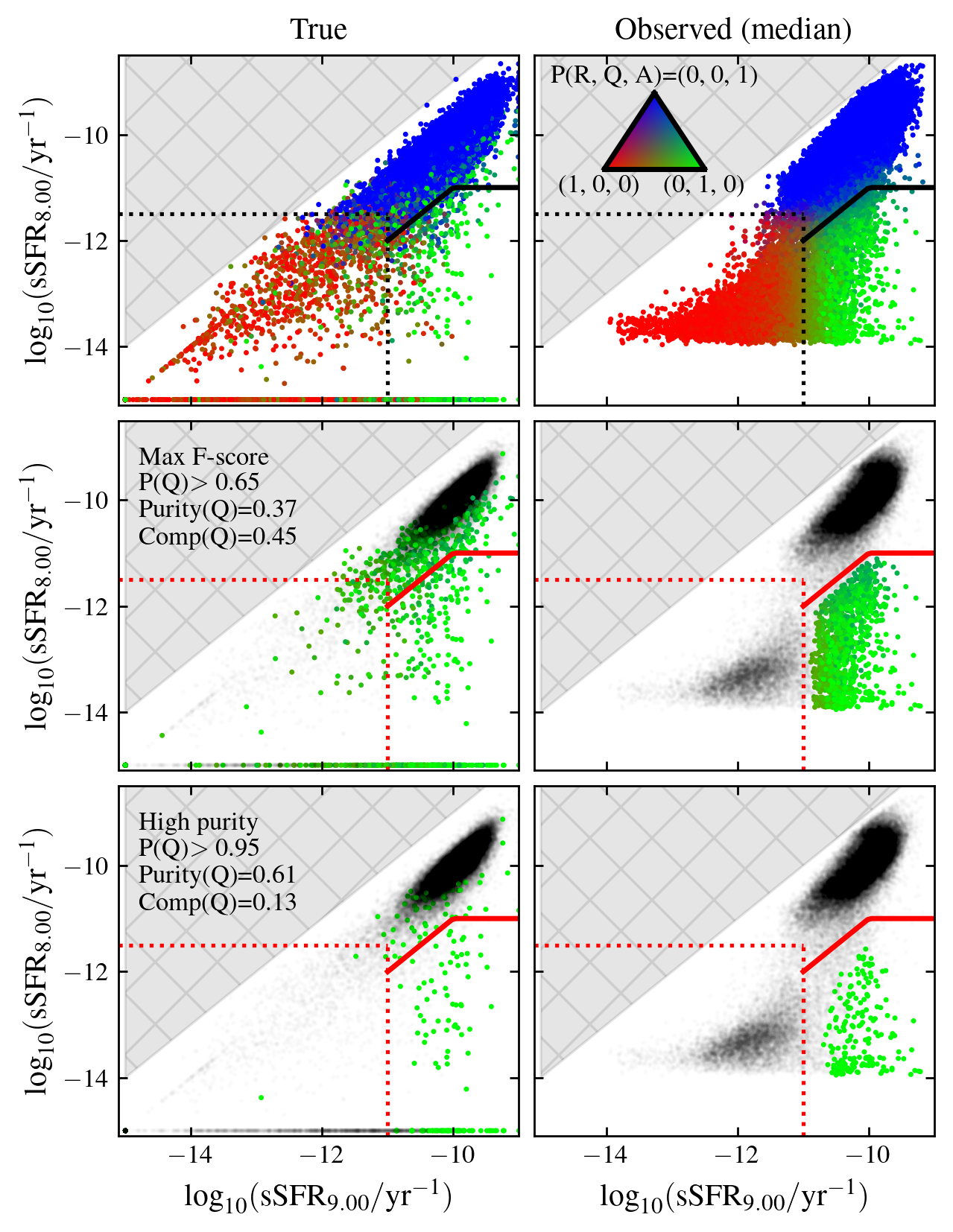}
        \caption{Distribution of IllustrisTNG galaxies across the \avssfrval{9} versus ~\avssfrval{8} plane. Left and right columns show the distribution based on the true (null values of \ssfr are arbitrarily set to $\log_{10}(\ssfr / \rm yr^{-1})=-15$) and inferred median values, respectively. Top row panels display the full sample RGB colour-coded by the probability of belonging to the retired, quenched, and ageing classes. Solid and dotted lines represent the demarcation lines given by Eqs.~\ref{eq:quenched_line}~and~\ref{eq:retired_line}, respectively. Middle and bottom rows illustrate the two proposed approaches for selecting quenched galaxies (coloured points) by colouring in black with low transparency those systems classified as retired or ageing.}
        \label{fig:tng_quenched_selection}
\end{figure}

In this section, we present two approaches for selecting recently quenched galaxies, each tailored to different priorities in classification performance.
The first approach emphasises a balance between purity and completeness, achieved by selecting the probability threshold $P_{\rm min}(\rm Quenched)$ that maximises the $F$-score, defined as
\begin{equation}
    F = \frac{2 N_{\rm TP}}{2N_{\rm TP}+N_{\rm FP}+N_{\rm FN}},
\end{equation}
where $N_{\rm TP}$, $N_{\rm FP}$, $N_{\rm TN}$, and $N_{\rm FN}$ represent the number of true positives, false positives, true negatives, and false negatives, respectively.

From the purity and completeness values shown in Fig.~\ref{fig:tng_vanilla_classes}, we determine the optimal $P_{\rm min}(\rm Quenched)$ values for the samples at $z=0.0$, $0.3$, and $0.6$ to be 0.63, 0.51, and 0.65, respectively.
When combining the three redshift samples, the optimal threshold becomes $P_{\rm min}(\rm Quenched)=0.65$, yielding a combined purity of 37\% and a completeness of 45\%.

The second approach prioritises maximizing the purity of the quenched galaxy sample, even at the expense of low completeness.
For this, we adopt a stricter threshold, $P_{\rm min}(\rm Quenched)=0.95$.
This choice results in sample purities of 35\%, 86\%, and 76\% at $z=0.0$, $0.3$, $0.60$, respectively, with a combined purity fraction of 61\%.

To evaluate the performance of these classification schemes, Fig.~\ref{fig:tng_quenched_selection} displays the true (left) and median-based (right) distribution of IllustrisTNG galaxies across the \avssfrval{9} versus \avssfrval{8} plane.
In the top row, all galaxies are colour-coded according to their probabilities of belonging to the retired, quenched, and ageing domains using an RGB palette.

The `balanced' classification approach, which maximises the $F$-score, is illustrated in the middle row.
Here, retired and ageing galaxies are shown in semi-transparent black to emphasise the spatial distribution of quenched systems.
As can be readily seen, this method misclassifies a significant fraction of ageing systems located below the main sequence (MS) locus, roughly centred at  $\logavssfrval{8,\,9}\approx-10$.
In addition, there is also a small fraction ($\simeq3\%$) of contaminants with very low values of \avssfrval{9}.

The classification that maximises sample purity is shown in the bottom row. This method significantly reduces contamination, with the remaining misclassified systems primarily consisting of ageing galaxies with moderate to low star-formation rates.
Overall, the selected population of quenched galaxies is now dominated by systems with higher values of \avssfrval{9}, implying a more vigorous star-forming activity in the past compared to the current rate.
These systems fall closer to the `post-starburst' definition: galaxies with strong stellar Balmer absorption, due to the preponderance of A-type stars, and very little to null nebular emission, indicative of the demise of O- and B-type stars, implying a quenching event that truncated star-formation after the burst of star-formation.
However, the biggest drawback of this classification is its low completeness, only reaching values close to 10\%.
Nevertheless, the vast cosmological volume probed by \Euclid will alleviate this problem by providing an immense wealth of galaxies.

\subsection{Model-driven classification \label{sec:kde}}

\begin{figure*}[htbp!]
        \includegraphics[width=\linewidth]{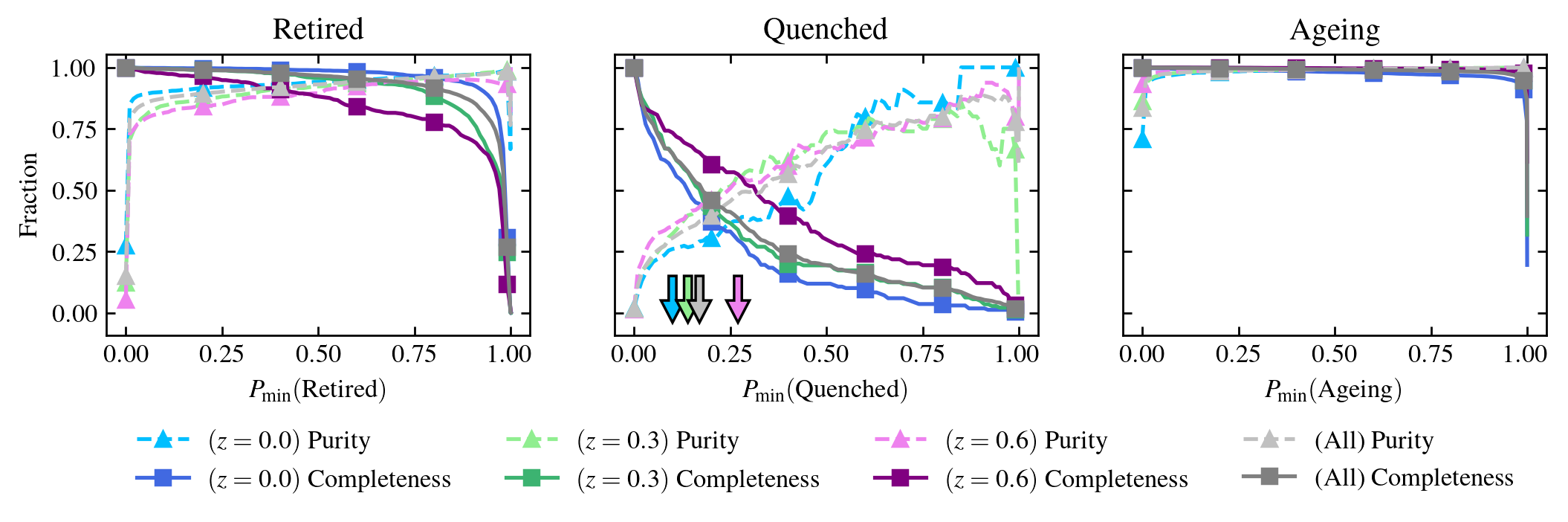}
        \caption{Same as Fig.~\ref{fig:tng_vanilla_classes}, using the classification presented in Sect.~\ref{sec:kde}.}
        \label{fig:tng_ad_classes}
\end{figure*}

\begin{figure}
        \includegraphics[width=\linewidth]{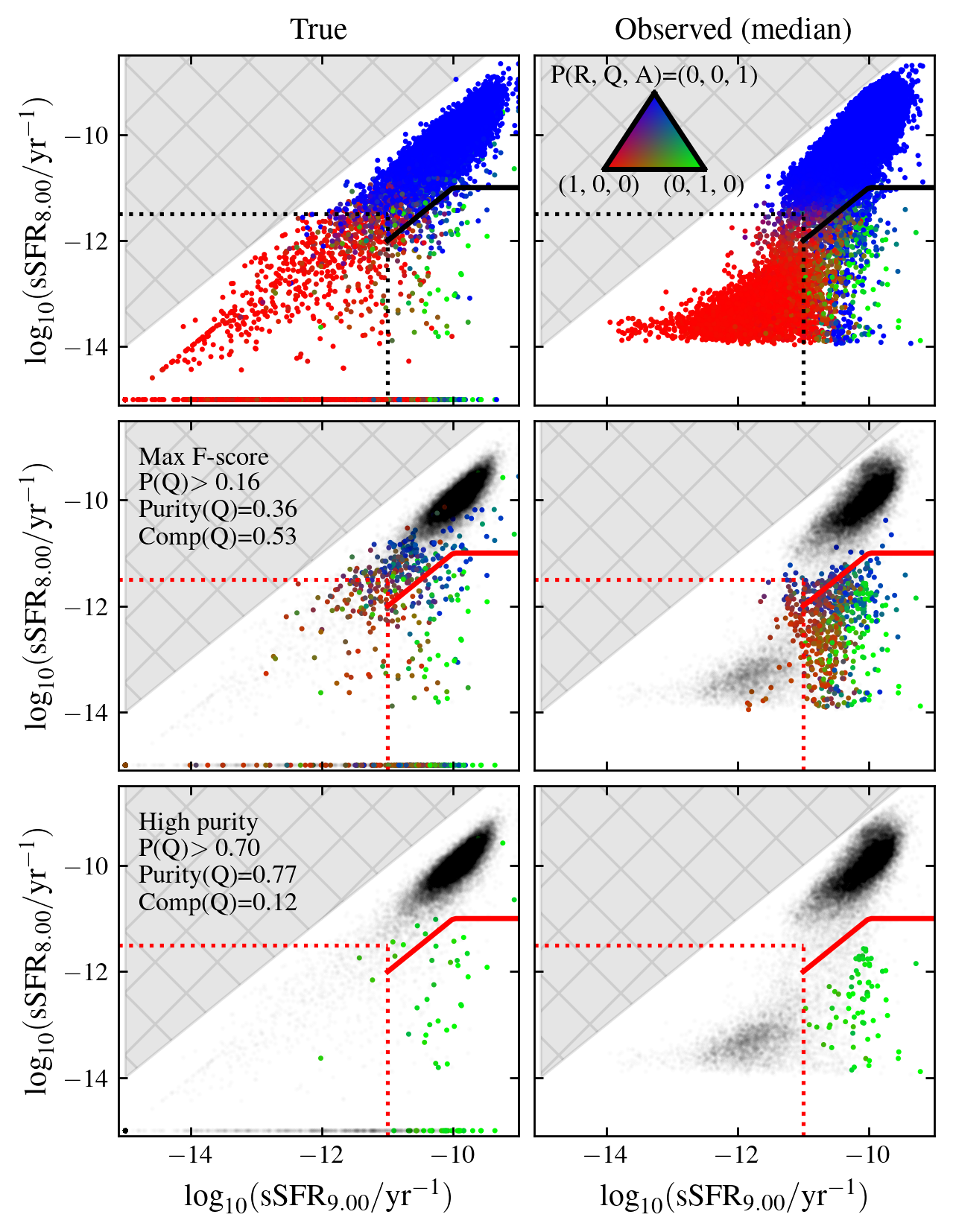}
        \caption{Same as Fig.~\ref{fig:tng_quenched_selection}, using the classification presented in Sect.~\ref{sec:kde}.}
        \label{fig:tng_quenched_kde_selection}
\end{figure}

The classification method presented in the previous section relies on a marginalised version of the posterior, constrained to the space defined by \avssfrval{9} and \avssfrval{8}.
While this approach is appealing due to its simplicity and ease of interpretation, the full posterior, $P(\avssfr | f)$, contains significantly richer information that can be used to refine the classification.
To this end, we have estimated the probability distribution, defined in terms of the 5th, 16th, 50th, 84th, and 95th percentiles of \avssfr, with $\log_{10}(\tau / {\rm yr})=\{8.00,\,8.48,\, 8.70,\, 9.00\}$, along with the correlation coefficients computed from the covariance matrix, $\rho\left(\avssfrval{\tau_1}, \avssfrval{\tau_2}\right)$, for the population of ageing, quenched and retired galaxies, respectively.
This distribution is more sensitive to subtle differences between the three populations, and can enhance the reliability of the classification.

To properly handle the 26-dimensional parameter space, denoted by the vector $\theta$, we have used a Gaussian KDE.
We denote the PDF associated with each AD class, $i$, as ${\rm KDE}_{i}(\theta)$.
To classify individual objects, we compute the probability of the galaxy being drawn from the ageing, quenched, or retired distribution.
The probability of belonging to each class is computed as
\begin{equation}
        P_{{\rm KDE},\,i}(\theta) = \frac{{\rm KDE}_{i}(\theta)}{{ \sum_{{i}\in \{\rm A,\,Q,\,R\}} {\rm KDE}_{i}(\theta) }}.
\end{equation}

The training process uses 80\% of the total sample, randomly selected, while the remaining 20\% forms the test sample to assess classification performance.
The kernel bandwidth is optimised to maximise the mean classification score across all three classes and fixed to 0.38.
Figure~\ref{fig:tng_ad_classes} presents the resulting purity and completeness scores for the test sample, demonstrating an improvement compared to the results in Fig.~\ref{fig:tng_vanilla_classes}.
For ageing and retired systems, purity and completeness exhibit reduced sensitivity to the threshold $P_{\rm min}$, and are systematically higher across all cuts. 
For quenched galaxies, the most significant improvement lies in the increase in purity across all redshift bins, nearly doubling for the $z=0$ sample.
However, completeness decreases consistently, falling below 30\% for $P_{\rm KDE,\,min}(\rm Quenched)\geq 0.5$.

Figure~\ref{fig:tng_quenched_kde_selection} illustrates the application of the KDE-based classification method, analogous to Fig.~\ref{fig:tng_quenched_selection}.
The number of false positives classified as quenched systems is significantly reduced, but this improvement comes at the cost of a smaller sample size and lower completeness.
For the KDE method, the threshold $P_{\rm KDE,\,min}(\rm Quenched)=0.16$ maximises the $F$-score, achieving purity and completeness values of 36\% and 53\%, respectively, which is a marginal improvement over the criteria proposed earlier.
However, as shown in the middle-row panels, many selected sources exhibit higher probabilities of being classified as ageing or retired systems.

On the other hand, this method excels the previous one when selecting a high-purity sample.
A stricter cut of $P_{\rm KDE,\,min}(\rm Quenched)\geq 0.7$, shown in the bottom panels, achieves a purity of 77\%, and completeness of 12\%.
This represents an improvement over the previous results, which yielded a purity and completeness of 61\% and 13\%, respectively.

While the KDE-based classification significantly enhances the identification of quenched systems, this improvement is strongly influenced by the training data.
The classification depends on the star-formation histories of IllustrisTNG galaxies, the recipes used to generate the synthetic observables (\texttt{PyPopStar}, \texttt{PST}) and the recovery of physical properties using (\texttt{BESTA}).
Careful validation is necessary when applying this method to other datasets to ensure its robustness.

\subsection{\label{sec:results_euclid}Ageing and quenching in Euclid}

\begin{figure*}[htbp!]
        \includegraphics[width=\linewidth]{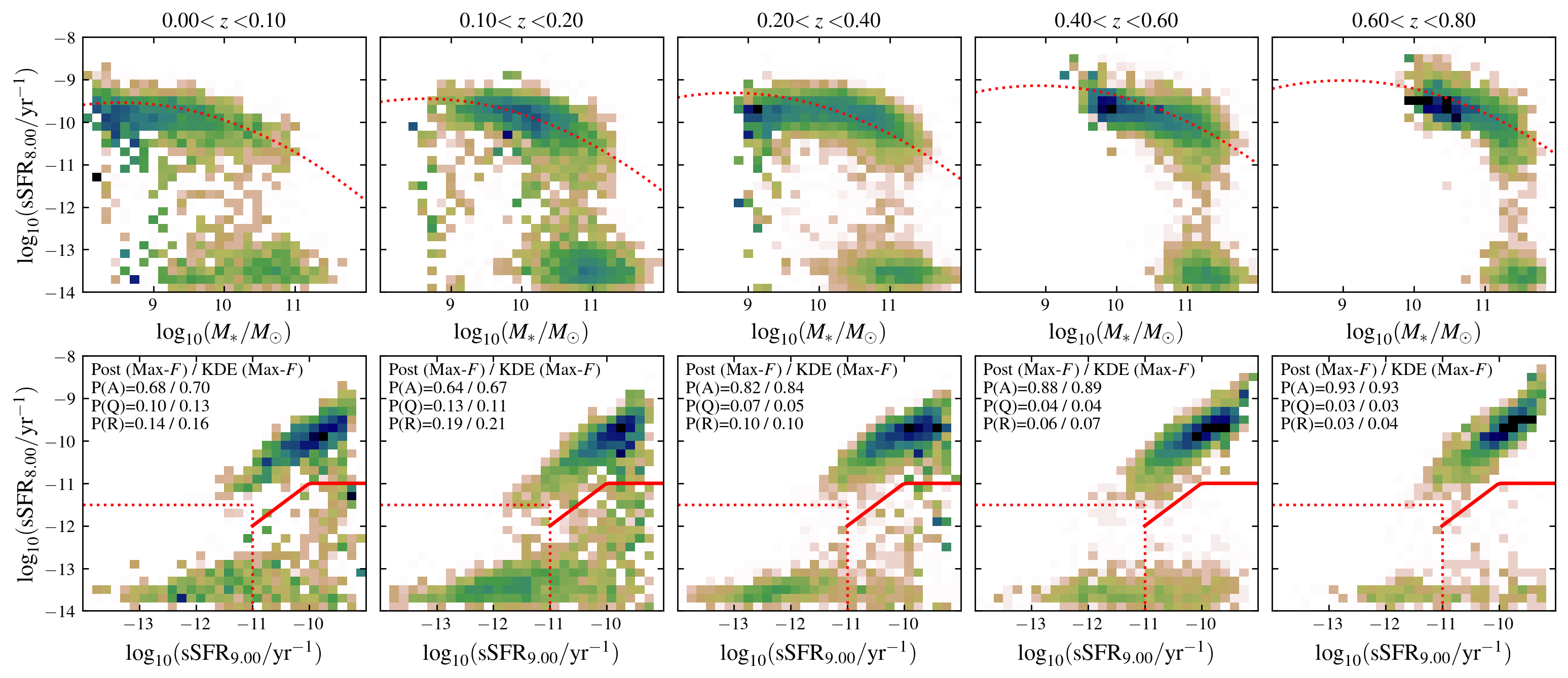}
        \caption{Top: Distribution of \Euclid galaxies across the stellar mass versus \avssfrval{8} plane. The dotted line denotes the definition of the SFMS by \citet{Popesso+23}. Bottom: Distribution of \Euclid galaxies across the \avssfrval{9} versus \avssfrval{8} plane. The dotted and solid lines denote the retired and quenched domain demarcation lines, respectively.}
        \label{fig:euc_ssfr_zbins}
\end{figure*}

Here we apply the classification methods described in the previous Sects.~\ref{sec:ad_prob} and \ref{sec:kde} to our sample of \Euclid galaxies, based on the results inferred using \besta (see Sect.~\ref{sec:bayes_inference}).
The distributions and fractions presented in this section are volume-corrected, accounting for the sample selection effects described in Appendix~\ref{appendix:sample_completeness}.

The top row of Fig.~\ref{fig:euc_ssfr_zbins} illustrates the distribution of galaxies across the plane defined by their inferred (median) total stellar mass and \avssfrval{8} across multiple redshift bins.
The location of galaxies along the star-forming main sequence aligns reasonably well with previously reported values in the literature \citep[e.g.,][]{Popesso+23} and the recent findings of \citet{Q1-SP031}.
In terms of mass completeness, dwarf galaxies with stellar masses below $10^9~\Msun$ are detected only in the lowest redshift bin ($z\leq0.1$).
Galaxies in the mass range ($10^{9} \leq \mstar / \,\Msun \leq 10^{10}$) are observed up to redshift of approximately $0.4$, while massive systems ($\mstar\geq 3\times10^{10}\,\Msun$) are detected over the whole redshift range.

In terms of star-forming activity, galaxies with stellar masses above $\mstar\gtrsim10^{10.5}\,\Msun$ generally exhibit very low \avssfrval{8} values, often corresponding to upper limits.
Conversely, the fraction of low-mass systems below the main sequence increases at lower redshifts.

The bottom row of Fig.~\ref{fig:euc_ssfr_zbins} presents the distribution of galaxies across the \avssfrval{9} versus \avssfrval{8} plane.
This parameter space offers a complementary view to the $\mstar$--$\avssfrval{8}$ plane, offering additional insights into the current evolutionary states of galaxies.
A well-defined sequence of ageing galaxies is observed, characterised by roughly constant \ssfr values over the past Gyr, ranging from $3\times 10^{-12}$ to $10^{-9}\,\rm yr^{-1}$, across all redshift bins.
Additionally, a distinct population of quenched galaxies, recently detached from the ageing sequence, is detected.
These two evolutionary pathways converge in the lower-left region of the diagram, forming a population of retired galaxies, with minimal star formation, whose light is completely dominated by old stellar populations.

The top-left corner of each panel indicates the volume-corrected fractions of ageing, quenched, and retired galaxies (see Fig.~\ref{fig:QuenchRet_completeness} for details about the sample completeness in terms of RGs), labelled using the posterior and KDE-based classification schemes with cuts that maximise the $F$-score (see Sects.~\ref{sec:quenched_selection} and \ref{sec:kde}).
The results from the two classifiers are consistent across all redshift bins.
Ageing galaxies consistently dominate the population, representing over $70\%$ in all bins.
Quenched galaxies become increasingly prevalent at lower redshifts, while the fraction of retired systems grows from $8$--$10$ up to $15$--$25\%$.
However, these trends are also affected by sample selection effects, since low-mass quenched and retired galaxies become more difficult to detect at high $z$ (see Sect.~\ref{sec:discussion_ad_fractions} for further discussion).

To contextualise these findings, we compare the fractions in the lowest redshift bin ($0.01<z<0.1$), where the sample is complete up to about $10^9\,\Msun$, with the results reported in \citetalias{Corcho-Caballero+23a}.
The authors estimated volume-corrected fractions of ageing, quenched, and retired galaxies, using both IllustrisTNG and the MaNGA survey \citep{MANGAoverview}, based the \ew versus \balbreak ageing diagram.
Table~\ref{tab:ad_fractions} summarises this comparison, listing the total fractions of these populations derived from the \Euclid and IllustrisTNG samples (the latter using the realisation at $z=0$).
We computed the fractions with both the Bayesian posterior and KDE-based classifiers, employing three criteria: the minimum probability threshold ($P_{\rm min}$) that maximises the $F$-score; the threshold maximizing purity; and the label with the highest probability.
This results in six fraction estimates for each sample.

\begin{table}
        \centering
        \caption{Fractions of retired (RG), quenched (QG), and ageing (AG) galaxies at $z<0.1$.
        }
        \label{tab:ad_fractions}
        \begin{tabular}{llll}
                \hline\hline
                Dataset and Method & F(RGs) & F(QGs) & F(AGs)\\[2pt]
                \hline\\[-8pt]
                \multicolumn{4}{c}{\Euclid ($z<0.1$)}\\[2pt]
                \hline\\[-8pt]
                Post. (max $F$-score) & 0.14 & 0.10 & 0.68 \\[2pt]
        Post. (high purity) & 0.14 & 0.04 & 0.68 \\[2pt]
        Post. (max prob.) & 0.14 & 0.17 & 0.68 \\[2pt]
        KDE (max $F$-score) & 0.19 & 0.08 & 0.72 \\[2pt]
        KDE (high purity) & 0.16 & 0.13 & 0.70 \\[2pt]
        KDE (max prob.) & 0.19 & 0.09 & 0.72 \\[2pt]
                \hline\\[-8pt]
                \multicolumn{4}{c}{IllustrisTNG ($z=0$)} \\[2pt]
                \hline\\[-8pt]
                Truth (true values) & 0.27 & 0.02 & 0.71 \\[2pt]
        Post. (max $F$-score) & 0.26 & 0.03 & 0.67 \\[2pt]
        Post. (high purity) & 0.26 & $<$0.01 & 0.67 \\[2pt]
        Post. (max prob.) & 0.26 & 0.06 & 0.67 \\[2pt]
        KDE (max $F$-score) & 0.27 & 0.03 & 0.70 \\[2pt]
        KDE (high purity) & 0.28 & $<$0.01 & 0.70 \\[2pt]
        KDE (max prob.) & 0.18 & 0.04 & 0.78 \\[2pt]
                \hline\\[-8pt]
                \multicolumn{4}{c}{\citetalias{Corcho-Caballero+23a} Results} \\[2pt]
                \hline\\[-8pt]
                IllustrisTNG & 0.23--0.25 & 0.11--0.14 & 0.61 \\[2pt]
                MaNGA & 0.13--0.15 & 0.08--0.10 & 0.72--0.74 \\[2pt]
        \hline
        \end{tabular}
    \tablefoot{The table lists estimated fractions for the \Euclid and IllustrisTNG samples, along with results from \citetalias{Corcho-Caballero+23a}. Fractions are estimated using Bayesian posterior and KDE-based classification schemes under three criteria: minimum probability threshold ($P_{\rm min}$) that maximises $F$-score, threshold maximizing purity, and label with the highest probability.  
    For IllustrisTNG, fractions are also provided using the true values of $\avssfrval{8}$ and $\avssfrval{9}$.}
\end{table}

Additionally, we can qualitatively compare our results with the star-forming and `quiescent'\footnote{Note that their definition of quiescent galaxies differs from ours; their quiescent population broadly corresponds to our QGs+RGs category.} fractions reported in \citet{Q1-SP031}, based on the colour-based classification criteria from\citet{Williams+09} and \citet{Ilbert+13}.
For $0.2<z<0.5$, $23\%$ galaxies are classified as quiescent, and the reminder as star-forming.
Our findings are in qualitative agreement, with QGs+RGs (AGs) making up 15--17\% (82--84\%) at $0.2<z<0.4$, and 10--11\% (88--89\%) in the $0.4<z<0.6$ bin.
At higher redshifts ($0.5<z<0.8$), \citet{Q1-SP031} find a quiescent fraction of $15\%$ ($85-86\%$ star-forming), whereas we estimate 6--7\% QGs+RGs and $93\%$ AGs, most likely driven by the lack of low-mass quenched and/or retired systems in our sample (as discussed in Appendix~\ref{appendix:sample_completeness}, our fractions are only representative above $10^{11}\,\Msun$ in this redshift bin; cf. Fig.~\ref{fig:euc_ad_frac_mass_zbins} below).

As noted earlier, the two classifiers (Bayesian posterior and KDE) produce consistent results, with more pronounced differences arising between classification criteria, particularly for ageing and retired fractions.
Overall, there is good qualitative agreement between this work and \citetalias{Corcho-Caballero+23a}.
However, when comparing results based on IllustrisTNG, the current analysis yields systematically higher fractions of ageing galaxies, while the fraction of quenched systems is reduced from 12\% to 3\%.

Beyond these methodological differences, selection effects and classification biases likely contribute to the observed discrepancies (see Sect.~\ref{sec:ad_prob} for a detailed discussion).
Despite these challenges, the comparison with observational data reveals significant agreement, with variations well within the expected systematic uncertainties.

\section{\label{sec:discussion}Discussion}

Understanding the formation mechanisms of galaxies dominated by old stellar populations, with little to no ongoing star formation, become a topic of considerable scientific interest in recent years \citep[e.g.,][]{Casado+15, Quai+18, Owers+19, Corcho-Caballero+21a, Corcho-Caballero+23a}.
The detection of such systems at high redshift challenges widely accepted formation models, exposing potential limitations in our cosmological framework \citep[e.g.,][]{Girelli+19, Merlin+19, Lovell+23, De_Lucia+24, Lagos+25}.
The classification of galaxies into categories such as passive, quiescent, quenched, or, following the nomenclature proposed in \citet{ Corcho-Caballero+21b, Corcho-Caballero+23a, Corcho-Caballero+23b}, retired (i.e. a heterogeneous population of galaxies influenced by secular, regulating, suppressing, and quenching processes), remains contentious.
This reflects not only the complexity of the varied and intertwined processes that regulate (and potentially halt) star formation, but also the challenges of inferring such low levels of star formation from observational data.

Quiescent galaxies are often identified by imposing sSFR thresholds over specific timescales \citep{Moustakas+13, Donnari+19, Tacchella+22}, as well as cuts using proxies of \avssfr such as broadband colours \citep[e.g.,][]{Peng+10, Schawinski+14, Weaver+23}.
These thresholds are typically calibrated to isolate systems below the star-forming main sequence, with some incorporating redshift evolution \citep[e.g.,][]{Tacchella+22}.
While this approach has provided valuable insights into galaxy evolution, such as trends in quiescent fractions and their dependence on stellar mass and environment \citep[e.g.,][]{Peng+10, Donnari+19, Weaver+23}, it often oversimplifies the transition to quiescence by focusing on single timescales and neglecting the diversity of star-formation histories.

To better understand the physical processes that regulate star formation, and quantify their relative contribution, one would need to use more sophisticated classifications than the usual quiescent versus star-forming scenario \citep[e.g.,][]{Casado+15, Moutard+16, Moutard+18, Quai+18, Owers+19, Iyer+20}.
In this work, we advocate for a more refined approach, that incorporates multiple timescales to classify galaxies based on their star-formation activity.
This methodology captures better the diversity of evolutionary pathways, distinguishing between recently quenched systems and those that have been evolving secularly for longer periods of time.

The discussion is organised as follows.
Section~\ref{sec:discussion_ad_fractions} investigates the dependence of galaxy populations on stellar mass in the \Euclid and IllustrisTNG samples in the nearby Universe ($z<0.1$).
Section~\ref{sec:discussion_redshift_evol} explores the evolution of these populations across redshift, highlighting trends within different mass bins.

\subsection{Ageing, quenched, and retired fractions at $z < 0.1$}\label{sec:discussion_ad_fractions}

\begin{figure*}
        \includegraphics[width=\linewidth]{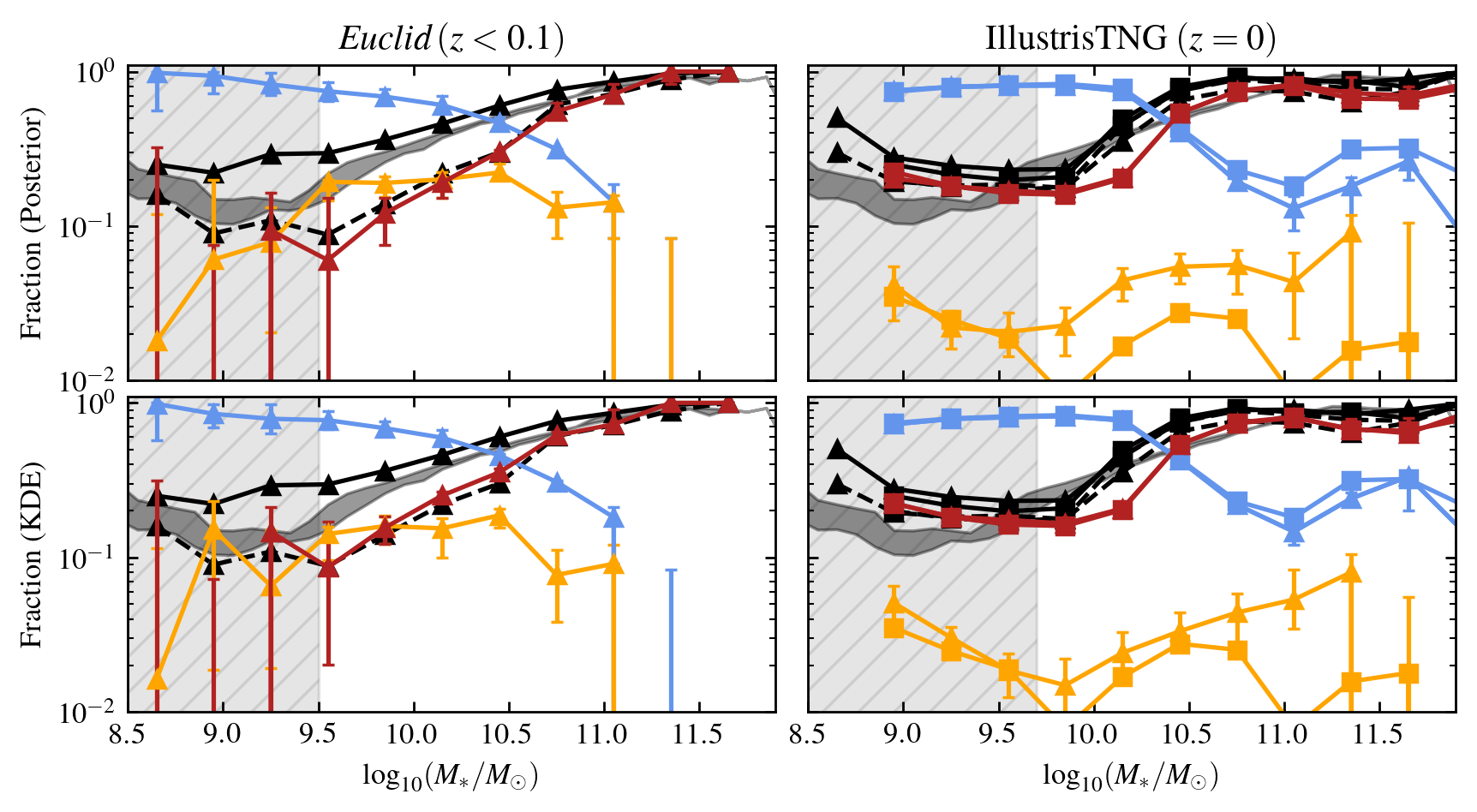}
        \caption{Fraction of galaxies in different classes as function of total stellar mass.
    Blue, orange, and red lines denote the fraction of ageing, quenched, and retired galaxies, respectively.
    Solid and dashed black lines denote the fraction of galaxies with values of \avssfrval{8} and \avssfrval{9} below $10^{-11}\,\rm yr^{-1}$.
    The grey-shaded region shows the fraction of galaxies with $\ssfr \leq 10^{-11}~\rm yr$ (estimated using nebular emission lines and optical photometry) from \citetalias{Corcho-Caballero+21a}.
    Top and bottom panels make use of the posterior- and KDE-based classification (see Sect.~\ref{sec:ad_prob},\,and~\ref{sec:kde}), respectively.
    Left panels show the fractions computed using a \Euclid sample of nearby galaxies ($z\leq0.1$), where the grey-hatched area denotes the mass-completeness limit (see Appendix~\ref{appendix:sample_completeness}). The error bars were estimated from 1000 random samples, each selecting 20\% of the total sample and computing the 5 and 95 percentiles.
    Right panels show the results from the IllustrisTNG sample at $z=0$, including the true values (square symbols), where
    the grey-hatched area denotes the retired mass resolution limit (see Sect~\ref{sec:classification}).}
        \label{fig:ad_frac_vs_mass_lowz}
\end{figure*}

This section examines the relationship between the fractions of galaxies classified as ageing, quenched, or retired see Sect.~\ref{sec:classification}) and their total stellar mass. Additionally, we compare our classification methodology to the traditional threshold-based approach that relies on a single average value of sSFR.

Figure~\ref{fig:ad_frac_vs_mass_lowz} presents the fraction of ageing, quenched, and retired galaxies as a function of stellar mass for nearby galaxies ($z<0.1$) in the \Euclid (left panels) and IllustrisTNG (right panels) samples.
The top and bottom panels compare the two classification schemes outlined in Sects.~\ref{sec:ad_prob} and \ref{sec:kde}, each optimised to maximise the $F$-score.
In addition, we include the fractions of galaxies below the star-formation rate threshold of $10^{-11}\,\rm yr^{-1}$ for \avssfrval{8} and \avssfrval{9} (solid and dashed black lines, respectively), and compare these with the results of \citetalias{Corcho-Caballero+21a} (grey shaded region), derived from SDSS and GAMA surveys.

The AD classification methods yield highly consistent fractions (coloured lines) for ageing and retired galaxies across both \Euclid and IllustrisTNG datasets.
However, the fraction of quenched galaxies shows greater sensitivity to the choice of classification method, though the overall trends with stellar mass remain qualitatively similar.
In the simulated data (right panels), the true fractions for each class are also included (square symbols).
As discussed earlier, the classification tends to slightly overestimate the fraction of quenched galaxies in IllustrisTNG, particularly at the high-mass end.

A comparison between the two datasets reveals some notable differences.
In the IllustrisTNG sample, retired galaxies dominate at stellar masses $\geq (2$--$3)\times10^{10}\,\rm \Msun$, comprising more than $75$\% of the population.
Below this threshold, ageing galaxies dominate, also accounting for over 75\% of the population, and the fraction remains nearly constant across these regimes.
Observational data from \Euclid, however, exhibits a smoother trend with stellar mass.
The fraction of ageing galaxies decreases steadily from approximately 90\% of the total sample at the low-mass end towards a negligible value at high masses.
In contrast, retired galaxies show a monotonic increase, reaching around 90\% for $\mstar \gtrsim 10^{11}\,\rm \Msun$.

The fraction of quenched galaxies in \Euclid remains approximately constant (10--20\%) between $3\times10^{9}$ to $10^{11}\,\Msun$, beyond which the fraction drops to null values.
The IllustrisTNG sample shows a more intricate structure: while a valley is evident at $\mstar \simeq 10^{10} \,\Msun$, an additional dip is observed at even higher stellar masses, around $\mstar \simeq 3\times 10^{11} \,\Msun$.
Note that, due numerical resolution effects (Sect.~\ref{sec:illustris}), the IllustrisTNG sample does not accurately probe the galaxy population below $\mstar \simeq 5\times10^{9} \,\Msun$, but we do observe a clear difference between the simulated and observed populations for $\mstar>5\times10^9\,\Msun$.
While most quiescent galaxies in \Euclid are classified as quenched, they belong to the retired class in IllustrisTNG.

As previously discussed in \citetalias{Corcho-Caballero+23b}, the choice of timescale for estimating the current star-formation rate significantly impacts the inferred fraction of passive galaxies.
For example, when using a longer timescale (e.g., \avssfrval{9}), it yields a steadily increasing fraction of passive galaxies with stellar mass, closely matching the definition of retired galaxies.
In contrast, shorter timescales such as \avssfrval{8}, result in a combined fraction that includes both quenched and retired populations, consistent with \citetalias{Corcho-Caballero+21a}.

The observed flat distribution of quenched galaxies suggests that the quenching mechanism(s), responsible for shutting down star formation on timescales of $\simeq500$ Myr \citepalias{Corcho-Caballero+23a}, are equally efficient at all stellar masses, across the range probed by the sample.
Conversely, the population of retired galaxies is primarily populated by massive systems, dominating the fraction of galaxies above $3\times10^{10}\, \Msun$, which can be interpreted in terms of an increase in the impact of internal, and slow, `regulating' or `suppression' processes such as AGN thermal feedback, virial shock heating, but also enhanced star formation efficiency \citep{Dekel+06, Peng+10, Tortora+10, Tortora+25}.
However, these interpretations remain qualitative.
A larger sample, more rigorous selection, and expanded analysis are needed to refine these conclusions and further elucidate the physical mechanisms governing galaxy quenching.
In summary, our classification provides a rich framework for studying and interpreting the evolutionary trends of galaxies in terms of their recent star-formation history.

\subsection{Redshift evolution up to $z=0.8$\label{sec:discussion_redshift_evol}}

\begin{figure*}[htbp!]
        \includegraphics[width=\linewidth]{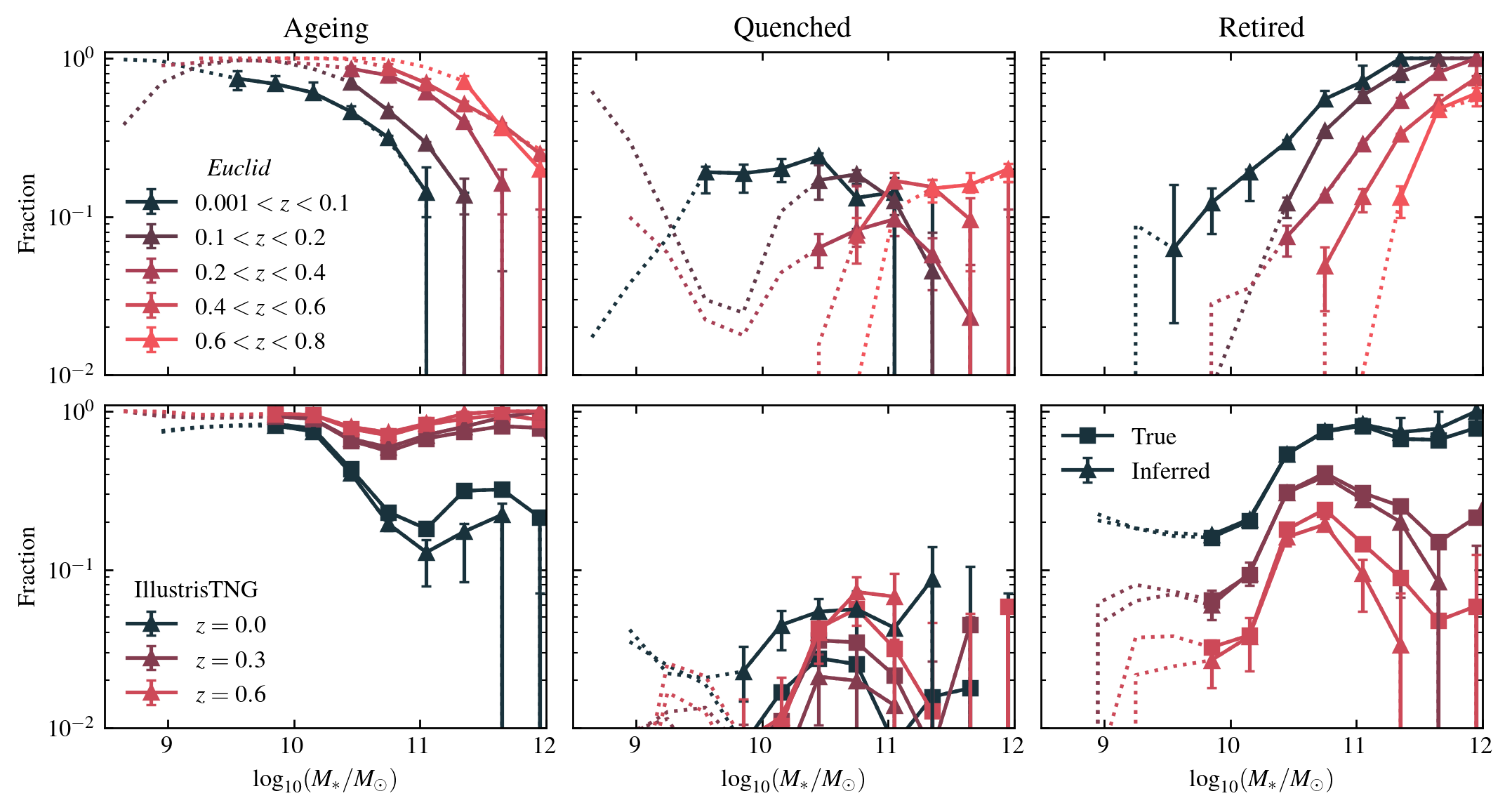}
        \caption{Fraction of ageing, quenched, and retired galaxies as function of stellar mass in redshift bins.
    Each row of panels shows fractions inferred using \besta for \Euclid (top) and IllustrisTNG (middle) samples, along with true fractions from IllustrisTNG (squares).
    The error bars were estimated from 1000 random samples, each selecting 20\% of the total sample and computing the 5 and 95 percentiles.
    In the top row, fractions for masses below the mass completeness limit (see Appendix~\ref{appendix:sample_completeness}) are shown as dotted lines.
    Similarly, in IllustrisTNG, masses below the retired resolution limit are represented as dotted lines.
    }
        \label{fig:euc_ad_frac_mass_zbins}
\end{figure*}

Following the results of the previous section, here we examine the evolution of the fraction of each galaxy population as a function of redshift and stellar mass.
Figure~\ref{fig:euc_ad_frac_mass_zbins} illustrates the fraction of ageing (left), quenched (middle), and retired (right) galaxies as a function of redshift, divided in bins of stellar mass.
The top and bottom rows display the results inferred using \besta for \Euclid and IllustrisTNG samples, including the fractions computed using the true labels of each IllustrisTNG galaxy.
The fraction of galaxies in each domain has been computed using the probability estimated from the posterior, adopting the criteria that maximises the $F$-score.
The results are rather insensitive to the choice of classifier (i.e. posterior- or KDE-based), presenting only a subtle change on the overall normalisation (see previous section).
For the inferred fractions, we have computed qualitative estimates of the Poisson errors by bootstrapping the distribution selecting 1000 random samples and computing the 5th, 50th, and 95th percentiles of the resulting distribution.

In general, the fractions of ageing and retired galaxies exhibit growth and decline, respectively, with increasing $z$ across all stellar mass ranges.
Notably, in \Euclid there is a strong correlation between the fraction of retired galaxies at a given redshift and stellar mass: more massive galaxies are significantly more likely to be in the retired domain compared to their lower-mass counterparts,in agreement with previous studies \citep[e.g.,][]{Moutard+16}.
In IllustrisTNG we observe that, although still present, the correlation with stellar mass is more complicated.
For example, in Fig.~\ref{fig:euc_ad_frac_mass_zbins} the fraction of retired galaxies presents a redshift-independent maximum at $3$--$5\times10^{10}\,\Msun$, after which the fraction decays.

Regarding the growth of the retired population with cosmic time, we find that the simulation predicts a roughly mass-independent logarithmic increase for $\mstar \leq 5\times10^{11}\,\Msun$, whereas in observational data, the fraction evolves more rapidly with decreasing stellar mass.
At low stellar masses, the growth of the retired population proceeds more slowly in the simulation compared to observations.
For \Euclid massive galaxies with $\mstar \geq 3\times10^{11}\,\Msun$, the retired population already comprised $
\gtrsim40\%$ of the total population at $z\simeq0.6$.
These findings are in qualitative agreement with the results presented in \citet{Q1-SP017}, where the authors studied the evolution of the fraction of passive galaxies as function of redshift and environment by using an time-dependent \ssfr threshold\footnote{Note that their definition of passive completely encompasses both quenched and retired galaxies, as well as the a fraction of ageing systems with mild star-formation levels.}.

Conversely, the simulation suggests that for many massive galaxies, the transition from the ageing to the retired state, via ageing or Quenching evolutionary channels, occurs at later cosmic times, predominantly within the last 3--4 Gyr.
This result is important as it might encode useful information for discriminating between the various physical processes responsible for regulating (or even quenching) star formation.
For example, the relatively high fraction of massive retired galaxies at $z\gtrsim0.8$ might support the idea of a merger-driven quenching scenario, while the simulation's preferred quenching mechanism is AGN-driven feedback \citep{Donnari+21}, which also manifests in the presence of a `characteristic mass' at $\simeq5\times10^{10}\,\Msun$.
Nevertheless, potential bias that might drive differences between both samples are observational selection effects, but also numerical issues due to the reduced size of the simulation box (about $100^3$ Mpc$^3$) and the backwards-modelling approach used to predict the fraction of IllustrisTNG galaxies at $z>0$ (see Sect.~\ref{sec:illustris}).

Let us now focus on the evolution of the fraction of quenched galaxies.
The population shows a weaker correlation with stellar mass with respect to the other two classes.
The fraction of quenched galaxies remains relatively constant, around $0.02$--$0.08$ in IllustrisTNG and $0.05$--$0.15$ in \Euclid.
Building on the previous discussion, we find that in IllustrisTNG most of the quenching occurs at intermediate masses, around $5\times10^{10}\,\Msun$, strongly suggesting that Quenching is the primary formation channel of the retired population.
In \Euclid, the distribution is less pronounced but we still find hints of a mode that evolves with redshift (i.e. more massive galaxies are more likely to have undergone quenching at earlier times).

To improve this analysis, a more careful treatment of the sample selection, as well as a more complete sample, is required.
Fortunately, future releases of \Euclid will provide an unprecedented wealth of data, that will facilitate this endeavour.

\subsection{Mass-size-metallicity relation\label{sec:mass_size}}

\begin{figure*}[htbp!]
        \includegraphics[width=\linewidth]{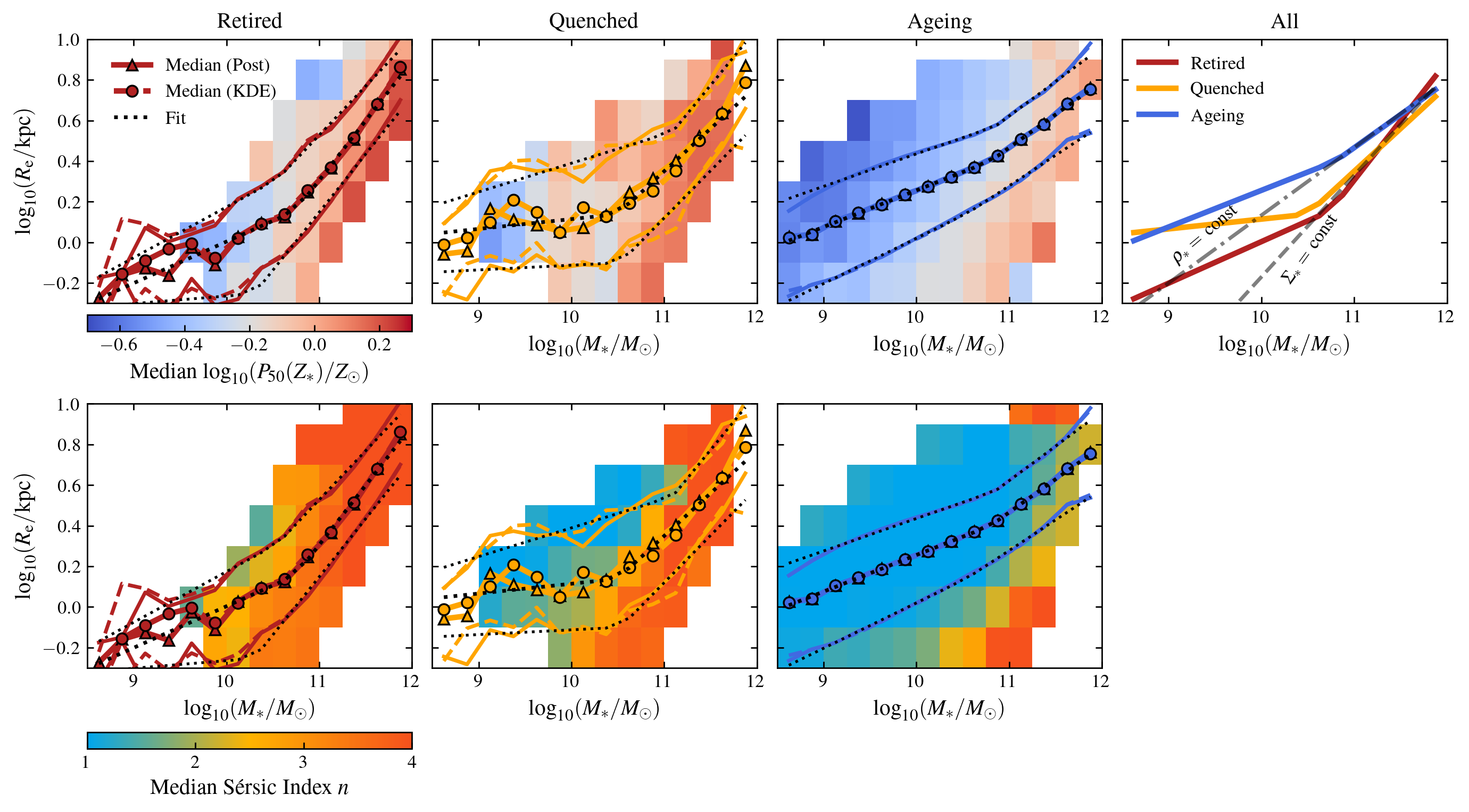}
        \caption{Stellar mass size relation for the full sample of \Euclid galaxies. From left to right, panels depict the distribution of retired, quenched, ageing, and the three combined classes across the parameter space.
    Solid and dashed lines denote the 16th, 50th, and 84th running percentiles computed using the Posterior- and KDE-based classifiers, respectively, adopting the criteria that maximises the $F$-score.
    The dotted dashed lines denotes the best-fit of Eq.~\eqref{eq:mass_size} to the percentiles.
    Top and bottom coloured background maps illustrate the median value per bin of the inferred median stellar metallicity and S\'ersic index, respectively, where only bins with more than ten galaxies have been considered.
        }
        \label{fig:mass_size_relation}
\end{figure*}

Figure~\ref{fig:mass_size_relation} shows the distribution of the full sample of \Euclid galaxies across the plane defined by their total stellar mass and effective radius, \re, used as a proxy for their physical size.
The values of \re are derived from S\'ersic surface brightness profile fits to VIS \Euclid images \citep[see][for details on the instrument and data-reduction pipeline]{EuclidSkyVIS, Q1-TP002}.
In addition to their excellent spatial resolution, the wide spectral coverage of the \IE band ($5500$--$9000$ \AA) provides a robust estimate of galaxy size, being less sensitive to mass-to-light ratio variations compared to optical $ugriz$ photometry.
From left to right, the panels present the running 16th, 50th, and 84th percentiles of \re as function of stellar mass, \mstar, for retired, quenched, and ageing galaxies, respectively.
The background colour maps shown in the top and bottom rows illustrate  the median stellar metallicity ($Z_\ast$) inferred by \besta and \IE-based S\'ersic index ($n$), respectively, restricted to bins with at least 10 galaxies.
To quantify the observed trends, we model the percentiles of $\re(\mstar)$ using a `broken' power-law model:
\begin{equation}
        \label{eq:mass_size}
        \re(\mstar) =\begin{cases}
                R_{\rm break} \left( \frac{\mstar}{M_{\rm break}} \right)^\alpha, & \,\mstar \leq M_{\rm break}, \\[1ex]
                R_{\rm break} \left( \frac{\mstar}{M_{\rm break}} \right)^\beta, & \,\mstar > M_{\rm break}, \\
        \end{cases}
\end{equation}
where $M_{\rm break}$ represents the characteristic mass where the logarithmic slope transitions between low-mass ($\alpha$) and high-mass ($\beta$) regimes, and $R_{\rm break}$ the predicted effective radius at $M_{\rm break}$.
Figure~\ref{fig:mass_size_relation} includes black dotted lines showing the best-fit models for each population percentile, with parameters summarised in Table~\ref{tab:mass_size_bestfit}. The rightmost panel compares the median relations across the three populations.

Retired galaxies exhibit the steepest correlation between \re and \mstar, maintaining consistent slopes ($\beta\simeq0.5$--$0.6$) across percentiles.
They also show systematically the highest values of $n$, aligning with previous results from the literature for early-type or bulge-dominated systems \citep[e.g.,][]{Sanchez-Almeida20}.
For $\mstar\gtrsim5\times 10^{10}~\Msun$, their distribution roughly follows a line of constant stellar surface density ($\Sigma_\ast\simeq7\times 10^3~\Msun\,\rm pc^{-2}$).
Below this mass, the slope flattens ($\alpha\simeq0.2$), resembling the distribution of ageing galaxies.
This turnover likely reflects a morphological transition from intermediate-mass red spirals to massive spheroidal systems.

\begin{table}
        \centering
        \caption{Best-fit values of the broken power-law model (Eq.~\ref{eq:mass_size}).}
        \label{tab:mass_size_bestfit}
        \begin{tabular}{ccccc}
                \hline\hline\\[-8pt]
                Percentile & $M_{\rm break}~[\Msun]$ & $\alpha$ & $\beta$ & $R_{\rm break}~[\rm kpc]$ \\[2pt]
                \hline\\[-8pt]
                \multicolumn{5}{c}{Retired} \\[2pt]
                \hline\\[-8pt]
                16 & $3.5\times10^{10}$ & 0.18 & 0.60 & 0.7 \\
                50 & $5.6\times10^{10}$ & 0.21 & 0.59 & 1.4 \\
                84 & $3.6\times10^{10}$ & 0.25 & 0.48 & 2.1 \\
                \hline\\[-8pt]
                \multicolumn{5}{c}{Quenched} \\[2pt]
                \hline\\[-8pt]
                16 & $3.3\times10^{10}$ & 0.02 & 0.46 & 0.8 \\
                50 & $3.2\times10^{10}$ & 0.05 & 0.43 & 1.4 \\
                84 & $2.4\times10^{10}$ & 0.09 & 0.29 & 2.5 \\
                \hline\\[-8pt]
                \multicolumn{5}{c}{Ageing} \\[2pt]
                \hline\\[-8pt]
                16 & $3.2\times10^{10}$ & 0.23 & 0.28 & 1.3 \\
                50 & $6.4\times10^{10}$ & 0.18 & 0.33 & 2.5 \\
                84 & $7.7\times10^{10}$ & 0.16 & 0.35 & 3.8\\
        \hline
        \end{tabular}
    \tablefoot{The results are presented in Fig.~\ref{fig:mass_size_relation}.}
\end{table}

The mass-size relation for ageing galaxies is flatter than that of retired galaxies across the whole mass range.
Below $(3$--$5) \times10^{10}~\Msun$, the median \re is parallel to, but offset from, the retired population by a factor of about $1.6$.
Their extended size and low $n$ indicates the presence of prominent discs, that allow ageing galaxies to sustain moderate-to-high levels of star-formation over large spatial (and temporal) scales.
Above $(3$--$5)\times10^{10}~\Msun$, the slope steepens to approximately $\re(\mstar) \propto \mstar^{1/3}$, implying constant volumetric stellar density ($\rho_\ast$) and characteristic time ($\sqrt{\re^3/G\mstar}$).

Quenched galaxies occupy an intermediate position between the other populations. 
Below around $3\times 10^{10}~\Msun$, \re shows little dependence on mass ($\alpha=0.04$--$0.013$), with characteristic sizes of $\re\sim1$--$1.6$ kpc, indicating higher density and compactness (illustrated by the higher values of $n$) than ageing galaxies.
At $3\times 10^{10}~\Msun$, the slope steepens ($\beta=0.36$--$0.47$) bridging the trends of ageing and retired populations.

Stellar metallicity (\zstar) complements our findings based on the morphology.
Retired galaxies show uniformly high \zstar, aligning with their compact morphology and dense stellar populations, tracing lines of constant $\Sigma_\ast$.
Ageing galaxies span a wide metallicity range.
They are more chemically primitive at lower masses, but we observe a tighter correlation between \zstar and $\Sigma_\ast$ than with \mstar.
Quenched galaxies exhibit once again an intermediate behaviour, now in terms of chemical enrichment.

In \citetalias{Corcho-Caballero+23a} and \citetalias{Corcho-Caballero+23b}, galaxy quenching was identified as a ubiquitous yet rare process, predominantly affecting low-mass systems in relatively dense environments.
Furthermore, \citetalias{Corcho-Caballero+23a} highlighted that present-day quenched galaxies often exhibit evidence of preceding `starburst' episodes before the cessation of star formation.
Combined with their more compact morphologies and richer chemical compositions compared to ageing galaxies, these findings point to environmental quenching mechanisms \citep[e.g.,][]{Wetzel+13}, such as ram-pressure stripping, as the dominant mechanism for shutting down star-formation in the local Universe.
The results of this work are consistent with these conclusions, further strengthening the observed links between morphological and chemical properties across the retired, quenched, and ageing galaxy populations.
A natural extension of this analysis would be to apply our method separately to centrals and satellites at fixed stellar mass, potentially revealing distinct quenching timescales and underlying processes in the two populations.

\section{\label{sec:conclusions}Summary and conclusions}

In this study, we have analysed the SFHs of a sample of galaxies from \Euclid Q1 \citep[][]{Q1-TP001} that span the redshift range $0 < z <0.8$.
Building upon previous work by \citet{Corcho-Caballero+23a,Corcho-Caballero+23b}, we developed a probabilistic classification framework based on the inferred SFH that leveraged two estimates of the average specific star-formation rate (\avssfr) measured over distinct timescales ($\tau=\{10^8,\ 10^9\}$ yr).
This framework classifies galaxies into three evolutionary classes: ageing (undergoing slow evolution), quenched (recently halted star-formation due to a quenching episode), and retired (former ageing or quenched systems, currently dominated by old stellar populations with minimal star formation).

To test the limits of our classification scheme, we also generated synthetic observations of galaxies from the IllustrisTNG simulation at multiple redshifts.
We inferred the SFHs by fitting optical-to-near-infrared photometry across $ugriz\YE\JE\HE$ bands using a Bayesian approach that captures the full posterior probability distribution (Sect.~\ref{sec:bayes_inference}).

We introduced two classification methods:
\begin{itemize}
        \item Probabilistic classification: This estimates the likelihood of a galaxy belonging to each class by integrating the posterior probability distribution, in terms of \avssfrval{8} and \avssfrval{9}, over the regions that delimit each class domain (Sect.~\ref{sec:classification}).
        \item Model-driven classification: Leveraging IllustrisTNG, this approach optimises sample purity and completeness for each class by exploring the parameter space including additional information from the posterior distribution (Sect.~\ref{sec:kde}).
\end{itemize}

Applying these methods to the \Euclid galaxy sample, we estimated the fractions of ageing, quenched, and retired galaxies at low redshift to be approximately 68--72\%, 8--17\%, and 14--19\%, respectively.
These findings align with the results of \citet{Corcho-Caballero+23a}, which were based on spectroscopic data, suggesting that photometric classifications can achieve similar reliability.

We studied the fraction of ageing, quenched, and retired galaxies as a function of stellar mass.
Our results suggest that, at the low mass-end, the number of quenched galaxies and retired systems is approximately the same.
Conversely, the fraction of retired galaxies, which shows a smooth increasing function with stellar mass, exceeds the fraction of quenched systems by a significant factor at masses $\gtrsim 3\times 10^{10}\,\Msun$. 

We then explored the evolution of the fraction of each class as a function of redshift in different stellar mass bins.
As expected, the fraction of ageing galaxies increases with increasing redshift, whereas the fraction of retired objects presents a strong dependence on stellar mass and redshift: more massive systems become retired earlier as compared to their low-mass counterparts, in line with the `downsizing' picture \citep{Cowie+96} and the results reported in \citet{Q1-SP031}.
In contrast, the fraction of quenched galaxies shows mild trends with mass or redshift, and ranges between 5\% and 15\%.
Our results show tentative evidence in favour of a scenario where more massive galaxies usually undergo quenching episodes at earlier times with respect to their low-mass counterparts.

We analysed the differences, in terms of the mass-size-metallicity relation, between ageing, quenched, and retired \Euclid galaxies.
Ageing galaxies are consistent with a population of late-type galaxies typically dominated by disc morphologies and low stellar metallicities.
Retired objects follow a tight sequence in terms of stellar mass and effective radius that matches with that of early-type objects.
Above $\mstar \simeq 3\times 10^{10}~\Msun$, they arrange along a sequence consistent with constant stellar mass surface density ($\Sigma_\ast\simeq7\times 10^3~\Msun\,\rm pc^{-2}$).
Finally, quenched galaxies appear as an intermediate population that is composed of relatively compact and more chemically evolved systems compared to their ageing counterparts at a given stellar mass.
This characterisation of ageing, quenched, and retired galaxies is in excellent agreement with previous results from \citet{Corcho-Caballero+23b}.

These results provide promising directions for understanding the physical mechanisms that drive quenching in galaxies.
Nonetheless, the heterogeneous sample used in this work might suffer from selection effects and more careful selections are needed to extract more robust conclusions.
Fortunately, our study represents a very small region of the sky ($\approx63$ deg$^2$) compared to the expected area surveyed by the \Euclid mission (about $\num{14000}$ deg$^2$).
The combination of \Euclid photometric data with either \Euclid grism spectra and/or ancillary information will enable a far more comprehensive analysis, and underscores \Euclid's transformative potential in unravelling galaxy evolution on cosmological scales.

\begin{acknowledgements}
  \AckEC
  Based on data from UNIONS, a scientific collaboration using three Hawaii-based telescopes: CFHT, Pan-STARRS and Subaru \url{www.skysurvey.cc}\,.
  Based on data from the Dark Energy Camera (DECam) on the Blanco 4-m Telescope at CTIO in Chile \url{https://www.darkenergysurvey.org}\,.
\end{acknowledgements}

%
%
\bibliographystyle{aa}
\bibliography{aanda}

\begin{appendix}

\clearpage

\section{Sample completeness and volume corrections}
\label{appendix:sample_completeness}

This appendix details the methodology used to assess the completeness of our sample and apply volume corrections to account for selection effects.
Our approach follows the procedure outlined in \citet{Favole+24}.

For each spectroscopic survey contributing to our sample, we estimate the maximum distance at which a galaxy would be detected by considering both a filter-dependent flux limit and the imposed signal-to-noise ratio (S/N) threshold.

To determine the maximum apparent magnitude, $m_{\rm lim}$, for each photometric band in each survey, we compute the 95th percentile of the apparent magnitude in redshift bins.
The final $m_{\rm lim}$ is obtained as the median of these percentiles.
Table~\ref{tab:ap_mag_lim} summarises the magnitude limits for each survey.

Galaxies with apparent magnitudes fainter than these limits are discarded.
The maximum distance at which a source would still be detected in each band is then given by the maximum distance modulus $\mu(z_{\lim})$, at which a source would be detected on each band as
\begin{equation}
\label{eq:max_dist_mod}
    \mu(z_{\lim}) = m_{\rm lim} - M - K(z_{\lim}),
\end{equation}
where $M$ is the absolute magnitude and $K(z_{\lim})$ is the K-correction at the maximum redshift.
The values of $K(z)$ for each source are estimated from the photometric properties of the stellar populations inferred using \besta (see Sect.~\ref{sec:bayes_inference}).

We also account for the selection based on S/N by computing the minimum detectable flux for a given noise level
\begin{equation}
    f_{\nu,\, \rm lim} = {\rm (S/N)_{min}}\,\sigma(f_{\nu,\,\rm obs}),
\end{equation}
which translates into a limiting apparent magnitude
\begin{equation}
    m_{\lim} = -2.5\log_{10}\left({\rm (S/N)_{lim}}\right) - 2.5\log_{10}\left(\sigma\right).
\end{equation}
By combining this with Eq.~\eqref{eq:max_dist_mod}, we estimate a second maximum detection distance.

For each galaxy in our sample, we compute two maximum redshifts for each available photometric band, with the final limiting redshift, $z_{\rm max}$, taken as the minimum value across all bands.
To estimate the statistical contribution of galaxies in a given redshift bin $(z_{\rm low}, z_{\rm up})$, we use the $1/V_{\rm max}$ formalism \citep{Schmidt68}
\begin{equation}
    V_{\rm max} = \frac{A}{3} \left(d_{\rm L}[{\rm min}(z_{\rm max}, z_{\rm up})]^3 - d_{\rm L}[z_{\rm low}]^3\right),
\end{equation}
where $A$ is the survey area in steradians and $d_{\rm L}(z)$ is the luminosity distance.

\begin{figure*}
	\includegraphics[width=\linewidth]{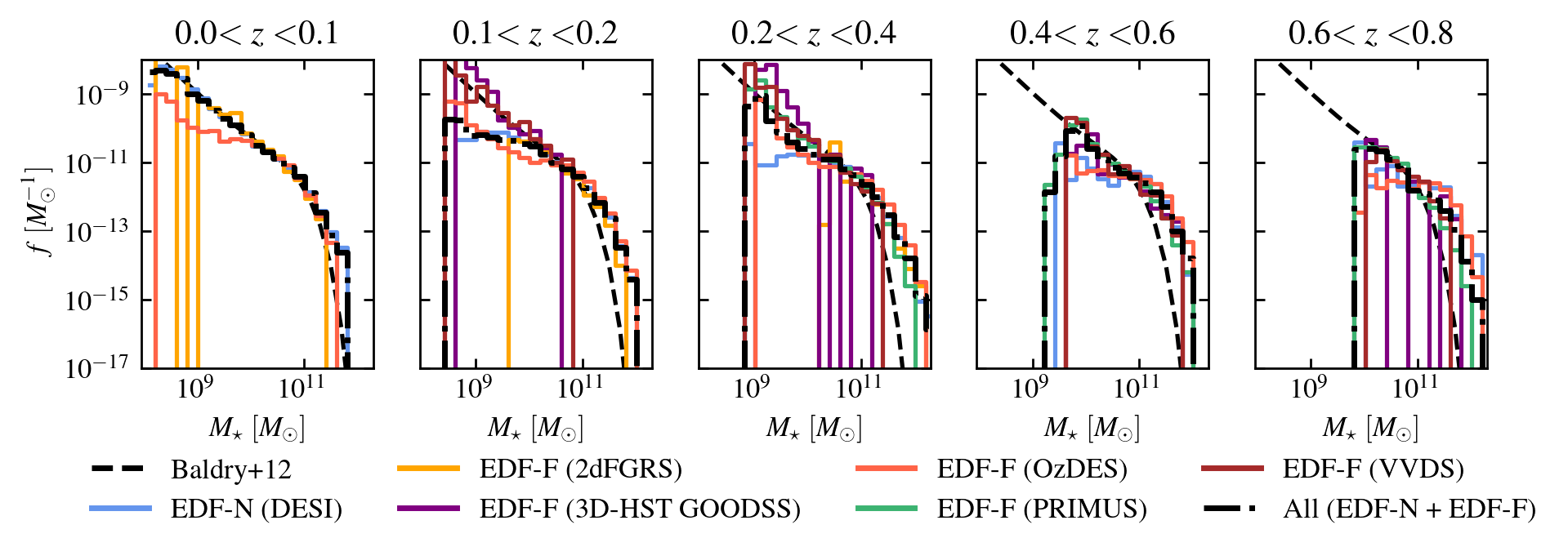}
	\caption{Stellar mass function in different redshift bins. To account for the different effective areas of the surveys, the distributions are normalised between the mass limits of $3\times10^{10}~\Msun$, and $10^{12}~\Msun$.
		Dashed lines denote the double-Schechter fit (restricted to $z\leq0.06$) from \citet{Baldry+12}. Coloured lines denote the resulting mass function from the different surveys that compose the sample. Black dashed-dotted illustrate the resulting mass function of the total sample.}
	\label{fig:mass_functions}
\end{figure*}

Since our primary interest lies in the relative contributions of different galaxy populations rather than their absolute number densities, we normalise the survey areas relative to the stellar mass function from \citet{Baldry+12}.
The resulting stellar mass functions in various redshift bins, derived from individual surveys (coloured lines) and the combined dataset (black dashed lines), are shown in Fig.~\ref{fig:mass_functions}.

\begin{figure}
    \includegraphics[width=\linewidth]{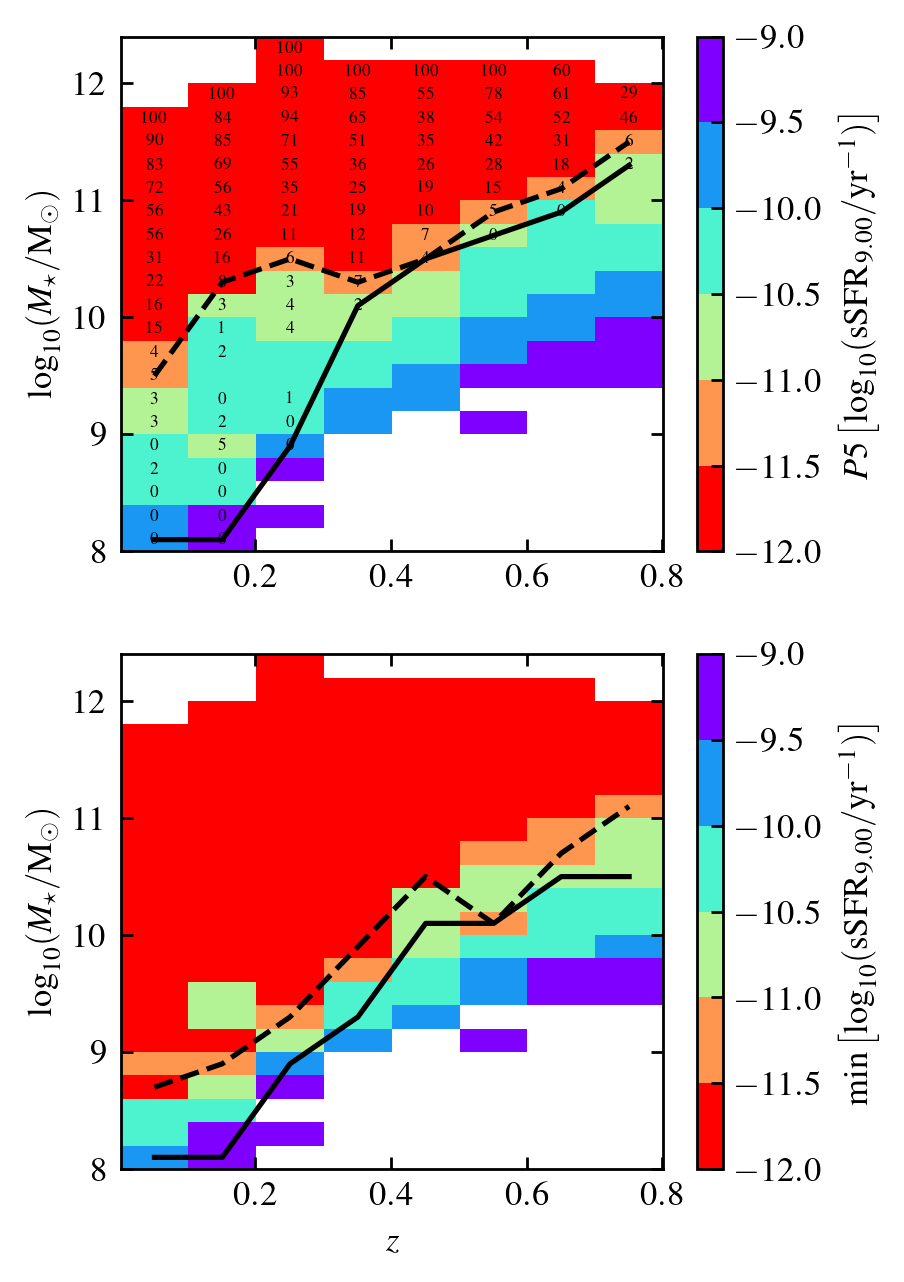}
    \caption{
    Completeness of retired galaxies in terms of $z$, and \mstar.
    The coloured map illustrates the 5th percentile of \avssfrval{9} on each $z$--\mstar bin.
    Each bin includes the (uncorrected) fraction of retired galaxies with respect to the total, based on the classification that maximises the $F$-score.
    The dashed and solid lines denote the mass completeness limit above which retired galaxies are detected. Each line tracks the lower envelope that satisfies $\avssfrval{9}<10^{-11}\,\rm yr^{-1}$ and $\avssfrval{8}<3\times10^{-11}\,\rm yr^{-1}$, based on the 5th and 1st percentiles, respectively.
    }
    \label{fig:QuenchRet_completeness}
\end{figure}

Although the $V_{\rm max}$ correction accounts for selection effects, it remains effective only if a fraction of the missing galaxy population is still detectable.
Figure~\ref{fig:QuenchRet_completeness} illustrates the completeness limits for quenched and retired galaxies as a function of redshift and stellar mass.

These limits are estimated using the 5th percentile and the minimum value of \avssfrval{8} and \avssfrval{9} in each redshift–stellar mass bin. The bottom panel of Figure~\ref{fig:QuenchRet_completeness} provides an absolute detection limit for each galaxy class, indicating the regime where volume corrections are still valid. The top panel presents a more conservative threshold.

\begin{table*}[hbt!]
	\centering
	\caption{Adopted apparent magnitude flux limit per survey and photometric band.}
	\label{tab:ap_mag_lim}
	\begin{tabular}{ccccccc}
		\hline\hline\\[-8pt]
		Band & DESI EDR & 2dFGRS & 3D-HST & OzDES & PRIMUS & VVDS \\[2pt]
		\hline\\[-8pt]
		$u$ & 22.48 & -- & -- & -- & -- & -- \\
		$g$ & 22.19 & 19.50 & 23.41 & 23.36 & 23.50 & 23.43 \\
		$r$ & 21.65 & 18.76 & 22.39 & 22.00 & 23.43 & 22.54 \\
		$i$ & 21.37 & 18.42 & 22.06 & 21.60 & 22.06 & 22.13 \\
		$z$ & 21.46 & 18.28 & 21.83 & 21.40 & 21.89 & 21.95 \\
		\hline\\[-8pt]
		No. sources & 9697 & 845 & 116 & 5471 & 9040 & 303 \\
		\hline
	\end{tabular}
\end{table*}

\FloatBarrier

\clearpage
\section{SFH inference performance as function of redshift\label{appendix:sfh_inference}}

\begin{figure*}[htb!]
	\includegraphics[width=\linewidth]{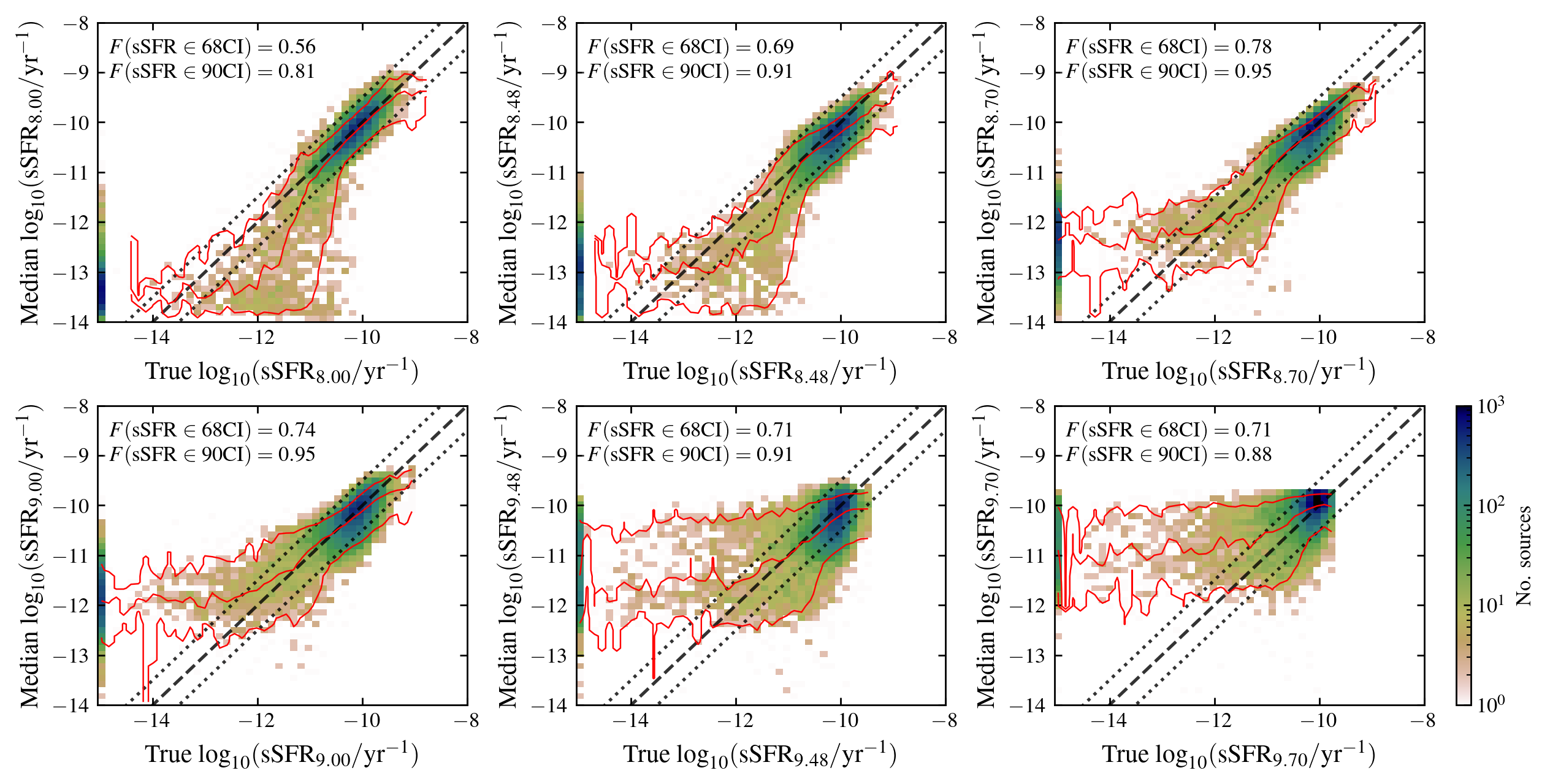}
	\caption{Same as Fig.~\ref{fig:illustris_ssfr_recovered_values} but restricted to the estimates at $z=0$.}
	\label{fig:illustris_ssfr_recovered_values_z00}
\end{figure*}

\begin{figure*}[htb!]
        \includegraphics[width=\linewidth]{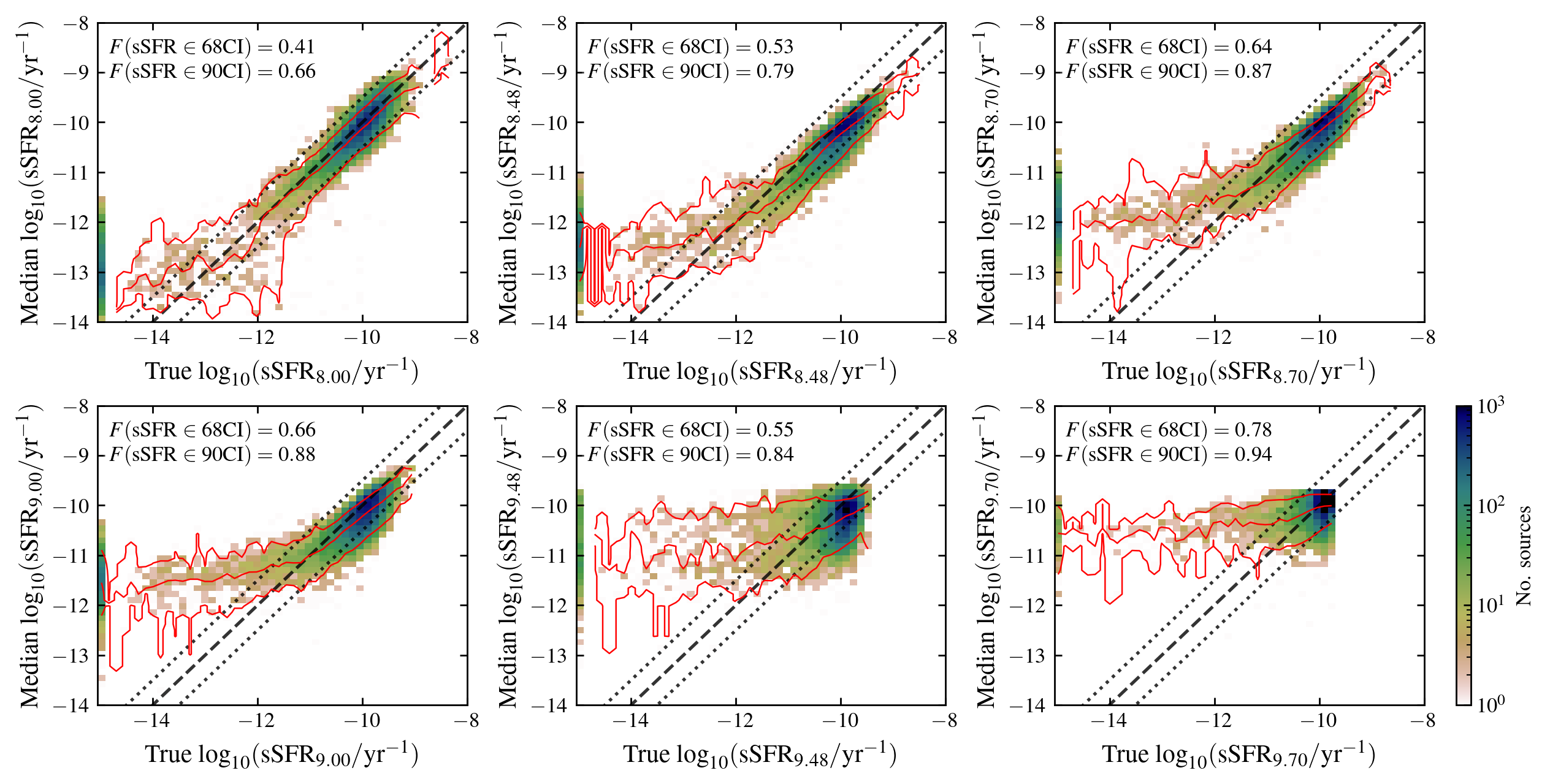}
        \caption{Same as Fig.~\ref{fig:illustris_ssfr_recovered_values} but restricted to the estimates at $z=0.3$.}
        \label{fig:illustris_ssfr_recovered_values_z03}
\end{figure*}

\begin{figure*}[htb!]
        \includegraphics[width=\linewidth]{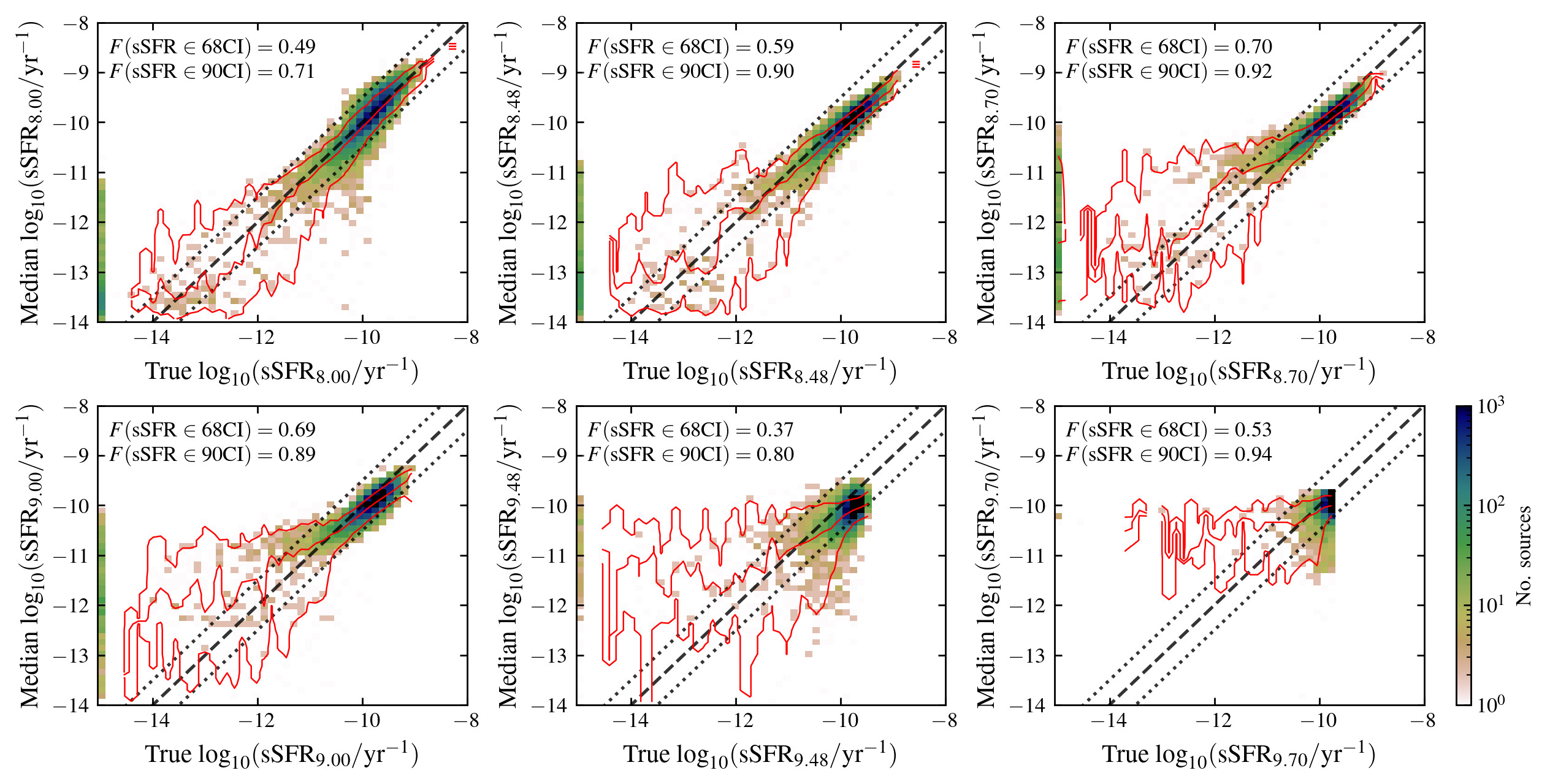}
        \caption{Same as Fig.~\ref{fig:illustris_ssfr_recovered_values} but restricted to the estimates at $z=0.6$.}
        \label{fig:illustris_ssfr_recovered_values_z06}
    \label{LastPage}
\end{figure*}

In this appendix, we present the results obtained from fitting the IllustrisTNG $ugriz\YE\JE\HE$ photometry with \besta as function of their observed redshift.
Figures~\ref{fig:illustris_ssfr_recovered_values_z00},~\ref{fig:illustris_ssfr_recovered_values_z03}, and \ref{fig:illustris_ssfr_recovered_values_z06} show the true input values of \avssfr versus the median value recovered by \besta for the realisations at $z=0.0,\,0.3$, and $0.6$, respectively, complementing Fig.~\ref{fig:illustris_ssfr_recovered_values}.

The effect of redshifting the SED of our sample of galaxies has a significant impact on the performance of the recovery of the values of \avssfr.
From $z=0$ to $z=0.6$, we find a progressive improvement in the performance of the fits illustrated by the reduction of the scatter along the one-to-one line.
The number of catastrophic outliers also seems to decrease, especially for $\tau\geq 10^9\,\rm yr$.
On the other hand, the fraction of sources whose input value of \avssfr is within the 68 and 90\% credible intervals decreases in most of the cases with respect to the realisation at $z=0$, implying that the estimate of the posterior probability distribution might be too sharp and the uncertainties are underestimated.

\FloatBarrier

\clearpage

\section{Comparison with alternative classification schemes\label{appendix:class_comparison}}

In this appendix, we present a direct comparison between our classification scheme (Sect.~\ref{sec:classification}) and the one proposed by \citet{Moutard+18}, which is based on the rest-frame $(NUV - r)$ versus $(r - K_s)$ colour-colour diagram \citep[hereafter $NUVrK$; see also][]{Arnouts+13, Moutard+16}.
The $NUVrK$ plane is used to distinguish between three primary galaxy populations -- ``star-forming'' (SF), ``green-valley'' (GV), and ``quiescent'' (Q) -- with each group further subdivided into ``young'' (Y) or ``old'' (O) categories based on their $r - K_s$ colour, yielding six distinct classes.

\begin{figure*}
	\centering
	\label{fig:classification_comparison}
	\includegraphics[width=0.49\linewidth]{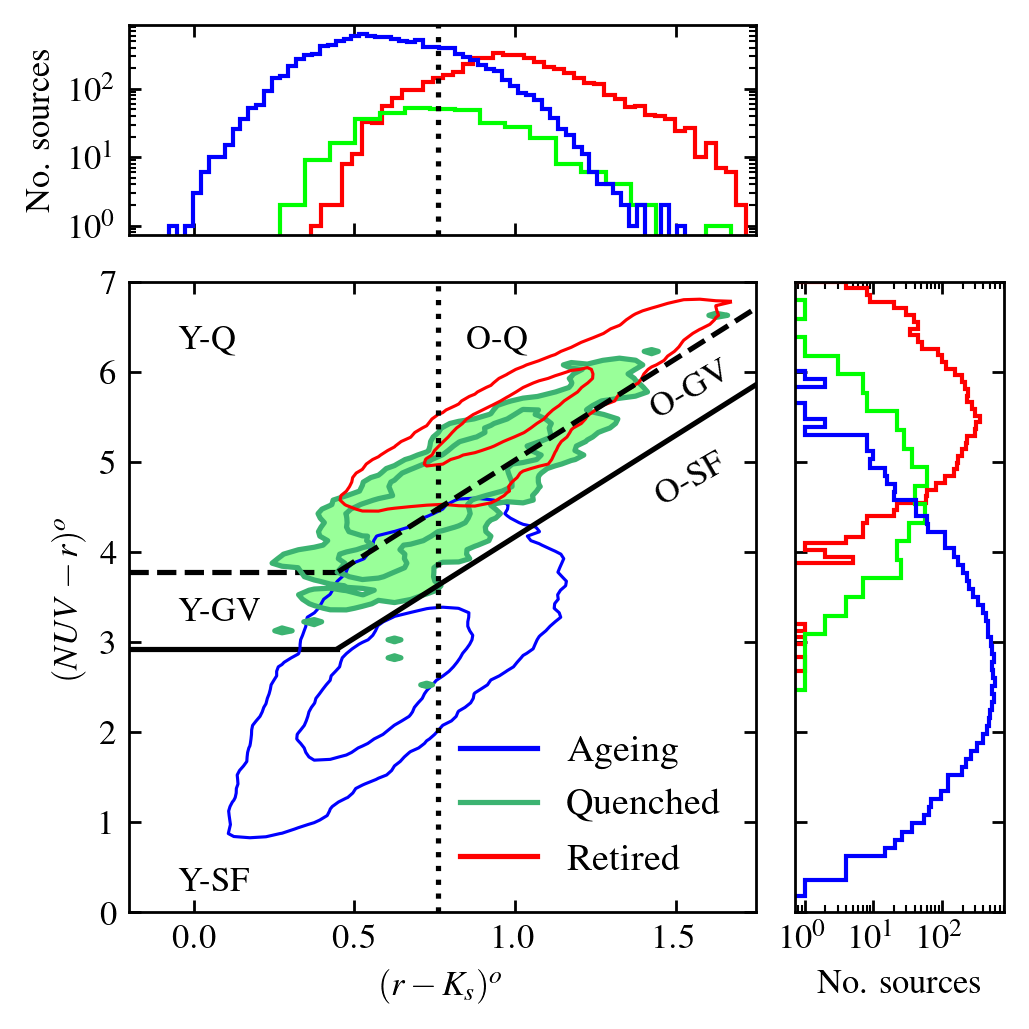}
	\includegraphics[width=0.49\linewidth]{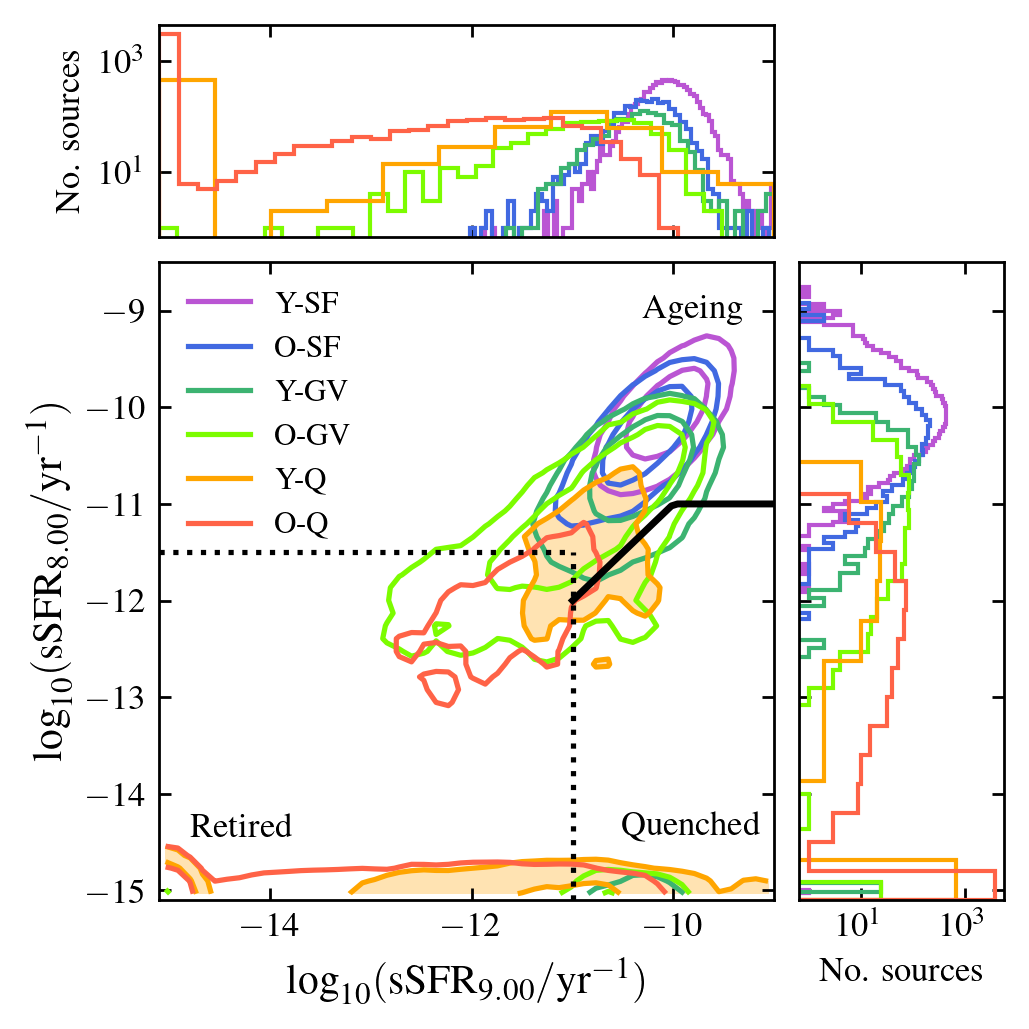}
	\caption{
		Left: Distribution of IllustrisTNG galaxies across the $NUVrK$ colour-colour diagram, coloured according to the classification defined in Sect.~\ref{sec:classification}; contours enclosing 68\% and 95\% of the ageing and retired galaxies are displayed as red and blue lines, respectively, whereas the quenched population is also highlighted by a green shaded area.
		Black lines indicate the demarcation criteria adopted by \citet{Moutard+18}.
		Vertical and horizontal histograms show the marginal distributions of each class in terms of $(NUV - r)$ and $(r - K_s)$, respectively.
		Right: Same galaxy sample shown in the \avssfrval{8} versus \avssfrval{9} plane, with colours corresponding to the six-class scheme of \citet{Moutard+18}; young (Y) / old (O), quiescent (Q) / green valley (GV) / and star-forming (SF).
		The yellow shaded area highlights the young quiescent (Y-Q) population.
		Vertical and horizontal histograms indicate the marginal distributions in \avssfrval{8} and \avssfrval{9}, respectively.
	}
\end{figure*}

To enable a direct comparison, we use synthetic rest-frame photometry from the IllustrisTNG simulation at $z=0$, using the CFHT $r$ band alongside GALEX $NUV$ \citep{Martin+05} and 2MASS $K_s$ \citep{Skrutskie+06} filters.
The left panel of Fig.~\ref{fig:classification_comparison} shows the distribution of our ageing (blue), quenched (green), and retired (red) galaxies across the $NUVrK$ diagram.
Black lines indicate the demarcation criteria from \citet{Moutard+18}.

In this diagram, retired galaxies predominantly lie above the quiescent boundary and are most frequently classified as ``old quiescent'' (84\%).
Ageing galaxies are mainly located within the star-forming region and are typically labelled as ``young star-forming'' (67\%), although a notable fraction (13\%) extends into the green-valley regime. 
Quenched galaxies span both the GV and quiescent regions, with a broad range in $r - K_s$ colour: 16\% are classified as Y-GV, 15\% as O-GV, as 33\% Y-Q, and 36\%  as O-Q.
Notably, the number of quenched galaxies dominates over retired systems in the range $r - K_s \lesssim 0.5$.

The right panel of Fig.~\ref{fig:classification_comparison} presents the distribution of the six $NUVrK$-based classes across the \avssfrval{8} versus \avssfrval{9} plane used in our classification.
The ageing domain encompasses the entirety of the SF population (both Y and O), the majority of GV galaxies (94\% of Y-GV and 70\% of O-GV), and small subset of the quiescent class (6\% of Y-Q and 1\% of O-Q).
The quenched region contains 5\%, 8\%, 15\%, and 3\% of the Y-GV, O-GV, Y-Q, and O-Q populations, respectively.
Finally, the retired domain includes 1\% of Y-GV, 22\% of O-GV, 79\% of Y-Q, and 96\% of O-Q galaxies.

\end{appendix}

\end{document}